\documentclass[%
 reprint,
nofootinbib,
 amsmath,amssymb,
 aps,
]{revtex4-2}
\usepackage{MnSymbol} 
\usepackage{graphicx}
\usepackage{dcolumn}
\usepackage{bm}
\usepackage{ORCIDinREVTeX}
\usepackage{hyperref}
\usepackage[capitalise]{cleveref}
\usepackage[compat=1.1.0]{tikz-feynman}
\usepackage{hyperref}
\usepackage{natbib}
\usepackage{bm}
\usetikzlibrary{shapes.misc}
\setcitestyle{author,square}
\tikzset{cross/.style={cross out, draw=black, minimum size=2*(#1-\pgflinewidth), inner sep=0pt, outer sep=0pt},
cross/.default={5pt}}
\begin{document}

\title{Running coupling constant in thermal $\phi^4$ theory up to two loop order}

\author{K. Arjun}
\orcid{0000-0001-9260-7050}
  \email{arjunk\underline{ }dop@uoc.ac.in}
\author{A. M. Vinodkumar}%
\orcid{0000-0002-8204-7800}
\author{Vishnu Mayya Bannur}
\affiliation{%
 Department of Physics \\ University of Calicut.\\ Kerala, India - 673635.
}%

\begin{abstract}
Using the imaginary time formalism in thermal field theory, we derive running coupling constant and running mass in two loop order. In the process, we express the imaginary time formalism of Feynman diagrams as the summation of non-thermal quantum field theory (QFT) Feynman diagrams with coefficients that depend on temperature and mass. 
Renormalization constants for thermal $\phi^4$ theory were derived using simple diagrammatic analysis. Our model links the non-thermal QFT and the imaginary time formalism by assuming both have the same mass scale $\mu$ and coupling constant \emph{g}. When these results are combined with the renormalization group equations (RGE) and applied simultaneously to thermal and non-thermal proper vertex functions, coupling constant and running mass with implicit temperature dependence are obtained. We evaluated pressure for scalar particles in two loop orders at zero external momentum limit by substituting the running mass result in the quasi-particle model.
\end{abstract}

\pacs{Valid PACS appear here}
\maketitle

\section{Introduction}
There are different quasi-particle models \cite{Goloviznin1993, Peshier1994,Schneider2001,Meisinger2004,Bannur2007b,Bannur2007,Bannur2007a,Koothottil2019,Koothottil2021} used to describe the state of the quark-gluon plasma (QGP). Many of these models depend upon the corresponding running coupling constant. 
Therefore the nature of coupling constant and associated renormalization group equations were always exciting topics in quantum field theory. \\

 Using the renormalization group equation, \citet{Collins1975} published a significant study on ultra-dense nuclear matter (the idea of quark soup at high density) in 1975. The concept of thermally dependent renormalization group equations(RGE) \cite{Kapusta1979, Matsumoto1984} considered mass and coupling constant as temperature functions.
In deriving coupling constant, changes in the vertex function results in variations in coupling constant \cite{Fujimoto1988, Baier1990}. This means, the coupling constant has a strong dependence on the vertex function \cite{Nakkagawa1987,Nakkagawa1988}. Also, the dependence of coupling constant on the gauge was investigated in \cite{Chaichian1996,Sasaki1997}.
\citet{Braaten1990} introduced significant ideas like hard thermal loops and, using it, they presented resummation of QCD thermal perturbation theory. A perturbative analysis of QCD at the static limit was shown in \cite{EIJCK1994}. The authors used both momentum and temperature dependent RGEs and derived a coupling constant that depends on momentum scale and temperature. During the same decade, several renormalization concepts were introduced, including environment friendly renormalization \cite{Eijck1994b,EIJCK1995} and idea of using more than one renormalization group for running mass and temperature \cite{Stephens1998,Stephens2004}. Recent developments include the research on magnetic field effects on coupling constant and expression for thermomagnetic coupling constant at one loop order \cite{Ayala2018,Ferrer2015}.\\

On the description of QGP \cite{Nadkarni1983,Schneider2001,Meisinger2004,Bannur2007b,Bannur2007,Bannur2007a}, most of the models take readily available coupling constants \cite{Deur2016} from quantum chromodynamics (QCD) and evaluate their results. The most famous coupling constants in QCD have inverse logarithmic \cite{Schneider2002,
Schneider2003} thermal dependency. \citet{Steffens2006}, investigates the non-logarithmic and logarithmic nature of the QCD coupling of the high-temperature limit. 
We can find works related to quark dynamics in a magnetic field which apply magnetic field-dependent coupling constants \cite{Ferrer2015,Ayala2018} in quasi-particle models \cite{Koothottil2019,Koothottil2021,Kurian2017,Kurian2019} to determine the equation of state. These models consider quarks and gluons as quasi-particles with thermal mass depending on various parameters like coupling constant, temperature, and magnetic field. \\
 
  In the context of non-thermal and thermal field theory, we pay special attention to the quartic($\phi^4$) interaction (also known as toy model) theory for two loop order. In this work, we use the simple diagrammatic analysis \cite{Kleinert2001} in imaginary time formalism(ITF) \cite{Kapusta2006} to derive a coupling constant that obeys RGE for both thermal and non-thermal field theory under the same mass scale $\mu$ and coupling constant \emph{g}. We take advantage of the analytically consistent theory \cite{Takao}, evaluated at external momentum equal to zero. We can extend the same approach to QCD.\\

The self-energy part of quartic interaction tadpole can be found in papers \cite{OCONNOR1993,Parwani1992,Arnold1994,Peshier1996, Peshier1998} under various approximations like massive and massless Lagrangian density. \cite{Frenkel1992,Arnold1994} calculates the free energy part of the $\phi^4$ interactions, and \cite{Braaten1995} contains the higher-order calculations. In our work, we define Lagrangian density as  
\begin{equation}
\mathcal{L}= \frac{1}{2} \left\lbrace \partial_\mu \phi \partial^{\mu} \phi - m^2 \phi^2 \right\rbrace - \frac{\lambda}{4 !} \phi^4
\end{equation} \\
and follow a systematic approach ($\lambda$ is the non-renormalized coupling constant).
The Lagrangian density is massive itself and consists of a quadratic and quartic interaction term. Some of the Feynman diagram integral results and formulae derived and used in this work may be found in other publications \cite{Kapusta2006, Bugrij1995, Andersen2000, Andersen2001a, Kleinert2001}. \\
We employ the dimensional regularization method \cite{Bollini1972, Hooft1972, Hooft1973}, which successfully regularizes non-Abelian gauge theory while preserving symmetries. The minimal subtraction scheme (MS scheme) follows after regularization and in which it corresponds to canceling out the pole parts ($\epsilon$) through the method of counter terms \cite{Bogoliubow1957}. Then we apply the corresponding RGE to finite proper vertex function (FPVF) in imaginary time formalism in which, the removal of divergences at $\epsilon \to 0$, the vertex function becomes finite \cite{Collins1974, Speer1974, Breitenlohner1977a, Caswell1982}.\\

In this work, we use subscripts QFT and ITF to represent Feynman diagrams expressed in \textit{non-thermal QFT} and \textit{imaginary time formalism}.
Abbreviations TLA, SMC, and FPVF, represent for the \textit{two loop approximation}, the \textit{same mass scale ($\mu$) and coupling constant ($g$)} and \textit{finite proper vertex function}, respectively. Throughout this paper, we used Euclidean momentum $K=[\omega_{n_k},k]$ to write $K^2=\omega_{n_k}^2+k^2$. One can see three $\beta$'s appearing in this work. $\beta$ denotes $1/T$. $\beta(g)$ denotes the beta function associated with RGE. $\beta_2$ and $\beta_3$ are the coefficients of $g^2$ and $g^3$ of beta function, respectively. \\

In the RGE, we have an equation connecting coupling constant with mass scale using $\beta(g) = \frac{dg}{d \ln \mu}$ and equation connecting running mass with mass-scale  $\gamma_m(g) = \frac{d \ln m}{d \ln \mu}$. In thermal and non-thermal field theory, $\beta(g)$ and $\gamma_m(g)$ have the same expression. Suppose we have an additional equation connecting the coupling constant with temperature. With the help of $\beta(g), \gamma_m(g)$, and this new temperature-dependent equation, it is possible to find the temperature-dependent running mass, mass scale, and coupling constant. \\

We introduce a new approach where we apply RGE simultaneously on thermal and non-thermal FPVF under SMC. It results in a polynomial expression in \emph{g} of order four. After a close examination, one can find that coefficients of $g^2,g,g^0$ are zero. So the non-trivial solution reduces to a linear equation in \emph{g} with coefficients that depends on running mass, temperature, and mass scale. A minor rearrangement brings out the relation between coupling, running mass, temperature, and mass scale. \\

We follow a systematic approach \cite{Kleinert2001}. To ease the calculation, we make relations between diagrams in thermal and non-thermal field theory. We try to express all the ITF diagrams as the sum of QFT diagrams with thermal coefficients. \\

In \cref{regularization}, we write down all essential diagrams, such as two and four-point functions in one and two loop order, in ITF and QFT. We express ITF diagrams as sums of QFT diagrams (with and without thermal coefficients). We derive the requisite counter term diagrams in \cref{counterterms} to remove the divergences from the corresponding proper vertex function and derive renormalization constants in ITF. In \cref{renormalizationcoefficients}, we calculate renormalization constants through necessary diagrams. \cref{RGEITF} discusses the idea and outcome of applying RGE on the FPVF difference and how it will lead to the temperature-dependent coupling constant under SMC. \cref{resultanddiscussion} includes plots revealing the temperature relations, such as the coupling constant vs. temperature, running mass vs. temperature, mass-scale vs. temperature, and pressure vs. temperature. In \cref{resultanddiscussion}, we verify the qualitative behavior of running mass and coupling results. We substitute running mass on the self-consistent quasi-particle model \cite{Bannur2007,Bannur2007a,Bannur2007b} pressure relation and check whether it will lead to the Stephen-Boltzmann limit at the higher temperature approximations.\\

Another essential nature is the flexibility of this model. It depends on the integration constants, and functions $\beta(g)$ and $\gamma_m(g)$. Once we have data points, we can fit running mass, coupling, pressure results to the data points by choosing appropriate values for the integral constants.

\section{Regularization}\label{regularization}
It is possible to write down all diagrams in the $\phi^4$ theory by a set of fundamental non-trivial Feynman diagrams known as one-particle irreducible (1 PI) diagrams. In other words, they are the smallest building block of diagrams. 
\begin{align}\label{baretwopoint}
\Gamma^{(2)}=\left( 
\begin{tikzpicture}
		\draw(-6.0,-12) -- (-5.5,-12);
	\end{tikzpicture}
	\right)^{-1} -\left( 
	\frac{1}{2} \begin{tikzpicture}
		\draw(-6,-11.75) circle(0.25);
		\draw(-6.5,-12) -- (-5.5,-12);
	\end{tikzpicture} 
	+ \frac{1}{4} \begin{tikzpicture}
		\draw(-6,-11.75) circle(0.25);
		\draw(-6.5,-12) -- (-5.5,-12);
		\draw(-6,-11.25) circle(0.25);
\end{tikzpicture}
+ \frac{1}{6} \begin{tikzpicture}
		\draw(-6,-12) circle(0.25);
		\draw(-6.5,-12) -- (-5.5,-12);
	\end{tikzpicture}
	\right)
\end{align}
\begin{align}\label{barefourpoint}
\Gamma^{(4)}=- 
\begin{tikzpicture} 
\draw(0.176776695,0.176776695) -- (-0.176776695,-0.176776695);
\draw(0.176776695,-0.176776695) -- (-0.176776695,0.176776695);
\end{tikzpicture} - \frac{3}{2} \ \begin{tikzpicture}
		\draw(-6,-11.75) circle(0.25);
		\draw(-6.25,-11.75) -- (-6.43,-11.6);
		\draw(-6.25,-11.75) -- (-6.43,-11.92);
		\draw(-5.75,-11.75) -- (-5.57,-11.6);
		\draw(-5.75,-11.75) -- (-5.57,-11.92);
\end{tikzpicture}- 3 \ \begin{tikzpicture}
		\draw(-6,-12) circle(0.25);
		\draw(-6.5,-12) -- (-5.5,-12);
		\draw(-6,-11.75) -- (-6.2,-11.65);
		\draw(-6,-11.75) -- (-5.8,-11.65);
	\end{tikzpicture} - \frac{3}{4} \begin{tikzpicture}
		\draw(-6,-11.75) circle(0.25);
		\draw(-6.25,-11.75) -- (-6.43,-11.6);
		\draw(-6.25,-11.75) -- (-6.43,-11.92);
		\draw(-5.5,-11.75) circle(0.25);		
		\draw(-5.25,-11.75) -- (-5.07,-11.6);
		\draw(-5.25,-11.75) -- (-5.07,-11.92);
	\end{tikzpicture} - \frac{3}{2} \begin{tikzpicture}
		\draw(-6,-11.375) circle(0.125);
		\draw(-6,-11.75) circle(0.25);
		\draw(-6.25,-11.75) -- (-6.43,-11.6);
		\draw(-6.25,-11.75) -- (-6.43,-11.92);
		\draw(-5.75,-11.75) -- (-5.57,-11.6);
		\draw(-5.75,-11.75) -- (-5.57,-11.92);
	\end{tikzpicture}  
\end{align}
\cref{baretwopoint,barefourpoint} are two and four-point vertex functions, expressed using corresponding 1 PI diagrams.

In $\phi^4$ ITF and non-thermal QFT, both these two and four-point functions and their corresponding integrals diverge. We follow the dimensional regularization procedure. In which we introduce a parameter to pinpoint the diverging term and separate the singularity \cite{Bollini1972, Hooft1972, Hooft1973}. In the following subsections \hyperref[oneloopdiagrams]{A} and \hyperref[twolooporderdiagrams]{B}, we express the two and four-point vertex functions in ITF as the summation of corresponding non-thermal QFT vertex functions with thermal coefficients. In the following evaluations, the remaining $g \mu^\epsilon$ will become $\lambda \approx g$ at $\epsilon \to 0$. Similarly in case of finite terms, $\lambda^2 \to (g \mu^\epsilon)^2=g^2$ as $\epsilon \to 0$.
\subsection{One Loop Order Diagrams}\label{oneloopdiagrams}
\subsubsection{Two-point function}
We define the tadpole diagram in imaginary time formalism as
\begin{equation}\label{Tadpole_first}
\begin{tikzpicture}
		\draw(-6,-11.75) circle(0.25);
		\draw(-6.5,-12) -- (-5.5,-12);
	\end{tikzpicture}_{ITF}= -\lambda \ T  \sum_{n} \int \frac{1}{p^2+{m}^2+{\omega_{n}^2}} \frac{d^3 p}{(2 \pi)^3}
\end{equation}
and that of non-thermal $\phi^4$ theory as
\begin{align}
\begin{tikzpicture}
		\draw(-6,-11.75) circle(0.25);
		\draw(-6.5,-12) -- (-5.5,-12);
	\end{tikzpicture}_{QFT}= -\lambda  \int \frac{1}{p^2+{m}^2} \frac{d^4 p}{(2 \pi)^4}
\end{align}
we use the result from Appendix \hyperref[A-1]{A-1} and express Tadpole in ITF in terms of Tadpole in non-thermal QFT as  
\begin{align}\label{Eq. Tadpole2a}
 \begin{tikzpicture}
		\draw(-6,-11.75) circle(0.25);
		\draw(-6.5,-12) -- (-5.5,-12);
	\end{tikzpicture}_{{ITF}}  &=  \begin{tikzpicture}
		\draw(-6,-11.75) circle(0.25);
		\draw(-6.5,-12) -- (-5.5,-12);
	\end{tikzpicture}_{{QFT}}  -\lambda  S_{1}(m,T)
	\end{align}
	where 
\begin{align}
S_1(m,T)&=\int \frac{n_B(\beta \epsilon_p)}{\epsilon_p} \frac{d^3p}{(2 \pi)^3}\\ &=\frac{1}{\pi} \sum_{n=1}^\infty \left( \frac{m}{2 \pi n \beta} \right)K_1(n \beta m) \nonumber
\end{align}
We can describe the divergence as a singularity of parameter $\epsilon$ by substituting $\lambda=g \mu^\epsilon$ at dimension \emph{N} (say) to $N-\epsilon$. We get the regularization result 
\begin{align}\label{tadpole_one}
 \begin{tikzpicture}
		\draw(-6,-11.75) circle(0.25);
		\draw(-6.5,-12) -- (-5.5,-12);
	\end{tikzpicture}_{ ITF}  &= \frac{m^2g}{(4 \pi)^2}\left[ \frac{2}{\epsilon}+\psi(2)+\ln \left( \frac{4 \pi \mu^2}{m^2}  \right) \right] \nonumber \\ 
	&-g \mu^{\epsilon} S_{1}(m,T) +\mathcal{O}(\epsilon)
\end{align}
In which $\psi(n)$ is Euler Digamma function, \emph{g} the dimensionless coupling constant, and $n_B(x)=(e^x-1)^{-1}$. As $\epsilon \rightarrow 0$, the first term diverges.  
\subsubsection{Four-point function}
The four-point function diagram definition for one loop order is
\begin{align}\label{scatter1}
\begin{tikzpicture}
		\draw(-6,-11.75) circle(0.25);
		\draw(-6.25,-11.75) -- (-6.43,-11.6);
		\draw(-6.25,-11.75) -- (-6.43,-11.92);
		\draw(-5.75,-11.75) -- (-5.57,-11.6);
		\draw(-5.75,-11.75) -- (-5.57,-11.92);
	\end{tikzpicture}_{ITF}=\int \frac{d^3p}{(2 \pi)^3} \sum_{n_p= -\infty}^\infty \frac{\lambda^2 T}{\omega_{n_p}^2+\varepsilon_p^2} \frac{1}{\omega_{n_p-n_q}^2+\varepsilon_{p-q}^2}
\end{align}
with $\varepsilon_p^2=p^2+m^2$, (Diagram is also known as scattering diagram). The corresponding diagram in the non-thermal QFT is
\begin{align}
\begin{tikzpicture}
		\draw(-6,-11.75) circle(0.25);
		\draw(-6.25,-11.75) -- (-6.43,-11.6);
		\draw(-6.25,-11.75) -- (-6.43,-11.92);
		\draw(-5.75,-11.75) -- (-5.57,-11.6);
		\draw(-5.75,-11.75) -- (-5.57,-11.92);
	\end{tikzpicture}_{QFT}=\int \frac{d^4P}{(2 \pi)^4}  \frac{\lambda^2 }{P^2+m^2} \frac{1}{(P-Q)^2+m^2}.
\end{align}
One can relate these two diagrams after regularization ($\lambda \to g \mu^\epsilon, \ \frac{d^3p}{(2 \pi)^3} \to \frac{d^{3-\epsilon}p}{(2 \pi)^{3 - \epsilon}}, \ \frac{d^4p}{(2 \pi)^4} \to \frac{d^{4-\epsilon}p}{(2 \pi)^{4-\epsilon}} $) as shown in Appendix \hyperref[A-2]{A-2} as
\begin{align}
 \begin{tikzpicture}
		\draw(-6,-11.75) circle(0.25);
		\draw(-6.25,-11.75) -- (-6.43,-11.6);
		\draw(-6.25,-11.75) -- (-6.43,-11.92);
		\draw(-5.75,-11.75) -- (-5.57,-11.6);
		\draw(-5.75,-11.75) -- (-5.57,-11.92);
	\end{tikzpicture}_{ITF}= 
\begin{tikzpicture}
		\draw(-6,-11.75) circle(0.25);
		\draw(-6.25,-11.75) -- (-6.43,-11.6);
		\draw(-6.25,-11.75) -- (-6.43,-11.92);
		\draw(-5.75,-11.75) -- (-5.57,-11.6);
		\draw(-5.75,-11.75) -- (-5.57,-11.92);
	\end{tikzpicture}_{QFT,Q_0=\omega_{n_q}} + (g \mu^\epsilon)^2 W(q,n_q)
\end{align}
with
\small
\begin{align}
W(r,n_r)=\int \frac{2n_B(\beta \varepsilon_p) \left(r^2+ 2 pr \cos \theta+\omega_{n_r}^2 \right)}{\varepsilon_p(\left(r^2+ 2 pr \cos \theta+\omega_{n_r}^2)^2+4 \varepsilon_p^2 \omega_{n_r}^2 \right)}  \frac{d^3p}{(2 \pi)^3}
\end{align}
\normalsize
Using dimensional regularization ($\lambda=g \mu^\epsilon$) and standard textbook results \cite{Kleinert2001}, we can write
\small
\begin{align}
\begin{tikzpicture}
		\draw(-6,-11.75) circle(0.25);
		\draw(-6.25,-11.75) -- (-6.43,-11.6);
		\draw(-6.25,-11.75) -- (-6.43,-11.92);
		\draw(-5.75,-11.75) -- (-5.57,-11.6);
		\draw(-5.75,-11.75) -- (-5.57,-11.92);
	\end{tikzpicture}_{QFT}&= \frac{g^2\mu^\epsilon}{(4 \pi)^2} \left( \frac{2}{\epsilon}+\psi(1)+\int_0^1 dx \ln \left[ \frac{4 \pi \mu^2}{Q^2x(1-x)+m^2} \right] \right) \nonumber \\ &+\mathcal{O}(\epsilon) 
\end{align}
\normalsize
with $Q=[\omega_{n_q},\vec{q}]$.

The external momentum $\vec{k}$ is the sum of the incoming momenta $\vec{k}_1+\vec{k}_2$ and the Matsubara summation parameter $\omega_{n_k}=\omega_{n_{k_1}}+\omega_{n_{k_2}}$, where $\omega_{n_k}=2 \pi n_k T$ with $n_k$ being an integer. For Matsubara summation, we use the symbol $\sumint_p=\sum_{n_p=-\infty}^{\infty} \int \frac{d^3p}{(2 \pi)^3}$.
\subsection{Two Loop Order Diagrams}\label{twolooporderdiagrams}
\subsubsection{Two point function}
In TLA, we have to evaluate the following diagrams and express them in terms of thermal independent QFT diagrams with temperature function coefficients. The integral expression of $\begin{tikzpicture}
		\draw(-6,-11.85) circle(0.15);
		\draw(-6.25,-12) -- (-5.75,-12);
		\draw(-6,-11.55) circle(0.15);
\end{tikzpicture}$ in ITF and QFT are
\small
\begin{align}
\sumint_{p_1,p_2} \frac{\lambda^2 T^2}{\varepsilon_{p1}^2+\omega_{n_{p_1}}^2} \left[ \frac{1}{\varepsilon_{p2}^2+\omega_{n_{p_2}}^2} \right]^2  
\end{align}
\normalsize
and
\small
\begin{align}
 \int \frac{\lambda^2}{\varepsilon_{p1}^2} \left[ \frac{1}{\varepsilon_{p2}^2} \right]^2 \frac{d^4p_1}{(2 \pi)^4} \frac{d^4p_2}{(2 \pi)^4} 
\end{align}
\normalsize
 respectively with $\varepsilon_p^2=p^2+m^2$. We follow the same convention as shown in \cref{oneloopdiagrams}. As specified in Appendices \hyperref[A-3]{A-3} and \hyperref[A-4]{A-4}, after regularization ($\lambda \to g \mu^\epsilon, \ \frac{d^3p}{(2 \pi)^3} \to \frac{d^{3-\epsilon}p}{(2 \pi)^{3 - \epsilon}}, \ \frac{d^4p}{(2 \pi)^4} \to \frac{d^{4-\epsilon}p}{(2 \pi)^{4-\epsilon}}, \epsilon \to 0$) we get the results
\begin{align}\label{doubletad}
\begin{tikzpicture}
		\draw(-6,-11.75) circle(0.25);
		\draw(-6.5,-12) -- (-5.5,-12);
		\draw(-6,-11.25) circle(0.25);
	\end{tikzpicture}_{ITF} &=  \begin{tikzpicture}
		\draw(-6,-11.75) circle(0.25);
		\draw(-6.5,-12) -- (-5.5,-12);
		\draw(-6,-11.25) circle(0.25);
	\end{tikzpicture}_{QFT } \\  \nonumber &- \left[ g\mu^\epsilon  \frac{S_{0}(m,T)}{4 \pi} \ \right]  \left[  \begin{tikzpicture}
		\draw(-6,-11.75) circle(0.25);
		\draw(-6.5,-12) -- (-5.5,-12);
	\end{tikzpicture}_{QFT}\right] & \\  \nonumber &+ g\mu^\epsilon  S_{1}(m,T) \frac{\partial}{\partial m^2}\left[{ \begin{tikzpicture}
		\draw(-6,-11.75) circle(0.25);
		\draw(-6.5,-12) -- (-5.5,-12);
	\end{tikzpicture} }_{QFT}\right] & \\  \nonumber & + (g\mu^\epsilon)^2 \frac{S_{1}(m,T) S_{0}(m,T)}{4 \pi} 
\end{align}
The expression for $\begin{tikzpicture}
		\draw(-6,-12) circle(0.15);
		\draw(-6.25,-12) -- (-5.75,-12);
	\end{tikzpicture}$ in ITF and QFT are 
	\small
\begin{align}
\sumint_{p_1,p_2} \frac{1}{\varepsilon_{p_1}^2+\omega_{n_{p_1}}^2} \frac{\lambda^2}{\varepsilon_{p_2}^2+\omega_{n_{p_2}}^2} \frac{T^2}{\varepsilon_{p_1+p_2+q}^2+\omega_{n_{p_1+p_2+q}}^2} 
\end{align}
	and 
\begin{align}
\int \frac{\lambda^2}{\varepsilon_{p_1}^2} \frac{1}{\varepsilon_{p_2}^2} \frac{1}{\varepsilon_{p_1+p_2+q}^2} \frac{d^4p_1}{(2 \pi)^4} \frac{d^4p_2}{(2 \pi)^4}
\end{align} 
\normalsize
respectively.
From the result of QFT \cite{Kleinert2001,Andersen2001a,Andersen2000} and  corresponding  ITF, after regularization ($\lambda \to g \mu^\epsilon$, $d^3p \to d^{3-\epsilon}p$, $d^4p \to d^{4-\epsilon}p$), we express the sunset/sunrise diagram  at external momentum zero as
\small
\begin{align}
&  \frac{1}{6}  \begin{tikzpicture}
		\draw(-6,-12) circle(0.25);
		\draw(-6.5,-12) -- (-5.5,-12);
	\end{tikzpicture}_{ITF}|_{K=0}=  \frac{1}{6} \begin{tikzpicture}
		\draw(-6,-12) circle(0.25);
		\draw(-6.5,-12) -- (-5.5,-12);
	\end{tikzpicture}_{QFT} |_{K=0}  \\ \nonumber
	&+ \frac{S_{1}(m,T)}{2}  \frac{g^2 \mu^{\epsilon}}{(4 \pi)^2} \left[\psi(1)+\ln \left( \frac{4 \pi \mu^2}{m^2} \right) +\frac{2}{\epsilon} \right] \\ \nonumber & +\frac{S_{1}(m,T)}{2} \frac{g^2 \mu^{\epsilon}}{(4 \pi)^2} \left( 2-\frac{\pi}{\sqrt{3}} \right)  +\frac{g^2m^2}{64 \pi^4}  Y(m,T)\\
	\nonumber &+\mathcal{O}(\epsilon)
\end{align} 
\normalsize
where
\small
\begin{align}
Y(m,T)&=\int_0^\infty \int_0^\infty U(x) U(y) G(x,y) \ dx \ dy \\
U(x)&=\frac{\sinh(x)}{\exp \left( \beta m \cosh(x) \right)-1}\\
	G(x,y)&=\ln \left( \frac{1+2 \cosh(x-y)}{1+2 \cosh(x+y)} \frac{1-2 \cosh(x+y)}{1-2 \cosh(x-y)} \right)  \\
S_N(m,T) &= \frac{1}{\pi}   \sum_{j=1}^\infty \left(\frac{m}{2 \pi j \beta}\right)^{N} K_N(j \beta m)
\end{align}
\normalsize \\

In these two cases, the remaining $g \mu^\epsilon$ will become $\lambda \approx g$ at $\epsilon \to 0$. Similarly in case of finite terms, $(g \mu^\epsilon)^2=\lambda^2=g^2$ as $\epsilon \to 0$. 
Detailed expansion and results are given in Appendices \hyperref[A-3]{A-3} and \hyperref[A-4]{A-4}.
\subsubsection{Four-point function}
This Section also expresses the other relevant $\phi^4$ four-point diagrams in TLA according to the same convention as \cref{oneloopdiagrams}.
We have regularized results
\begin{align}
 \begin{tikzpicture}
		\draw(-6,-11.75) circle(0.25);
		\draw(-6.25,-11.75) -- (-6.43,-11.6);
		\draw(-6.25,-11.75) -- (-6.43,-11.92);
		\draw(-5.5,-11.75) circle(0.25);		
		\draw(-5.25,-11.75) -- (-5.07,-11.6);
		\draw(-5.25,-11.75) -- (-5.07,-11.92);
	\end{tikzpicture}_{ITF}&=  \begin{tikzpicture}
		\draw(-6,-11.75) circle(0.25);
		\draw(-6.25,-11.75) -- (-6.43,-11.6);
		\draw(-6.25,-11.75) -- (-6.43,-11.92);
		\draw(-5.5,-11.75) circle(0.25);		
		\draw(-5.25,-11.75) -- (-5.07,-11.6);
		\draw(-5.25,-11.75) -- (-5.07,-11.92);
	\end{tikzpicture}_{QFT,k_0=\omega_{n_k}}\\ &-2 g \mu^\epsilon W(k,n_k) \left[  \begin{tikzpicture}
		\draw(-6,-11.75) circle(0.25);
		\draw(-6.25,-11.75) -- (-6.43,-11.6);
		\draw(-6.25,-11.75) -- (-6.43,-11.92);
		\draw(-5.75,-11.75) -- (-5.57,-11.6);
		\draw(-5.75,-11.75) -- (-5.57,-11.92);
	\end{tikzpicture}_{QFT,k_0=\omega_{n_k}} \right] \nonumber \\ &-\left( g \mu^\epsilon \right)^3W^2(k,n_k) \nonumber
\end{align}
where 
\begin{align}
-\lambda^3 \left( \sumint_{p} \frac{T}{\varepsilon_{p-k}^2+\omega_{n_{p-k}}^2} \frac{1}{\varepsilon_p^2+\omega_{n_p}^2}\right)^2 \end{align} and \begin{align}-\lambda^3 \left( \int  \frac{1}{\varepsilon_{p-k}^2} \frac{1}{\varepsilon_p^2} \frac{d^4p}{(2 \pi)^4} \right)^2 \end{align}  are corresponding non-regularized integral expressions for ITF and QFT. Here $[\omega_{n_k},\vec{k}]$ denotes either of three momentum combinations $[\omega_{n_{k_1}}+\omega_{n_{k_2}},\vec{k_1}+\vec{k_2}]$, $[\omega_{n_{k_1}}+\omega_{n_{k_3}},\vec{k_1}+\vec{k_3}]$, and $[\omega_{n_{k_1}}+\omega_{n_{k_4}},\vec{k_1}+\vec{k_4}]$. Detailed expansion and derivation are given in Appendices \hyperref[A-5]{A-5} and \hyperref[A-6]{A-6}. 
Other required diagram, and its ITF, QFT relations can be expressed as 
\begin{align}
 \begin{tikzpicture}
		\draw(-6,-11.375) circle(0.125);
		\draw(-6,-11.75) circle(0.25);
		\draw(-6.25,-11.75) -- (-6.43,-11.6);
		\draw(-6.25,-11.75) -- (-6.43,-11.92);
		\draw(-5.75,-11.75) -- (-5.57,-11.6);
		\draw(-5.75,-11.75) -- (-5.57,-11.92);
	\end{tikzpicture}_{ITF}&= \begin{tikzpicture}
		\draw(-6,-11.375) circle(0.125);
		\draw(-6,-11.75) circle(0.25);
		\draw(-6.25,-11.75) -- (-6.43,-11.6);
		\draw(-6.25,-11.75) -- (-6.43,-11.92);
		\draw(-5.75,-11.75) -- (-5.57,-11.6);
		\draw(-5.75,-11.75) -- (-5.57,-11.92);
	\end{tikzpicture}_{QFT,k_0=\omega_{n_k}} \\
&+\frac{g\mu^\epsilon S_1(m,T)}{2} \frac{\partial}{\partial m^2} \left[  \begin{tikzpicture}
		\draw(-6,-11.75) circle(0.25);
		\draw(-6.25,-11.75) -- (-6.43,-11.6);
		\draw(-6.25,-11.75) -- (-6.43,-11.92);
		\draw(-5.75,-11.75) -- (-5.57,-11.6);
		\draw(-5.75,-11.75) -- (-5.57,-11.92);
	\end{tikzpicture}_{QFT,k_0=\omega_{n_k}} \right] \nonumber \\
&	-\frac{(g \mu^\epsilon)^2}{2} \frac{\partial \ W(k,n_k)}{\partial m^2} \left[  \begin{tikzpicture}
		\draw(-6,-11.75) circle(0.25);
		\draw(-6.5,-12) -- (-5.5,-12);
	\end{tikzpicture}_{QFT}  \right] \nonumber \\
	&+(g \mu^\epsilon)^3 \frac{S_1(m,T)}{2}  \frac{\partial}{\partial m^2} W(k,n_k) \nonumber.
\end{align}
 The integral expression of above diagram is \begin{align}
-\lambda^3 T^2 \sumint_p \frac{1}{\varepsilon_{p-k}^2+{\omega_{n_p-n_k}^2}} \ \frac{1}{(\varepsilon_p^2+\omega_{n_p}^2)^2} \sumint_q  \frac{1}{\varepsilon_q^2+\omega_{n_q}^2}
\end{align} and \begin{align}-\lambda^3 \int \frac{d^{4} p}{(2 \pi)^{4}} \frac{1}{\varepsilon_{p-k}^2} \ \frac{1}{(\varepsilon_p^2)^2} \int \frac{d^{4} q}{(2 \pi)^{4}} \frac{1}{\varepsilon_q^2}\end{align} for ITF and non thermal QFT respectively. 
By defining operator $\mathcal{K}$ \cite{Kleinert2001}, which picks up the diverging terms from the corresponding graphs on which it has applied, (i.e $\mathcal{K} \left(\frac{A}{\epsilon^n}+B +c \epsilon \right)=\frac{A}{\epsilon^n}$), another complex diagram result can be summarized as 
\begin{align}
\mathcal{K} \left[ \begin{tikzpicture}
		\draw(-6,-12) circle(0.25);
		\draw(-6.5,-12) -- (-5.5,-12);
		\draw(-6,-11.75) -- (-6.2,-11.65);
		\draw(-6,-11.75) -- (-5.8,-11.65);
	\end{tikzpicture}_{ITF} \right]&=\mathcal{K} \left[ \begin{tikzpicture}
		\draw(-6,-12) circle(0.25);
		\draw(-6.5,-12) -- (-5.5,-12);
		\draw(-6,-11.75) -- (-6.2,-11.65);
		\draw(-6,-11.75) -- (-5.8,-11.65);
	\end{tikzpicture}_{QFT,k_{0i}=\omega_{n_i}} \right]\\ &-g\mu^{\epsilon} W(k_i,n_{k_i}) \mathcal{K} \left(
\begin{tikzpicture}
		\draw(-6,-11.75) circle(0.25);
		\draw(-6.25,-11.75) -- (-6.43,-11.6);
		\draw(-6.25,-11.75) -- (-6.43,-11.92);
		\draw(-5.75,-11.75) -- (-5.57,-11.6);
		\draw(-5.75,-11.75) -- (-5.57,-11.92);
	\end{tikzpicture}_{QFT}\right) \nonumber
\end{align}
with
\small
\begin{align} \int \frac{1}{\varepsilon_p^2}\frac{-\lambda^3}{\varepsilon_{k_1+k_2-p}^2} \frac{1}{\varepsilon_q^2} \frac{1}{\varepsilon_{p-q+k_3}^2} \frac{d^{4}p}{(2 \pi)^{4}} \frac{d^{4}q}{(2 \pi)^{4}}
\end{align} and
\footnotesize 
\begin{align}
\sumint_{p,q} \frac{1}{\varepsilon_p^2+\omega_{n_p}^2}\frac{-\lambda^3 T^2}{\varepsilon_{k_1+k_2-p}^2+{\omega^2}_{n_{k_1+k_2-p}}} \frac{1}{\varepsilon_q^2+\omega_{n_q}^2} \frac{1}{\varepsilon_{p-q+k_3}^2+\omega_{n_{p-q+k_3}}^2}
\end{align} 
\normalsize 
These are the diagram's non-regularized integral expressions in QFT and ITF, respectively.
The complete derivation of the above diagram is in Appendix \hyperref[A-7]{A-7}.
\normalsize
\section{Counter terms and  minimal subtraction scheme}\label{counterterms}
Counter term diagrams are those that make the vertex function finite when added with the vertex function.
The previously derived diagrams in \cref{regularization} contain terms that diverge for $\epsilon \to 0$. 
We redefine the proper vertex function to divergence removed proper vertex function(FPVF). We use the MS-scheme to deal with diverging terms. Here, operator $\mathcal{K}$ separates the diverging terms. \\
i.e.; 
\begin{equation}
\widetilde{\Gamma}^{(n)}=\Gamma^{(n)}-\mathcal{K} \left(\Gamma^{(n)} \right).
\end{equation}
 By this definition
\begin{align}
\widetilde{\Gamma}^{(2)}&=\Gamma^{(2)}-\mathcal{K} \left( \Gamma^{(2)} \right) \\
\widetilde{\Gamma}^{(4)}&=\Gamma^{(4)}-\mathcal{K} \left( \Gamma^{(4)} \right)
\end{align}
\subsection{One loop calculation}
The counter terms in one loop order are usually the pole term in diagrams with a negative sign.
\subsubsection{Two-point function}
\label{twolooprenormalizationtwopointfunctions}
We have to find the counter term for first-order \emph{g}, and we follow \cite{Kleinert2001} as the reference text, then
\begin{equation}\label{twopointoneloop}
\begin{split}
 \widetilde{\Gamma}^{(2)}= (\begin{tikzpicture}
		\draw(-6.5,-12) -- (-5.5,-12);
	\end{tikzpicture})^{-1}_{ITF} &-\left\{ \frac{1}{2} \begin{tikzpicture}
		\draw(-6,-11.75) circle(0.25);
		\draw(-6.5,-12) -- (-5.5,-12);
	\end{tikzpicture}_{ITF}  \right. \\
	&\left. \begin{tikzpicture}
\draw(0.0,0) -- (1.0,0);
\draw (0.5,0) node[cross,rotate=0] {};
\end{tikzpicture}+\begin{tikzpicture}
\draw(0.0,0) -- (1.0,0);
\draw (.5,0) circle (3pt);
\end{tikzpicture} \right\}+\mathcal{O}(g^2)
\end{split}
\end{equation}
where $\begin{tikzpicture}
\draw(0.0,0) -- (1.0,0);
\draw (0.5,0) node[cross,rotate=0] {};
\end{tikzpicture}$ represents the contribution of mass counter term, and $\begin{tikzpicture}
\draw(0.0,0) -- (1.0,0);
\draw (.5,0) circle (3pt);
\end{tikzpicture}$ represents the field contribution, and 
$(\begin{tikzpicture}
		\draw(-6.5,-12) -- (-5.5,-12);
\end{tikzpicture})^{-1}_{ITF} =K^2+m^2|_{k_0= \omega_{n_k}}$ under Euclidean momentum representation.
 In imaginary time formalism \cite{Arnold1994,Bugrij1995}, the tadpole diagram's diverging term is the same as that of non-thermal QFT. The counter term needed to cancel the tadpole divergence is proportional to $m^2$ and is
\begin{align}\label{countertermpole1}
\begin{tikzpicture}
\draw(0.0,0) -- (1.0,0);
\draw (0.5,0) node[cross,rotate=0] {};
\end{tikzpicture}=-m^2 c_{m^2}^1&=-\frac{1}{2} \mathcal{K} \left( \begin{tikzpicture}
		\draw(-6,-11.75) circle(0.25);
		\draw(-6.5,-12) -- (-5.5,-12);
	\end{tikzpicture}_{ITF} \right)\\
	&\text{from Appendix \hyperref[A-1]{A-1}} \nonumber
	\\
	&=-\frac{1}{2} \mathcal{K} \left( \begin{tikzpicture}
		\draw(-6,-11.75) circle(0.25);
		\draw(-6.5,-12) -- (-5.5,-12);
	\end{tikzpicture}_{QFT} \right) \nonumber \\
	&=-m^2 \frac{g}{(4 \pi)^2} \frac{1}{\epsilon} \nonumber
\end{align}

and the counter term proportional to $K^2$ in first-order \emph{g} is zero, so
\begin{align}
-K^2c_{\phi}^1=\begin{tikzpicture}
\draw(0.0,0) -- (1.0,0);
\draw (.5,0) circle (3pt);
\end{tikzpicture}=0
\end{align}
Thus the finite proper vertex function, which is finite at $\epsilon \to 0$,
\small
\begin{align}
	\widetilde{\Gamma}^{(2)}=\left(\begin{tikzpicture}
\draw(0.0,0) -- (1.0,0);
\end{tikzpicture}\right)^{-1}&-\left( \frac{1}{2} \begin{tikzpicture}
		\draw(-6,-11.75) circle(0.25);
		\draw(-6.5,-12) -- (-5.5,-12);
	\end{tikzpicture}_{ITF} -\frac{1}{2} \mathcal{K} \left( \begin{tikzpicture}
		\draw(-6,-11.75) circle(0.25);
		\draw(-6.5,-12) -- (-5.5,-12);
	\end{tikzpicture}_{ITF} \right)
 \right) \nonumber \\ &+\mathcal{O}(g^2) \nonumber \\
 =\left(\begin{tikzpicture}
\draw(0.0,0) -- (1.0,0);
\end{tikzpicture}\right)^{-1}&-\left( \frac{1}{2} \begin{tikzpicture}
		\draw(-6,-11.75) circle(0.25);
		\draw(-6.5,-12) -- (-5.5,-12);
	\end{tikzpicture}_{QFT} -\frac{1}{2} \mathcal{K} \left( \begin{tikzpicture}
		\draw(-6,-11.75) circle(0.25);
		\draw(-6.5,-12) -- (-5.5,-12);
	\end{tikzpicture}_{QFT} \right)\right) \nonumber \\
	&+\frac{g}{2}S_1(m,T)+\mathcal{O}(g^2) 
\end{align}
\normalsize.
\subsubsection{Four-point function}
Similar to the two-point function case, four-point finite proper vertex function is
\begin{align}\label{fourpointoneloop}
	\widetilde{\Gamma}^{(4)}=-\left( 
\begin{tikzpicture} 
\draw(0.176776695,0.176776695) -- (-0.176776695,-0.176776695);
\draw(0.176776695,-0.176776695) -- (-0.176776695,0.176776695);
\end{tikzpicture} + \frac{3}{2} \ \begin{tikzpicture}
		\draw(-6,-11.75) circle(0.25);
		\draw(-6.25,-11.75) -- (-6.43,-11.6);
		\draw(-6.25,-11.75) -- (-6.43,-11.92);
		\draw(-5.75,-11.75) -- (-5.57,-11.6);
		\draw(-5.75,-11.75) -- (-5.57,-11.92);
\end{tikzpicture}_{ITF}+\begin{tikzpicture}
\draw[black,fill=black] (-6,-12) circle(0.5ex);
\draw (-6,-12) node[cross,rotate=0] {};
\end{tikzpicture}_{ITF}
\right)+\mathcal{O}(g^3)
\end{align}
The counter term is the pole term of the scattering diagram with a negative sign. In both ITF and QFT, the pole term is the same. Scattering diagrams in ITF and QFT differ by a finite thermal term.
\begin{align}\label{scattercounter}
\begin{tikzpicture}
\draw[black,fill=black] (-6,-12) circle(0.5ex);
\draw (-6,-12) node[cross,rotate=0] {};
\end{tikzpicture}_{ITF}&=- \mu^\epsilon g c_g^1=-\frac{3}{2} \mathcal{K} \left(\begin{tikzpicture}
		\draw(-6,-11.75) circle(0.25);
		\draw(-6.25,-11.75) -- (-6.43,-11.6);
		\draw(-6.25,-11.75) -- (-6.43,-11.92);
		\draw(-5.75,-11.75) -- (-5.57,-11.6);
		\draw(-5.75,-11.75) -- (-5.57,-11.92);
\end{tikzpicture}_{ITF} \right) \\
&\text{from Appendix \hyperref[B-1]{B-1}}
\nonumber \\
&=-\frac{3}{2} \mathcal{K} \left(\begin{tikzpicture}
		\draw(-6,-11.75) circle(0.25);
		\draw(-6.25,-11.75) -- (-6.43,-11.6);
		\draw(-6.25,-11.75) -- (-6.43,-11.92);
		\draw(-5.75,-11.75) -- (-5.57,-11.6);
		\draw(-5.75,-11.75) -- (-5.57,-11.92);
\end{tikzpicture}_{QFT} \right) \nonumber \\
&=\begin{tikzpicture}
\draw[black,fill=black] (-6,-12) circle(0.5ex);
\draw (-6,-12) node[cross,rotate=0] {};
\end{tikzpicture}_{QFT}=-\mu^\epsilon g \frac{3g}{(4 \pi)^2} \frac{1}{\epsilon} \nonumber
\end{align}
\subsection{Two loop calculation}\label{twoloopcalculation}
The redefinition of proper vertex function in one loop order (\cref{twopointoneloop,fourpointoneloop}) to make it finite  causes more counter-term diagrams to emerge in the two loop order calculation.
\subsubsection{Two-point function}
The finite two-point function up to TLA for ITF is
\begin{align}
\widetilde{\Gamma}^{(2)}&=(
\begin{tikzpicture}
		\draw(0,0) -- (0.5,0);
\end{tikzpicture}
)^{-1}-\left( \frac{1}{2} \begin{tikzpicture}
		\draw(-6,-11.75) circle(0.25);
		\draw(-6.5,-12) -- (-5.5,-12);
	\end{tikzpicture}_{ITF} +\begin{tikzpicture}
\draw(0.0,0) -- (1.0,0);
\draw (0.5,0) node[cross,rotate=0] {};
\end{tikzpicture}+\begin{tikzpicture}
\draw(0.0,0) -- (1.0,0);
\draw (.5,0) circle (3pt);
\end{tikzpicture}  \right) \\
&- \left(\frac{1}{4} \begin{tikzpicture}
		\draw(-6,-11.75) circle(0.25);
		\draw(-6.5,-12) -- (-5.5,-12);
		\draw(-6,-11.25) circle(0.25);
\end{tikzpicture}_{ITF} + \frac{1}{2} \begin{tikzpicture}
		\draw(-6,-11.75) circle(0.25);
		\draw(-6.5,-12) -- (-5.5,-12);
		\draw (-6,-11.5) node[cross,rotate=0] {};
\end{tikzpicture}_{ITF} \right) \nonumber \\
&- \left( \frac{1}{6} \begin{tikzpicture}
		\draw(-6,-12) circle(0.25);
		\draw(-6.5,-12) -- (-5.5,-12);
	\end{tikzpicture}_{ITF}+ \frac{1}{2} \begin{tikzpicture}
		\draw(-6,-11.75) circle(0.25);
		\draw(-6.5,-12) -- (-5.5,-12);
		\draw[black,fill=black] (-6,-12) circle(0.5ex);
	\end{tikzpicture}_{ITF} \right)+\mathcal{O}(g^3) \nonumber
\end{align}

When compared to \cref{baretwopoint}, the additional terms that appear here are the counter term diagrams in two loop order to make the proper vertex function finite. \\

The pole term ($1/\epsilon^n$, $n>0$) can have both thermal and non-thermal coefficients. Substituting $-m^2 c_{m^2}^1$ of \cref{countertermpole1} for $-g \mu^\epsilon$ in \cref{scatter1} scattering diagram at $n_k\text{,}k=0$ gives one of the counter diagrams required. Appendix \hyperref[B-2]{B-2} contains the detailed derivation.
\begin{align}
\frac{1}{2} \begin{tikzpicture}
		\draw(-6,-11.75) circle(0.25);
		\draw(-6.5,-12) -- (-5.5,-12);
		\draw (-6,-11.5) node[cross,rotate=0] {};
\end{tikzpicture}_{ITF}&=\frac{1}{2}\begin{tikzpicture}
		\draw(-6,-11.75) circle(0.25);
		\draw(-6.5,-12) -- (-5.5,-12);
		\draw (-6,-11.5) node[cross,rotate=0] {};
\end{tikzpicture}_{QFT}\\&+\frac{g}{4 \pi} \frac{S_0(m,T)}{4} \mathcal{K}\left[  \begin{tikzpicture}
		\draw(-6,-11.75) circle(0.25);
		\draw(-6.5,-12) -- (-5.5,-12);
	\end{tikzpicture}_{QFT} \right] \nonumber
\end{align}
It is possible to express the counter-term diagram as the sum of similar diagrams in QFT with non-thermal coefficients and other diagrams with thermal coefficients, as shown above. \\
As given in Appendix \hyperref[B-3]{B-3}, if we replace $-\mu^\epsilon g$ in Tadpole of \cref{tadpole_one}, in ITF by $-\mu^\epsilon g c_g^1$ of \cref{scattercounter} we get  
\begin{align}
\frac{1}{2} \begin{tikzpicture}
		\draw(-6,-11.75) circle(0.25);
		\draw(-6.5,-12) -- (-5.5,-12);
		\draw[black,fill=black] (-6,-12) circle(0.5ex);
	\end{tikzpicture}_{ITF}&=\frac{1}{2}\begin{tikzpicture}
		\draw(-6,-11.75) circle(0.25);
		\draw(-6.5,-12) -- (-5.5,-12);
		\draw[black,fill=black] (-6,-12) circle(0.5ex);
	\end{tikzpicture}_{QFT}\\ &-\frac{3g}{4}
S_{1}(m,T) \ \frac{\partial}{\partial m^2} \mathcal{K} \left(  \begin{tikzpicture}
		\draw(-6,-11.75) circle(0.25);
		\draw(-6.5,-12) -- (-5.5,-12);
	\end{tikzpicture}_{QFT} \right) \nonumber
\end{align}
with
\begin{equation}
	S_N(m,T)= \frac{1}{\pi}   \sum_{j=1}^\infty \left(\frac{m}{2 \pi j \beta}\right)^{N} K_N(j \beta m) 
\end{equation}
To find the remaining diverging terms in the vertex function, we write down the diverging terms of each diagram expressed in the two loop vertex function.

As stated in Appendix \hyperref[A-3]{A-3} the diverging terms in the diagram in \cref{doubletad} expressed in terms of QFT poles with thermal and non-thermal coefficients are
\begin{align}
\frac{1}{4} \mathcal{K} \left(	\begin{tikzpicture}
		\draw(-6,-11.75) circle(0.25);
		\draw(-6.5,-12) -- (-5.5,-12);
		\draw(-6,-11.25) circle(0.25);
\end{tikzpicture}_{ITF} \right) &= \frac{1}{4} \mathcal{K} \left( \begin{tikzpicture}
		\draw(-6,-11.75) circle(0.25);
		\draw(-6.5,-12) -- (-5.5,-12);
		\draw(-6,-11.25) circle(0.25);
\end{tikzpicture}_{QFT} \right) \\
	&-\frac{g}{4 \pi}S_0(m,T) \ \frac{1}{4} \mathcal{K} \left[  \begin{tikzpicture}
		\draw(-6,-11.75) circle(0.25);
		\draw(-6.5,-12) -- (-5.5,-12);
	\end{tikzpicture}_{QFT} \right] \nonumber \\ &+ \frac{g S_1(m,T)}{4} \frac{\partial}{\partial m^2} \mathcal{K} \left(  \begin{tikzpicture}
		\draw(-6,-11.75) circle(0.25);
		\draw(-6.5,-12) -- (-5.5,-12);
	\end{tikzpicture}_{QFT} \right) \nonumber
\end{align}
The diverging part of the sunset/sunrise diagram in vertex function is 
\begin{align}
\frac{1}{6} \mathcal{K} \left( \begin{tikzpicture}
		\draw(-6,-12) circle(0.25);
		\draw(-6.5,-12) -- (-5.5,-12);
	\end{tikzpicture}_{ITF} \right)
&=\frac{1}{6} \mathcal{K} \left( \begin{tikzpicture}
		\draw(-6,-12) circle(0.25);
		\draw(-6.5,-12) -- (-5.5,-12);
	\end{tikzpicture}_{QFT,k_0=\omega_{n_k}} \right)\\ &+\frac{g}{2}S_{1}(m,T) \ \frac{\partial}{\partial m^2} \mathcal{K} \left(  \begin{tikzpicture}
		\draw(-6,-11.75) circle(0.25);
		\draw(-6.5,-12) -- (-5.5,-12);
	\end{tikzpicture}_{QFT} \right) \nonumber
\end{align}
as described in Appendix \hyperref[A-4]{A-4}.

Interestingly ITF two-point functions total divergence is the same as that of QFT at $k_0=\omega_{n_k}$.
\\ i.e.,
\begin{align}\label{importantmassrenormform}
&\frac{1}{4} \mathcal{K} \left(	\begin{tikzpicture}
		\draw(-6,-11.75) circle(0.25);
		\draw(-6.5,-12) -- (-5.5,-12);
		\draw(-6,-11.25) circle(0.25);
\end{tikzpicture}_{ITF} \right)+\frac{1}{6} \mathcal{K} \left( \begin{tikzpicture}
		\draw(-6,-12) circle(0.25);
		\draw(-6.5,-12) -- (-5.5,-12);
	\end{tikzpicture}_{ITF} \right)\\
	&+\frac{1}{2} \begin{tikzpicture}
		\draw(-6,-11.75) circle(0.25);
		\draw(-6.5,-12) -- (-5.5,-12);
		\draw (-6,-11.5) node[cross,rotate=0] {};
\end{tikzpicture}_{ITF}+\frac{1}{2} \begin{tikzpicture}
		\draw(-6,-11.75) circle(0.25);
		\draw(-6.5,-12) -- (-5.5,-12);
		\draw[black,fill=black] (-6,-12) circle(0.5ex);
	\end{tikzpicture}_{ITF} \nonumber \\
	&=\frac{1}{4} \mathcal{K} \left(	\begin{tikzpicture}
		\draw(-6,-11.75) circle(0.25);
		\draw(-6.5,-12) -- (-5.5,-12);
		\draw(-6,-11.25) circle(0.25);
\end{tikzpicture}_{QFT} \right)+\frac{1}{6} \mathcal{K} \left( \begin{tikzpicture}
		\draw(-6,-12) circle(0.25);
		\draw(-6.5,-12) -- (-5.5,-12);
	\end{tikzpicture}_{QFT,k_0=\omega_{n_k}} \right) \nonumber \\
	&+\frac{1}{2} \begin{tikzpicture}
		\draw(-6,-11.75) circle(0.25);
		\draw(-6.5,-12) -- (-5.5,-12);
		\draw (-6,-11.5) node[cross,rotate=0] {};
\end{tikzpicture}_{QFT}+\frac{1}{2} \begin{tikzpicture}
		\draw(-6,-11.75) circle(0.25);
		\draw(-6.5,-12) -- (-5.5,-12);
		\draw[black,fill=black] (-6,-12) circle(0.5ex);
	\end{tikzpicture}_{QFT} \nonumber
\end{align}
All other terms cancel with each other. Nevertheless, still, divergence exists in \cref{importantmassrenormform}. $c_{m^2}$ and $c_\phi$ absorb the remaining divergences. So the total contribution to counter terms $c_{m^2}$ and $c_\phi$ up to $g^2$ in ITF is identical to that of QFT at $k_0=\omega_{n_k}$.
\begin{equation}
\begin{split}
(\begin{tikzpicture}
\draw(0.0,0) -- (1.0,0);
\draw (0.5,0) node[cross,rotate=0] {};
\end{tikzpicture}+\begin{tikzpicture}
\draw(0.0,0) -- (1.0,0);
\draw (.5,0) circle (3pt);
\end{tikzpicture})_{ITF}&=-\mathcal{K} \left\{ \frac{1}{2} \begin{tikzpicture}
		\draw(-6,-11.75) circle(0.25);
		\draw(-6.5,-12) -- (-5.5,-12);
	\end{tikzpicture}_{ITF} + \frac{1}{4} \begin{tikzpicture}
		\draw(-6,-11.75) circle(0.25);
		\draw(-6.5,-12) -- (-5.5,-12);
		\draw(-6,-11.25) circle(0.25);
\end{tikzpicture}_{ITF} \right. \\
& \left. +\frac{1}{6}  \begin{tikzpicture}
		\draw(-6,-12) circle(0.25);
		\draw(-6.5,-12) -- (-5.5,-12);
	\end{tikzpicture}_{ITF} +\frac{1}{2} \begin{tikzpicture}
		\draw(-6,-11.75) circle(0.25);
		\draw(-6.5,-12) -- (-5.5,-12);
		\draw (-6,-11.5) node[cross,rotate=0] {};
\end{tikzpicture}_{ITF} \right. \\
&\left. +\frac{1}{2} \begin{tikzpicture}
		\draw(-6,-11.75) circle(0.25);
		\draw(-6.5,-12) -- (-5.5,-12);
		\draw[black,fill=black] (-6,-12) circle(0.5ex);
	\end{tikzpicture}_{ITF} \right\} \\
	&=(\begin{tikzpicture}
\draw(0.0,0) -- (1.0,0);
\draw (0.5,0) node[cross,rotate=0] {};
\end{tikzpicture}+\begin{tikzpicture}
\draw(0.0,0) -- (1.0,0);
\draw (.5,0) circle (3pt);
\end{tikzpicture})_{QFT,K=[\omega_{n_k},\vec{k}]}
\end{split}
\end{equation}
\\
\begin{equation}
=- \left[ \frac{g}{(4 \pi)^2} \frac{m^2}{\epsilon} +\frac{g^2}{(4 \pi)^4} \left( \frac{2m^2}{\epsilon^2}-\frac{m^2}{2 \epsilon}-\frac{K^2}{12 \epsilon} \right) \right]\nonumber
\end{equation}
\normalsize
with $K^2=\omega_{n_k}^2+\vec{k}^2$. \\

i.e., in ITF, if we follow the textbook procedure (\cite{Kleinert2001}), the sum of counter terms are identical to that of QFT, with $k_0=\omega_{n_k}$. If we extract the polynomial with coefficient $m^2$ and $K^2$, we get (\cite{Kleinert2001}) \begin{align}\label{zm2}
	m^2(c_{m^2}^1+c_{m^2}^2)=m^2 \left[ \frac{g}{(4 \pi)^2} \frac{1}{\epsilon}+\frac{g^2}{(4 \pi)^4} \left( \frac{2}{\epsilon^2}-\frac{1}{2 \epsilon} \right) \right] 
\end{align}
For field renormalization, we have to consider the term proportional to $K^2$, so
\begin{align}\label{fieldrenormalization}
K^2 c_{\phi}^2=\frac{1}{6} \mathcal{K} \left( \begin{tikzpicture}
		\draw(-6,-12) circle(0.25);
		\draw(-6.5,-12) -- (-5.5,-12);
\end{tikzpicture}_{QFT} \right)|_{m=0,k_0=\omega_{n_k}} = - \frac{g^2}{(4 \pi)^4} \frac{K^2}{12 \epsilon}
\end{align} 
\subsubsection{Four-point function}
To proceed further, we have to remove the divergences of the four-point function using the renormalization procedure \cite{Kleinert2001}. The result is the same as that of non-thermal QFT. i.e.,
\begin{widetext}
\begin{equation}
\begin{split}
\widetilde{\Gamma}^{(4)} = & -\left\{ 
\begin{tikzpicture} 
\draw(0.176776695,0.176776695) -- (-0.176776695,-0.176776695);
\draw(0.176776695,-0.176776695) -- (-0.176776695,0.176776695);
\end{tikzpicture} + \frac{3}{2} \ \begin{tikzpicture}
		\draw(-6,-11.75) circle(0.25);
		\draw(-6.25,-11.75) -- (-6.43,-11.6);
		\draw(-6.25,-11.75) -- (-6.43,-11.92);
		\draw(-5.75,-11.75) -- (-5.57,-11.6);
		\draw(-5.75,-11.75) -- (-5.57,-11.92);
\end{tikzpicture}_{ITF}+\begin{tikzpicture}
\draw[black,fill=black] (-6,-12) circle(0.5ex);
\draw (-6,-12) node[cross,rotate=0] {};
\end{tikzpicture}_{ITF}  
    + 3 \ \begin{tikzpicture}
		\draw(-6,-12) circle(0.25);
		\draw(-6.5,-12) -- (-5.5,-12);
		\draw(-6,-11.75) -- (-6.2,-11.65);
		\draw(-6,-11.75) -- (-5.8,-11.65);
	\end{tikzpicture} + \frac{3}{4} \begin{tikzpicture}
		\draw(-6,-11.75) circle(0.25);
		\draw(-6.25,-11.75) -- (-6.43,-11.6);
		\draw(-6.25,-11.75) -- (-6.43,-11.92);
		\draw(-5.5,-11.75) circle(0.25);		
		\draw(-5.25,-11.75) -- (-5.07,-11.6);
		\draw(-5.25,-11.75) -- (-5.07,-11.92);
	\end{tikzpicture}_{ITF} + \frac{3}{2} \begin{tikzpicture}
		\draw(-6,-11.375) circle(0.125);
		\draw(-6,-11.75) circle(0.25);
		\draw(-6.25,-11.75) -- (-6.43,-11.6);
		\draw(-6.25,-11.75) -- (-6.43,-11.92);
		\draw(-5.75,-11.75) -- (-5.57,-11.6);
		\draw(-5.75,-11.75) -- (-5.57,-11.92);
	\end{tikzpicture}_{ITF} 
	  + 3 \ \begin{tikzpicture}
		\draw(-6,-11.75) circle(0.25);
		\draw(-6.25,-11.75) -- (-6.43,-11.6);
		\draw(-6.25,-11.75) -- (-6.43,-11.92);
		\draw(-5.75,-11.75) -- (-5.57,-11.6);
		\draw(-5.75,-11.75) -- (-5.57,-11.92);
		\draw[black,fill=black] (-5.75,-11.75) circle(0.5ex);
	\end{tikzpicture}_{ITF} + 3 \ \begin{tikzpicture}
		\draw(-6,-11.75) circle(0.25);
		\draw(-6.25,-11.75) -- (-6.43,-11.6);
		\draw(-6.25,-11.75) -- (-6.43,-11.92);
		\draw(-5.75,-11.75) -- (-5.57,-11.6);
		\draw(-5.75,-11.75) -- (-5.57,-11.92);
		\draw (-6,-11.5) node[cross,black]  {};
	\end{tikzpicture}_{ITF}
			\right\} \\ &+\mathcal{O}(g^4)
\end{split}
\end{equation}
From Appendix, one can verify the above results with complete derivation. Taking those results, we write from Appendix \hyperref[A-7]{A-7}
\begin{align}\label{53}
3 \mathcal{K} \left( \begin{tikzpicture}
		\draw(-6,-12) circle(0.25);
		\draw(-6.5,-12) -- (-5.5,-12);
		\draw(-6,-11.75) -- (-6.2,-11.65);
		\draw(-6,-11.75) -- (-5.8,-11.65);
	\end{tikzpicture}_{ITF} \right)=3 \mathcal{K} \left( \begin{tikzpicture}
		\draw(-6,-12) circle(0.25);
		\draw(-6.5,-12) -- (-5.5,-12);
		\draw(-6,-11.75) -- (-6.2,-11.65);
		\draw(-6,-11.75) -- (-5.8,-11.65);
	\end{tikzpicture}_{QFT,k_0=\omega_{n_k}} \right)-3 \ g \ W(k,n_k) \ \mathcal{K} \left( \begin{tikzpicture}
		\draw(-6,-11.75) circle(0.25);
		\draw(-6.25,-11.75) -- (-6.43,-11.6);
		\draw(-6.25,-11.75) -- (-6.43,-11.92);
		\draw(-5.75,-11.75) -- (-5.57,-11.6);
		\draw(-5.75,-11.75) -- (-5.57,-11.92);
	\end{tikzpicture}_{QFT} \right)
\end{align}
from Appendix \hyperref[A-5]{A-5} as
\begin{align}\label{54}
\frac{3}{4} \mathcal{K} \left( \begin{tikzpicture}
		\draw(-6,-11.75) circle(0.25);
		\draw(-6.25,-11.75) -- (-6.43,-11.6);
		\draw(-6.25,-11.75) -- (-6.43,-11.92);
		\draw(-5.5,-11.75) circle(0.25);		
		\draw(-5.25,-11.75) -- (-5.07,-11.6);
		\draw(-5.25,-11.75) -- (-5.07,-11.92);
	\end{tikzpicture}_{ITF} \right)=\frac{3}{4} \mathcal{K} \left( \begin{tikzpicture}
		\draw(-6,-11.75) circle(0.25);
		\draw(-6.25,-11.75) -- (-6.43,-11.6);
		\draw(-6.25,-11.75) -- (-6.43,-11.92);
		\draw(-5.5,-11.75) circle(0.25);		
		\draw(-5.25,-11.75) -- (-5.07,-11.6);
		\draw(-5.25,-11.75) -- (-5.07,-11.92);
	\end{tikzpicture}_{QFT,k_0=\omega_{n_k}} \right) -\frac{3}{2} \ g \ W(k,n_k) \ \mathcal{K} \left( \begin{tikzpicture}
		\draw(-6,-11.75) circle(0.25);
		\draw(-6.25,-11.75) -- (-6.43,-11.6);
		\draw(-6.25,-11.75) -- (-6.43,-11.92);
		\draw(-5.75,-11.75) -- (-5.57,-11.6);
		\draw(-5.75,-11.75) -- (-5.57,-11.92);
	\end{tikzpicture}_{QFT} \right)
\end{align}
from Appendix \hyperref[A-6]{A-6} as 
\begin{align}\label{55}
\frac{3}{2} \mathcal{K} \left( \begin{tikzpicture}
		\draw(-6,-11.375) circle(0.125);
		\draw(-6,-11.75) circle(0.25);
		\draw(-6.25,-11.75) -- (-6.43,-11.6);
		\draw(-6.25,-11.75) -- (-6.43,-11.92);
		\draw(-5.75,-11.75) -- (-5.57,-11.6);
		\draw(-5.75,-11.75) -- (-5.57,-11.92);
	\end{tikzpicture}_{ITF} \right)=\frac{3}{2} \mathcal{K} \left( \begin{tikzpicture}
		\draw(-6,-11.375) circle(0.125);
		\draw(-6,-11.75) circle(0.25);
		\draw(-6.25,-11.75) -- (-6.43,-11.6);
		\draw(-6.25,-11.75) -- (-6.43,-11.92);
		\draw(-5.75,-11.75) -- (-5.57,-11.6);
		\draw(-5.75,-11.75) -- (-5.57,-11.92);
	\end{tikzpicture}_{QFT,k_0=\omega_{n_k}} \right) 
	-\frac{3g^2}{4} \frac{\partial \ W(k,n_k)}{\partial m^2} \ \mathcal{K}\left[  \begin{tikzpicture}
		\draw(-6,-11.75) circle(0.25);
		\draw(-6.5,-12) -- (-5.5,-12);
	\end{tikzpicture}_{QFT}  \right]
\end{align}
The counter term derived for four-point function from Appendix \hyperref[B-4]{B-4} as
\begin{align}\label{56}
 3 \mathcal{K} \left( \begin{tikzpicture}
		\draw(-6,-11.75) circle(0.25);
		\draw(-6.25,-11.75) -- (-6.43,-11.6);
		\draw(-6.25,-11.75) -- (-6.43,-11.92);
		\draw(-5.75,-11.75) -- (-5.57,-11.6);
		\draw(-5.75,-11.75) -- (-5.57,-11.92);
		\draw[black,fill=black] (-5.75,-11.75) circle(0.5ex);
\end{tikzpicture}_{ITF} \right)= 3 \mathcal{K} \left( \begin{tikzpicture}
		\draw(-6,-11.75) circle(0.25);
		\draw(-6.25,-11.75) -- (-6.43,-11.6);
		\draw(-6.25,-11.75) -- (-6.43,-11.92);
		\draw(-5.75,-11.75) -- (-5.57,-11.6);
		\draw(-5.75,-11.75) -- (-5.57,-11.92);
		\draw[black,fill=black] (-5.75,-11.75) circle(0.5ex);
\end{tikzpicture}_{QFT,k_0=\omega_{n_k}=0} \right)+ \frac{9}{2} \ g \ W(k,n_k) \  \mathcal{K} \left(  \begin{tikzpicture}
		\draw(-6,-11.75) circle(0.25);
		\draw(-6.25,-11.75) -- (-6.43,-11.6);
		\draw(-6.25,-11.75) -- (-6.43,-11.92);
		\draw(-5.75,-11.75) -- (-5.57,-11.6);
		\draw(-5.75,-11.75) -- (-5.57,-11.92);
	\end{tikzpicture}_{QFT} \right)
\end{align}
the other counter term for four-point function from Appendix \hyperref[B-5]{B-5}
\begin{align}\label{57}
3 \mathcal{K} \left( \begin{tikzpicture}
		\draw(-6,-11.75) circle(0.25);
		\draw(-6.25,-11.75) -- (-6.43,-11.6);
		\draw(-6.25,-11.75) -- (-6.43,-11.92);
		\draw(-5.75,-11.75) -- (-5.57,-11.6);
		\draw(-5.75,-11.75) -- (-5.57,-11.92);
		\draw (-6,-11.5) node[cross,black]  {};
	\end{tikzpicture}_{ITF} \right) = 3 \mathcal{K} \left( \begin{tikzpicture}
		\draw(-6,-11.75) circle(0.25);
		\draw(-6.25,-11.75) -- (-6.43,-11.6);
		\draw(-6.25,-11.75) -- (-6.43,-11.92);
		\draw(-5.75,-11.75) -- (-5.57,-11.6);
		\draw(-5.75,-11.75) -- (-5.57,-11.92);
		\draw (-6,-11.5) node[cross,black]  {};
	\end{tikzpicture}_{QFT,k_0=\omega_{n_k}} \right) + \frac{3}{4} g^2  \left( \frac{\partial W(k,n_k)}{\partial m^2} \right)  \mathcal{K}\left[  \begin{tikzpicture}
		\draw(-6,-11.75) circle(0.25);
		\draw(-6.5,-12) -- (-5.5,-12);
\end{tikzpicture}_{QFT}  \right]
\end{align}
Adding \cref{53,54,55,56,57}, we get
\begin{align}
\begin{split}
\left\{ 3 \mathcal{K} \left( \begin{tikzpicture}
		\draw(-6,-12) circle(0.25);
		\draw(-6.5,-12) -- (-5.5,-12);
		\draw(-6,-11.75) -- (-6.2,-11.65);
		\draw(-6,-11.75) -- (-5.8,-11.65);
	\end{tikzpicture}_{ITF} \right)+\frac{3}{4} \mathcal{K} \left( \begin{tikzpicture}
		\draw(-6,-11.75) circle(0.25);
		\draw(-6.25,-11.75) -- (-6.43,-11.6);
		\draw(-6.25,-11.75) -- (-6.43,-11.92);
		\draw(-5.5,-11.75) circle(0.25);		
		\draw(-5.25,-11.75) -- (-5.07,-11.6);
		\draw(-5.25,-11.75) -- (-5.07,-11.92);
	\end{tikzpicture}_{ITF} \right)+\frac{3}{2} \mathcal{K} \left( \begin{tikzpicture}
		\draw(-6,-11.375) circle(0.125);
		\draw(-6,-11.75) circle(0.25);
		\draw(-6.25,-11.75) -- (-6.43,-11.6);
		\draw(-6.25,-11.75) -- (-6.43,-11.92);
		\draw(-5.75,-11.75) -- (-5.57,-11.6);
		\draw(-5.75,-11.75) -- (-5.57,-11.92);
	\end{tikzpicture}_{ITF} \right)  + 3 \mathcal{K} \left( \begin{tikzpicture}
		\draw(-6,-11.75) circle(0.25);
		\draw(-6.25,-11.75) -- (-6.43,-11.6);
		\draw(-6.25,-11.75) -- (-6.43,-11.92);
		\draw(-5.75,-11.75) -- (-5.57,-11.6);
		\draw(-5.75,-11.75) -- (-5.57,-11.92);
		\draw[black,fill=black] (-5.75,-11.75) circle(0.5ex);
\end{tikzpicture}_{ITF} \right)+3 \mathcal{K} \left( \begin{tikzpicture}
		\draw(-6,-11.75) circle(0.25);
		\draw(-6.25,-11.75) -- (-6.43,-11.6);
		\draw(-6.25,-11.75) -- (-6.43,-11.92);
		\draw(-5.75,-11.75) -- (-5.57,-11.6);
		\draw(-5.75,-11.75) -- (-5.57,-11.92);
		\draw (-6,-11.5) node[cross,black]  {};
	\end{tikzpicture}_{ITF} \right) \right\} 
\end{split}
\\
\begin{split}
= \left\{ 3 \mathcal{K} \left( \begin{tikzpicture}
		\draw(-6,-12) circle(0.25);
		\draw(-6.5,-12) -- (-5.5,-12);
		\draw(-6,-11.75) -- (-6.2,-11.65);
		\draw(-6,-11.75) -- (-5.8,-11.65);
	\end{tikzpicture}_{QFT} \right)+\frac{3}{4} \mathcal{K} \left( \begin{tikzpicture}
		\draw(-6,-11.75) circle(0.25);
		\draw(-6.25,-11.75) -- (-6.43,-11.6);
		\draw(-6.25,-11.75) -- (-6.43,-11.92);
		\draw(-5.5,-11.75) circle(0.25);		
		\draw(-5.25,-11.75) -- (-5.07,-11.6);
		\draw(-5.25,-11.75) -- (-5.07,-11.92);
	\end{tikzpicture}_{QFT} \right)+\frac{3}{2} \mathcal{K} \left( \begin{tikzpicture}
		\draw(-6,-11.375) circle(0.125);
		\draw(-6,-11.75) circle(0.25);
		\draw(-6.25,-11.75) -- (-6.43,-11.6);
		\draw(-6.25,-11.75) -- (-6.43,-11.92);
		\draw(-5.75,-11.75) -- (-5.57,-11.6);
		\draw(-5.75,-11.75) -- (-5.57,-11.92);
	\end{tikzpicture}_{QFT} \right)  + 3 \mathcal{K} \left( \begin{tikzpicture}
		\draw(-6,-11.75) circle(0.25);
		\draw(-6.25,-11.75) -- (-6.43,-11.6);
		\draw(-6.25,-11.75) -- (-6.43,-11.92);
		\draw(-5.75,-11.75) -- (-5.57,-11.6);
		\draw(-5.75,-11.75) -- (-5.57,-11.92);
		\draw[black,fill=black] (-5.75,-11.75) circle(0.5ex);
\end{tikzpicture}_{QFT} \right)+3 \mathcal{K} \left( \begin{tikzpicture}
		\draw(-6,-11.75) circle(0.25);
		\draw(-6.25,-11.75) -- (-6.43,-11.6);
		\draw(-6.25,-11.75) -- (-6.43,-11.92);
		\draw(-5.75,-11.75) -- (-5.57,-11.6);
		\draw(-5.75,-11.75) -- (-5.57,-11.92);
		\draw (-6,-11.5) node[cross,black]  {};
\end{tikzpicture}_{QFT} \right) \right\}|_{k_0=\omega_{n_k}} 
\end{split}
\end{align}
\normalsize
\section{Renormalization constants}\label{renormalizationcoefficients}
Therefore, we get the renormalization constants as 
\begin{align}
\begin{split}
Z_g(g,\epsilon^{-1})=1+\frac{1}{g \mu^\epsilon} \left\{ \frac{3}{2} \mathcal{K} \left( \begin{tikzpicture}
		\draw(-6,-11.75) circle(0.25);
		\draw(-6.25,-11.75) -- (-6.43,-11.6);
		\draw(-6.25,-11.75) -- (-6.43,-11.92);
		\draw(-5.75,-11.75) -- (-5.57,-11.6);
		\draw(-5.75,-11.75) -- (-5.57,-11.92);
\end{tikzpicture}_{ITF} \right) +3 \mathcal{K} \left( \begin{tikzpicture}
		\draw(-6,-12) circle(0.25);
		\draw(-6.5,-12) -- (-5.5,-12);
		\draw(-6,-11.75) -- (-6.2,-11.65);
		\draw(-6,-11.75) -- (-5.8,-11.65);
	\end{tikzpicture}_{ITF} \right) +\frac{3}{4} \mathcal{K} \left( \begin{tikzpicture}
		\draw(-6,-11.75) circle(0.25);
		\draw(-6.25,-11.75) -- (-6.43,-11.6);
		\draw(-6.25,-11.75) -- (-6.43,-11.92);
		\draw(-5.5,-11.75) circle(0.25);		
		\draw(-5.25,-11.75) -- (-5.07,-11.6);
		\draw(-5.25,-11.75) -- (-5.07,-11.92);
	\end{tikzpicture}_{ITF} \right) +\frac{3}{2} \mathcal{K} \left( \begin{tikzpicture}
		\draw(-6,-11.375) circle(0.125);
		\draw(-6,-11.75) circle(0.25);
		\draw(-6.25,-11.75) -- (-6.43,-11.6);
		\draw(-6.25,-11.75) -- (-6.43,-11.92);
		\draw(-5.75,-11.75) -- (-5.57,-11.6);
		\draw(-5.75,-11.75) -- (-5.57,-11.92);
	\end{tikzpicture}_{ITF} \right) \right. \\
 \left. + 3 \mathcal{K} \left( \begin{tikzpicture}
		\draw(-6,-11.75) circle(0.25);
		\draw(-6.25,-11.75) -- (-6.43,-11.6);
		\draw(-6.25,-11.75) -- (-6.43,-11.92);
		\draw(-5.75,-11.75) -- (-5.57,-11.6);
		\draw(-5.75,-11.75) -- (-5.57,-11.92);
		\draw[black,fill=black] (-5.75,-11.75) circle(0.5ex);
\end{tikzpicture}_{ITF} \right) 3 \mathcal{K} \left( \begin{tikzpicture}
		\draw(-6,-11.75) circle(0.25);
		\draw(-6.25,-11.75) -- (-6.43,-11.6);
		\draw(-6.25,-11.75) -- (-6.43,-11.92);
		\draw(-5.75,-11.75) -- (-5.57,-11.6);
		\draw(-5.75,-11.75) -- (-5.57,-11.92);
		\draw (-6,-11.5) node[cross,black]  {};
	\end{tikzpicture}_{ITF} \right) \right\} \\
	=1+\frac{1}{g \mu^\epsilon} \left\{ \frac{3}{2} \mathcal{K} \left( \begin{tikzpicture}
		\draw(-6,-11.75) circle(0.25);
		\draw(-6.25,-11.75) -- (-6.43,-11.6);
		\draw(-6.25,-11.75) -- (-6.43,-11.92);
		\draw(-5.75,-11.75) -- (-5.57,-11.6);
		\draw(-5.75,-11.75) -- (-5.57,-11.92);
\end{tikzpicture}_{QFT} \right) +3 \mathcal{K} \left( \begin{tikzpicture}
		\draw(-6,-12) circle(0.25);
		\draw(-6.5,-12) -- (-5.5,-12);
		\draw(-6,-11.75) -- (-6.2,-11.65);
		\draw(-6,-11.75) -- (-5.8,-11.65);
	\end{tikzpicture}_{QFT} \right)  +\frac{3}{4} \mathcal{K} \left( \begin{tikzpicture}
		\draw(-6,-11.75) circle(0.25);
		\draw(-6.25,-11.75) -- (-6.43,-11.6);
		\draw(-6.25,-11.75) -- (-6.43,-11.92);
		\draw(-5.5,-11.75) circle(0.25);		
		\draw(-5.25,-11.75) -- (-5.07,-11.6);
		\draw(-5.25,-11.75) -- (-5.07,-11.92);
	\end{tikzpicture}_{QFT} \right) +\frac{3}{2} \mathcal{K} \left( \begin{tikzpicture}
		\draw(-6,-11.375) circle(0.125);
		\draw(-6,-11.75) circle(0.25);
		\draw(-6.25,-11.75) -- (-6.43,-11.6);
		\draw(-6.25,-11.75) -- (-6.43,-11.92);
		\draw(-5.75,-11.75) -- (-5.57,-11.6);
		\draw(-5.75,-11.75) -- (-5.57,-11.92);
	\end{tikzpicture}_{QFT} \right) \right. \\
\left. + 3 \mathcal{K} \left( \begin{tikzpicture}
		\draw(-6,-11.75) circle(0.25);
		\draw(-6.25,-11.75) -- (-6.43,-11.6);
		\draw(-6.25,-11.75) -- (-6.43,-11.92);
		\draw(-5.75,-11.75) -- (-5.57,-11.6);
		\draw(-5.75,-11.75) -- (-5.57,-11.92);
		\draw[black,fill=black] (-5.75,-11.75) circle(0.5ex);
\end{tikzpicture}_{QFT} \right) 3 \mathcal{K} \left( \begin{tikzpicture}
		\draw(-6,-11.75) circle(0.25);
		\draw(-6.25,-11.75) -- (-6.43,-11.6);
		\draw(-6.25,-11.75) -- (-6.43,-11.92);
		\draw(-5.75,-11.75) -- (-5.57,-11.6);
		\draw(-5.75,-11.75) -- (-5.57,-11.92);
		\draw (-6,-11.5) node[cross,black]  {};
	\end{tikzpicture}_{QFT} \right) \right\}|_{k_0=\omega_{n_k}} 
\end{split}
\end{align}
So from standard result of QFT \cite{Kleinert2001} under SMC,
\begin{align}
Z_g(g,\epsilon^{-1})&=1+\frac{g}{(4 \pi)^2} \frac{3}{\epsilon}+\frac{g^2}{(4 \pi)^4} \left( \frac{9}{\epsilon^2}-\frac{3}{ \epsilon} \right)   
\end{align}

Similarly, from
\cref{importantmassrenormform,zm2}, Appendix \hyperref[A-1]{A-1}, and \cref{twoloopcalculation,twolooprenormalizationtwopointfunctions},
\begin{align}
\begin{split}
Z_{m^2}=1+\frac{1}{m^2} \left\{ \frac{1}{2} \mathcal{K} \left(\begin{tikzpicture}
		\draw(-6,-11.75) circle(0.25);
		\draw(-6.5,-12) -- (-5.5,-12);
	\end{tikzpicture}_{ITF} \right) +\frac{1}{4} \mathcal{K} \left(	\begin{tikzpicture}
		\draw(-6,-11.75) circle(0.25);
		\draw(-6.5,-12) -- (-5.5,-12);
		\draw(-6,-11.25) circle(0.25);
\end{tikzpicture}_{ITF} \right)+\frac{1}{6} \mathcal{K} \left( \begin{tikzpicture}
		\draw(-6,-12) circle(0.25);
		\draw(-6.5,-12) -- (-5.5,-12);
	\end{tikzpicture}_{ITF,K^2=0} \right) \right. \\
\left. +\frac{1}{2} \mathcal{K} \left(\begin{tikzpicture}
		\draw(-6,-11.75) circle(0.25);
		\draw(-6.5,-12) -- (-5.5,-12);
		\draw (-6,-11.5) node[cross,rotate=0] {};
\end{tikzpicture}_{ITF} \right)+\frac{1}{2} \mathcal{K} \left(\begin{tikzpicture}
		\draw(-6,-11.75) circle(0.25);
		\draw(-6.5,-12) -- (-5.5,-12);
		\draw[black,fill=black] (-6,-12) circle(0.5ex);
	\end{tikzpicture}_{ITF} \right) \right\}
\end{split} \\
\begin{split}
=1+\frac{1}{m^2} \left\{ \frac{1}{2} \mathcal{K} \left(\begin{tikzpicture}
		\draw(-6,-11.75) circle(0.25);
		\draw(-6.5,-12) -- (-5.5,-12);
	\end{tikzpicture}_{QFT} \right) +\frac{1}{4} \mathcal{K} \left(	\begin{tikzpicture}
		\draw(-6,-11.75) circle(0.25);
		\draw(-6.5,-12) -- (-5.5,-12);
		\draw(-6,-11.25) circle(0.25);
\end{tikzpicture}_{QFT} \right)+\frac{1}{6} \mathcal{K} \left( \begin{tikzpicture}
		\draw(-6,-12) circle(0.25);
		\draw(-6.5,-12) -- (-5.5,-12);
	\end{tikzpicture}_{QFT,K^2=0} \right) \right. \\
\left. +\frac{1}{2} \mathcal{K} \left(\begin{tikzpicture}
		\draw(-6,-11.75) circle(0.25);
		\draw(-6.5,-12) -- (-5.5,-12);
		\draw (-6,-11.5) node[cross,rotate=0] {};
\end{tikzpicture}_{QFT} \right)+\frac{1}{2} \mathcal{K} \left(\begin{tikzpicture}
		\draw(-6,-11.75) circle(0.25);
		\draw(-6.5,-12) -- (-5.5,-12);
		\draw[black,fill=black] (-6,-12) circle(0.5ex);
	\end{tikzpicture}_{QFT} \right) \right\} \nonumber \\
\end{split}
\end{align}
\begin{align}
Z_{m^2}(g,\epsilon^{-1})&=1+\frac{g}{(4 \pi)^2} \frac{1}{\epsilon}+\frac{g^2}{(4 \pi)^4} \left( \frac{2}{\epsilon^2}-\frac{1}{2 \epsilon} \right)
\end{align}
From \cref{twolooprenormalizationtwopointfunctions,fieldrenormalization}, we get
\begin{align}\label{cphi}
Z_{\phi}&=1+\frac{1}{K^2} \frac{1}{6} \mathcal{K} \left( \begin{tikzpicture}
		\draw(-6,-12) circle(0.25);
		\draw(-6.5,-12) -- (-5.5,-12);
	\end{tikzpicture}_{QFT} \right)|_{m^2=0,k_0=\omega_{n_k}}\\ 
	&=1+c_\phi  \nonumber\\ 
	&=1-\frac{g^2}{(4 \pi)^4}\frac{1}{12 \epsilon} \nonumber
\end{align} 
\end{widetext}
The above results show that for two, and four-point functions at TLA, the values of $Z_\phi, \ Z_g, \ \text{and } Z_{m^2}$ for ITF are the same as for non-thermal $\phi^4$ theory \cite{Kleinert2001}. So we demand that the renormalization group equation(RGE) for QFT must also be true for ITF. 
\section{Renormalization Group Equation in Two loop order two-point function}\label{RGEITF}
The previous results show that neither $Z_\phi$, $Z_g$, nor $Z_{m^2}$ are explicitly temperature-dependent, and they agree with the non-thermal field theory. However, a non-explicit reliance may exist via the coupling constant \emph{g} as $g(\mu(T))$. Furthermore, we use RGE for TLA to see if such relationships are possible. i.e., we demand \cref{EQ. Imp1,EQ. Imp1-1,EQ. Imp1-2,EQ. Imp1-3,EQ. Imp1-4} to be true for TLA under SMC. Here the RGE is explicitly temperature independent, but the vertex function in ITF is temperature-dependent.
\begin{widetext}
\begin{align}
\label{EQ. 3} \frac{d}{d(\ln \mu)} \widetilde{\Gamma}^{(n)} \left( m, g, T, \mu \right) = &\left[ \mu \frac{\partial}{\partial \mu}+ \beta(g) \frac{\partial}{\partial g}-n \gamma(g) + \gamma_m m \frac{\partial}{\partial m}  \right] \widetilde{\Gamma}^{(n)} \left( m, g, T, \mu \right) \approx_{TLA} 0 \\ \label{EQ. Imp1}
&\left[ \mu \frac{\partial}{\partial \mu}+ \beta(g) \frac{\partial}{\partial g}-2 \gamma(g) + \gamma_m m \frac{\partial}{\partial m}  \right] \widetilde{\Gamma}^{(2)}_{ITF} \left( m, g, T, \mu \right) \approx_{TLA} 0 \\ \label{EQ. Imp1-1}
&\left[ \mu \frac{\partial}{\partial \mu}+ \beta(g) \frac{\partial}{\partial g}-2 \gamma(g) + \gamma_m m \frac{\partial}{\partial m}  \right] \widetilde{\Gamma}^{(2)}_{QFT} \left( m, g, T, \mu \right) \approx_{TLA} 0 \\ \label{EQ. Imp1-2}
& \frac{d  g}{d  \ln(\mu)}= \beta(g) \approx \beta_2 g^2+ \beta_3 g^3 \\
\label{EQ. Imp1-3}
&\frac{d \ln(m(\mu))}{d \ln(\mu)}=\gamma_m(g) \approx \gamma_{m_1} g+\gamma_{m_2} g^2 \\
& \gamma(g) \approx \gamma_2 g^2 \label{EQ. Imp1-4}
\end{align}
\end{widetext}
In the previous section, we expressed ITF diagrams as the sum of corresponding QFT diagrams with subdiagrams having thermal coefficients. Therefore we can write the finite proper vertex function (FPVF) in ITF as the sum of FPVF of QFT with some subdiagrams having thermal coefficients under SMC.\\
At external momentum zero, we can write
\begin{align}
\widetilde{\Gamma}^{(2) ITF}_{finite,K=0}=\widetilde{\Gamma}^{(2) QFT}_{finite,K=0}+\Gamma^{\text{diff}}
\end{align}
where $\widetilde{\Gamma}^{(2)}_{finite}=\Gamma^{(2)}-\mathcal{K} \left( \Gamma^{(2)} \right)$ .\\
Thus any RGE which is valid for both ITF and QFT will also be true for their differences $\left(\widetilde{\Gamma}^{(2)}_{ITF}-\widetilde{\Gamma}^{(2)}_{QFT}\right)$.
 Then \cref{EQ. Imp1,EQ. Imp1-1} leads to 
\begin{align}\label{rge1}
\left\lbrace \frac{\partial}{\partial ln(\mu)}+\beta(g) \frac{\partial}{\partial g}-2 \gamma(g)+\gamma_m \frac{\partial}{\partial ln(m)} \right\rbrace \Gamma^{\text{diff}}=0
\end{align}
From the previous section, it is clear that the individual Feynman diagram evaluated in ITF and QFT at $k_0=\omega_{n_k}$ is different. Still, the two-point and four-point vertex divergences are of the same form. Thus the renormalization constants ($Z_{m^2},Z_{g},...c_{m^2},c_g$) are in the same structure. Therefore we can use the results of non-thermal $\phi^4$ theory
for finite functions $\beta(g), \ \gamma(g), \text{and} \ \gamma_m(g)$ from \cite{Kleinert2001} for ITF also.

At $K \neq 0$, the functions are
\begin{align}
\gamma(g)&=\frac{g^2}{(4 \pi)^4} \frac{1}{12}=\gamma_2 g^2\\
\gamma_m(g)&=\frac{1}{2} \frac{g}{(4 \pi)^2}-\frac{5}{12} \frac{g^2}{(4 \pi)^4} \nonumber = g \gamma_{m1} + g^2 \gamma_{m2} \\
\beta(g)&=-\epsilon g +\frac{3g^2}{(4 \pi)^2}-\frac{17g^3}{3(4 \pi)^4} =g \beta_1 + g^2 \beta_2 + g^3 \beta_3 \nonumber
\end{align}
When $K \to 0$, $\epsilon \to 0$, as shown in \cref{rmcoeff}, the functions change to
\begin{align}
\gamma(g)&=\gamma_2 g^2=0 \\
\gamma_m(g)&=\frac{1}{2} \frac{g}{(4 \pi)^2}-\frac{1}{2} \frac{g^2}{(4 \pi)^4} \nonumber =g \gamma_{m1} + g^2 \gamma_{m2}\\
\beta(g)&=\frac{3g^2}{(4 \pi)^2}-\frac{6g^3}{(4 \pi)^4}= g^2 \beta_2 + g^3 \beta_3 \nonumber
\end{align}

The general form of the two-point FPVF at TLA is
\begin{align}
\widetilde{\Gamma}^{(2)} &= (\begin{tikzpicture}
		\draw(-6.5,-12) -- (-5.5,-12);
	\end{tikzpicture})^{-1}  -\frac{1}{2}\left[ \begin{tikzpicture}
		\draw(-6,-11.75) circle(0.25);
		\draw(-6.5,-12) -- (-5.5,-12);
	\end{tikzpicture} - \mathcal{K} \left(  \begin{tikzpicture}
		\draw(-6,-11.75) circle(0.25);
		\draw(-6.5,-12) -- (-5.5,-12);
	\end{tikzpicture} \right) \right] \\ \nonumber & -\frac{1}{4} \left[ 
\begin{tikzpicture}
		\draw(-6,-11.75) circle(0.25);
		\draw(-6.5,-12) -- (-5.5,-12);
		\draw(-6,-11.25) circle(0.25);
	\end{tikzpicture} -  \mathcal{K} \left( 
\begin{tikzpicture}
		\draw(-6,-11.75) circle(0.25);
		\draw(-6.5,-12) -- (-5.5,-12);
		\draw(-6,-11.25) circle(0.25);
	\end{tikzpicture}  \right) \right] \\ \nonumber
	&- \frac{1}{6} \left[   \begin{tikzpicture}
		\draw(-6,-12) circle(0.25);
		\draw(-6.5,-12) -- (-5.5,-12);
	\end{tikzpicture}-\mathcal{K} \left(  \begin{tikzpicture}
		\draw(-6,-12) circle(0.25);
		\draw(-6.5,-12) -- (-5.5,-12);
	\end{tikzpicture}  \right) \right]
\end{align}
In order to simplify these equations, let us define an operator 
\begin{align}
\Delta (\mathbb{A}) &=\mathbb{A}_{ITF}|_{k,\omega_{n_k}=0}-\mathcal{K} \left( \mathbb{A}_{ITF}\right)|_{k,\omega_{n_k}=0}\\ 
&-\mathbb{A}_{QFT}|_{k_0,k=0}+\mathcal{K} \left( \mathbb{A}_{QFT} \right)|_{k_0,k=0} \nonumber
\end{align} where $\mathbb{A}$ represents the appropriate diagram.  \\
Since we defined $\Gamma^{\text{diff}}_{n_k,\vec{k}=0}=\Gamma^{(2)ITF}|_{n_k,k=0}-\Gamma^{(2)QFT}|_{k_0,k=0}$ we get 
\begin{align}
-\Gamma^{\text{diff}}_{n_k,\vec{k}=0}=\frac{\Delta}{2}  \left( \begin{tikzpicture}
		\draw(-6,-11.75) circle(0.25);
		\draw(-6.5,-12) -- (-5.5,-12);
	\end{tikzpicture} \right)+\frac{\Delta}{4}  \left( \begin{tikzpicture}
		\draw(-6,-11.75) circle(0.25);
		\draw(-6.5,-12) -- (-5.5,-12);
		\draw(-6,-11.25) circle(0.25);
	\end{tikzpicture} \right)+\frac{\Delta}{6}  \left( \begin{tikzpicture}
		\draw(-6,-12) circle(0.25);
		\draw(-6.5,-12) -- (-5.5,-12);
	\end{tikzpicture} \right)
\end{align}
From Appendix \hyperref[A-1]{A-1}
\begin{align}
\frac{1}{2} \Delta \left( \begin{tikzpicture}
		\draw(-6,-11.75) circle(0.25);
		\draw(-6.5,-12) -- (-5.5,-12);
	\end{tikzpicture} \right)=-\frac{g}{2}S_1(m,T)
\end{align}
From Appendix \hyperref[A-3]{A-3}, we get
\begin{align}
\frac{1}{4} \Delta \left( \begin{tikzpicture}
		\draw(-6,-11.75) circle(0.25);
		\draw(-6.5,-12) -- (-5.5,-12);
		\draw(-6,-11.25) circle(0.25);
	\end{tikzpicture} \right)&=\frac{g^2}{16 \pi}S_0(m,T)S_1(m,T)\\
	&-\frac{g^2 m^2 S_0(m,T)}{4 (4\pi)^3} \left[\psi(2)+\ln \left( \frac{4 \pi \mu^2}{m^2} \right) \right] \nonumber \\
	&+\frac{g^2 S_1(m,T)}{4(4 \pi)^2} \left[\psi(1)+\ln \left( \frac{4 \pi \mu^2}{m^2} \right) \right] \nonumber
\end{align}
From Appendix \hyperref[A-4]{A-4}, we get
\begin{align}
\frac{\Delta}{6} \left( \begin{tikzpicture}
		\draw(-6,-12) circle(0.25);
		\draw(-6.5,-12) -- (-5.5,-12);
	\end{tikzpicture} \right)&=\frac{g^2 S_1(m,T)}{2(4 \pi)^2} \left(\psi(1)+\ln \left( \frac{4 \pi \mu^2}{m^2} \right)\right) \\
	&+\frac{g^2 S_1(m,T)}{2(4 \pi)^2}\left(2-\frac{\sqrt{3} \pi}{3} \right) \nonumber \\
	&+\frac{g^2m^2}{64 \pi^4}  Y(m,T) \nonumber
\end{align}
with
\begin{align}
Y(m,T)&=\int_0^\infty \int_0^\infty U(x) U(y) G(x,y) \ dx \ dy \\
U(x)&=\frac{\sinh(x)}{\exp \left( \beta m \cosh(x) \right)-1}\\
	G(x,y)&=\ln \left( \frac{1+2 \cosh(x-y)}{1+2 \cosh(x+y)} \frac{1-2 \cosh(x+y)}{1-2 \cosh(x-y)} \right) 
\end{align}

Therefore by combining the above results, we get
\begin{align}\label{Gammadiff}
\begin{split}
\Gamma^{\text{diff}}_{n_k,\vec{k}=0}&= \  \frac{g}{2} \ S_1(m,T) \\ 
  &- \frac{3g^2}{4} \frac{S_1(m,T)}{(4 \pi)^2}   \left[ \psi(1)+\ln\left( \frac{4 \pi \mu^2}{m^2} \right) \right]  \\
  &- \frac{g^2}{4(4\pi)}S_0(m,T) S_1(m,T) \\
  & +\frac{g^2 m^2}{4 (4 \pi)^3} S_0(m,T) \left[ \psi(2)+\ln \left( \frac{4 \pi \mu^2}{m^2} \right) \right] \\ 
	&- \frac{g^2m^2}{64 \pi^4} Y(m,T) \\
	& -\frac{g^2}{32 \pi^2} S_1(m,T)  \left[ 2-\frac{\pi}{\sqrt{3}}   \right] 
\end{split}
\end{align}
Thus combining \cref{Gammadiff} and \cref{rge1} and expanding it as a polynomial in \emph{g} looks like a polynomial of order four. However, as shown in \cref{coupling derivation}, the coefficients of $g^2,g$ and $g^0$ are zero and lead to first-degree polynomial. When we rearrange the linear equation in \emph{g}, we get
\begin{equation}\label{coupling}
g(m,T,\mu)=\frac{A(m,T) \ln(\mu)+B(m,T)}{C(m,T) \ln(\mu)+D(m,T)}
\end{equation}
with
\begin{align}
A&=\left[-\gamma_{m1}V_{2, \ln m} -2 \beta_2 V_2(m,T) \right] \\
C&=\left[2(\beta_3-\gamma_2)V_2+\gamma_{m2}V_{2,\ln m} \right] \\
B&=(2 \gamma_2-\beta_3)T_1- 2 \beta_2 V_1(m,T) \\
&- \gamma_{m1} V_{1, \ln m} - \gamma_{m2}T_{1, \ln m} \nonumber \\
D&=2(\beta_3-\gamma_2)V_1+\gamma_{m2}V_{1,\ln m}  
\end{align}

\begin{align}
T_1 &= \frac{1}{2} S_1(m,T) \\
T_{1, \ln m} &= -\frac{m^2}{4 \pi} S_0(m,T)
\end{align}
\begin{align}
\begin{split}
	V_1(m,T) &= \frac{m^2}{4(4 \pi)^3}S_0(m,T) \left[ \psi(2)+ \ln \left( \frac{4 \pi}{m^2}\right) \right] \\
	& -\frac{3}{4} \frac{S_1(m,T)}{(4 \pi)^2} \left[ \psi(1)+ \ln \left( \frac{4 \pi}{m^2} \right)  \right] \\
&-\frac{1}{4(4 \pi)} S_0(m,T) S_1(m,T) -\frac{m^2}{64 \pi^4} Y(m,T) \\
&-\frac{S_1(m,T)}{32 \pi^2} \left[ 2- \frac{\sqrt{3} \pi}{3} \right]  \\
\end{split}
\end{align}
\begin{equation}
V_2(m,T) = \left( \frac{m^2}{2(4 \pi)^3}S_0(m,T) - \frac{3S_1(m,T)}{2(4 \pi)^2} \right) 
\end{equation}
\begin{equation}
	V_{2, \ln m} = \frac{4m^2 S_0(m,T)}{(4 \pi)^3} - \frac{m^4 S_{-1}(m,T)}{(4 \pi)^4}
\end{equation}
\begin{align}
	V_{1,\ln m} &=\frac{2 m^2}{(4 \pi)^3} S_0(m,T) \left[ \psi(1)+\ln \left( \frac{4 \pi}{m^2} \right) \right]  \\
	&-\frac{m^4}{2(4 \pi)^4} S_{-1}(m,T) \left[ \psi(2)+\ln \left(\frac{4 \pi}{m^2} \right) \right] \nonumber \\
	& + \frac{3}{2} \frac{S_1(m,T)}{(4 \pi)^2}+\frac{m^2 S_0^2(m,T)}{2(4 \pi)^2} \nonumber \\
	&+ \frac{m^2}{2(4 \pi)^2}S_1(m,T)S_{-1}(m,T) \nonumber \\
	&+\frac{m^2 S_0(m,T)}{(4 \pi)^3} \left[ 2 - \frac{\sqrt{3} \pi}{3} \right]  - \frac{m^2}{32 \pi^4} Y(m,T) \nonumber\\
	&-\frac{m^4}{32 \pi^4} \frac{\partial Y(m,T)}{\partial m^2} \nonumber
\end{align}
The other trivial value of \emph{g} that satisfies these equations is zero, so we do not care about those solutions. Instead, choosing another physically possible solution (as in \cref{coupling}) leads to a thermal-dependent coupling constant.
If we combine with beta coupling constant relation such as
\begin{equation}\label{mass-scale0}
\frac{d \ g(\mu)}{d \ \ln(\mu)} = \beta_2 g^2+ \beta_3 g^3
\end{equation}
give rise to the result
\begin{align}\label{mass-scale}
\ln(\mu) &=\int^{g} \frac{1}{\beta_2 t^2+ \beta_3 t^3} dt \\ \nonumber &= - \frac{1}{\beta_2 \ g}+ \frac{\beta_3}{\beta_2^2} \ln \left(\beta_3+ \frac{\beta_2}{g} \right)+\ln \mu_0
\end{align}
Similarly, the corresponding running mass coupling relation is
\begin{equation}\label{runningmass}
\frac{d \ \ln(m)}{d \ \ln(\mu)}=\gamma_m (g)
\end{equation}
When combined with \cref{mass-scale0}, it results in
\begin{align}
&\frac{\partial  \ \ln(m)}{\partial g } \frac{d g}{d \ \ln(\mu)}=\gamma_m(g) \\ 
&\implies \frac{\partial \ \ln(m)}{\partial g}=\frac{\gamma_m(g)}{\beta(g)} = \frac{\gamma_{m1}+\gamma_{m2} g}{\beta_2 g+ \beta_3 g^2}  \nonumber \\
& \ln \left( \frac{m}{m_0} \right) =\chi_2 + \frac{\gamma_{m1}}{\beta_2} \ln (g) + \left( \frac{\gamma_{m2}}{\beta_3}-\frac{\gamma_{m1}}{\beta_2} \right)  \ln(\beta_3 \ g+\beta_2) \label{mass-coupling}
\end{align}
where $\ln \mu_0$, $m_0$, and $\chi_2$ are the respective integration constants. \\

Solving \cref{mass-scale,mass-coupling,coupling} simultaneously, we get temperature-dependent running mass and coupling constant. \\
 
 Combining the above results with the quasi-particle model of Bannur \cite{Bannur2007a, Bannur2007b, Bannur2007}, we get expression for energy density and pressure as 
 \begin{align}\label{energydensity}
\varepsilon(T) &= g_f \frac{ m^4}{2 \pi^2} \sum_{n=1}^\infty \left[  \frac{3K_2(\frac{nm}{T})}{(\frac{nm}{T})^2}+\frac{K_1(\frac{nm}{T})}{\frac{nm}{T}} \right] \\
& = g_f \frac{m^4}{16 \pi^2} \sum_{n=1}^\infty \left[ K_4 \left(\frac{nm}{T} \right)-K_0 \left(\frac{nm}{T} \right) \right]
\end{align}
 and 
 \begin{align}\label{pressure}
 \frac{P}{T}-\frac{P_0}{T_0}=\int_{T_0}^T \frac{\varepsilon(m(T),T)}{T^2} \ dT
 \end{align}
 
 \section{Results and Discussion}\label{resultanddiscussion}
We have derived an equation relating running mass, mass scale, coupling, and temperature, i.e., \cref{coupling}. Similarly, we have the equation connecting running coupling with mass-scale known as beta function equation as in \cref{mass-scale}. Furthermore, we have an equation connecting running mass and coupling as in \cref{mass-coupling}. Solving \cref{coupling,mass-scale,mass-coupling} for each temperature value will give us the results as shown in the figures.
If one wishes to keep the $\mu$ independent from this set of equations, then as pointed out in \cite{Kleinert2001}, one can redefine it as $\mu \to \sigma \mu$, in that case, our above expressions will be a particular case of $\sigma=1$. One will have to rewrite all equations of RGE concerning these changes, and a new differential equation between $\sigma$ and $\mu$ will appear, as shown in \cite{Kleinert2001}.\\
%
\makeatletter
\renewcommand{\fnum@figure}{Fig. \thefigure}
\makeatother
\begin{figure}[h]\label{fig1}
\includegraphics[scale=0.4]{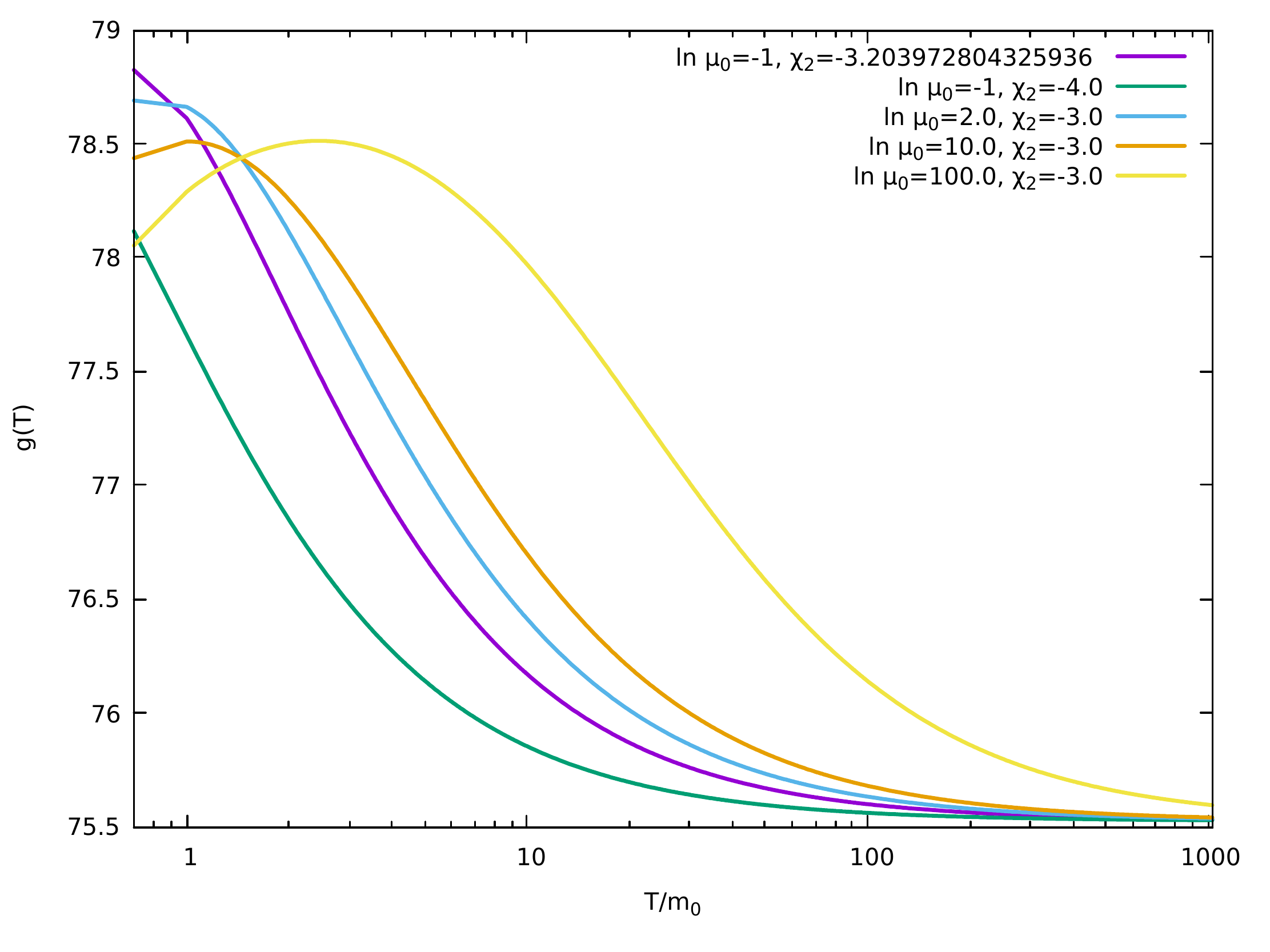}
\caption{Two loop coupling constant results. \emph{g} against $T/m_0$ plotted with varying values of integration constants $\ln \mu_0$ and $\chi_2$ with $m_0 \approx 1$}
 \end{figure}
 \\
 \begin{figure}[h]\label{fig2}
\includegraphics[scale=0.4]{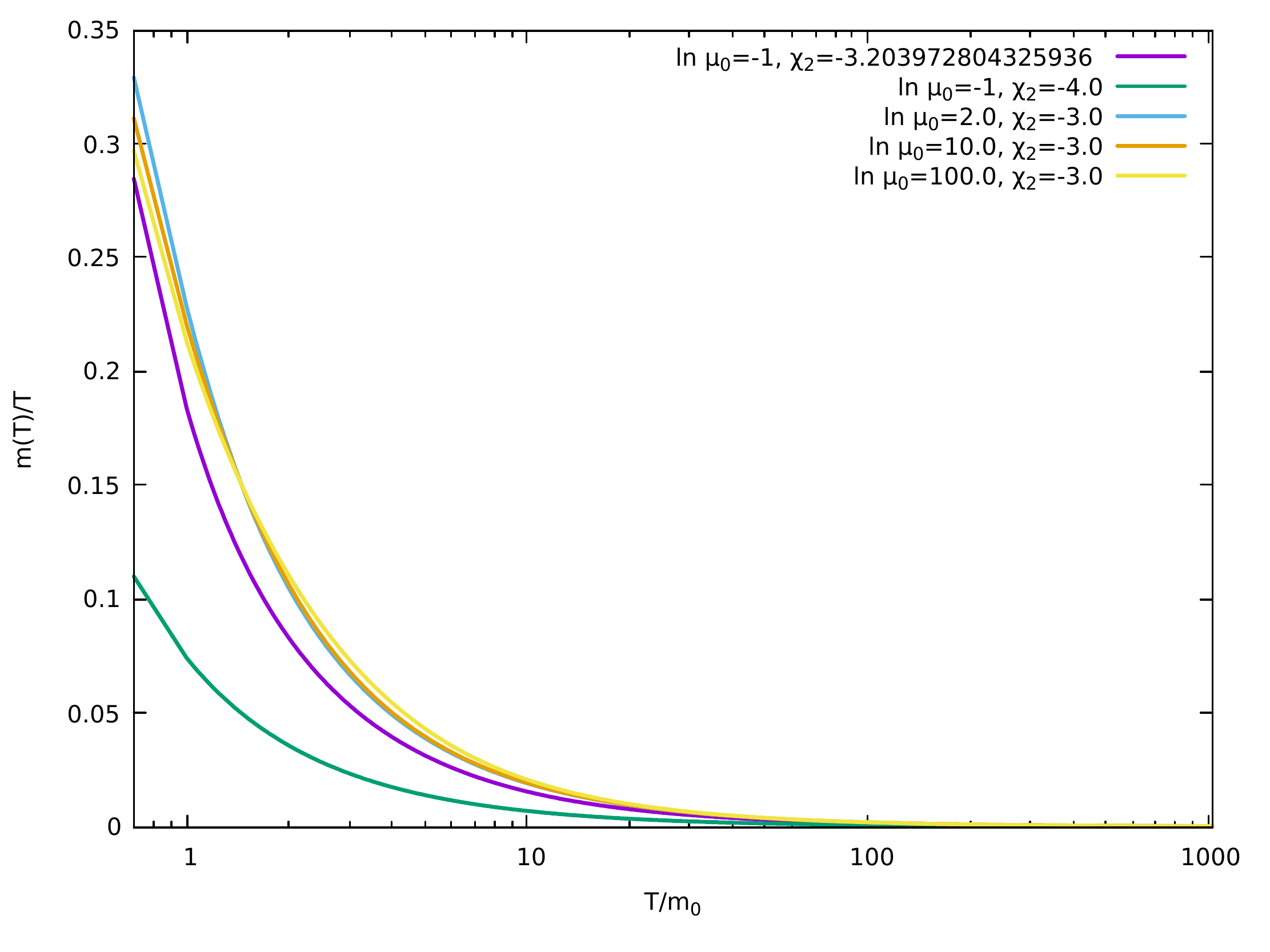}
\caption{Two loop running mass results. The difference between the curves is due to the different integration constants, as shown in the figure.}
 \end{figure}
  \begin{figure}[h]\label{fig3}
\includegraphics[scale=0.4]{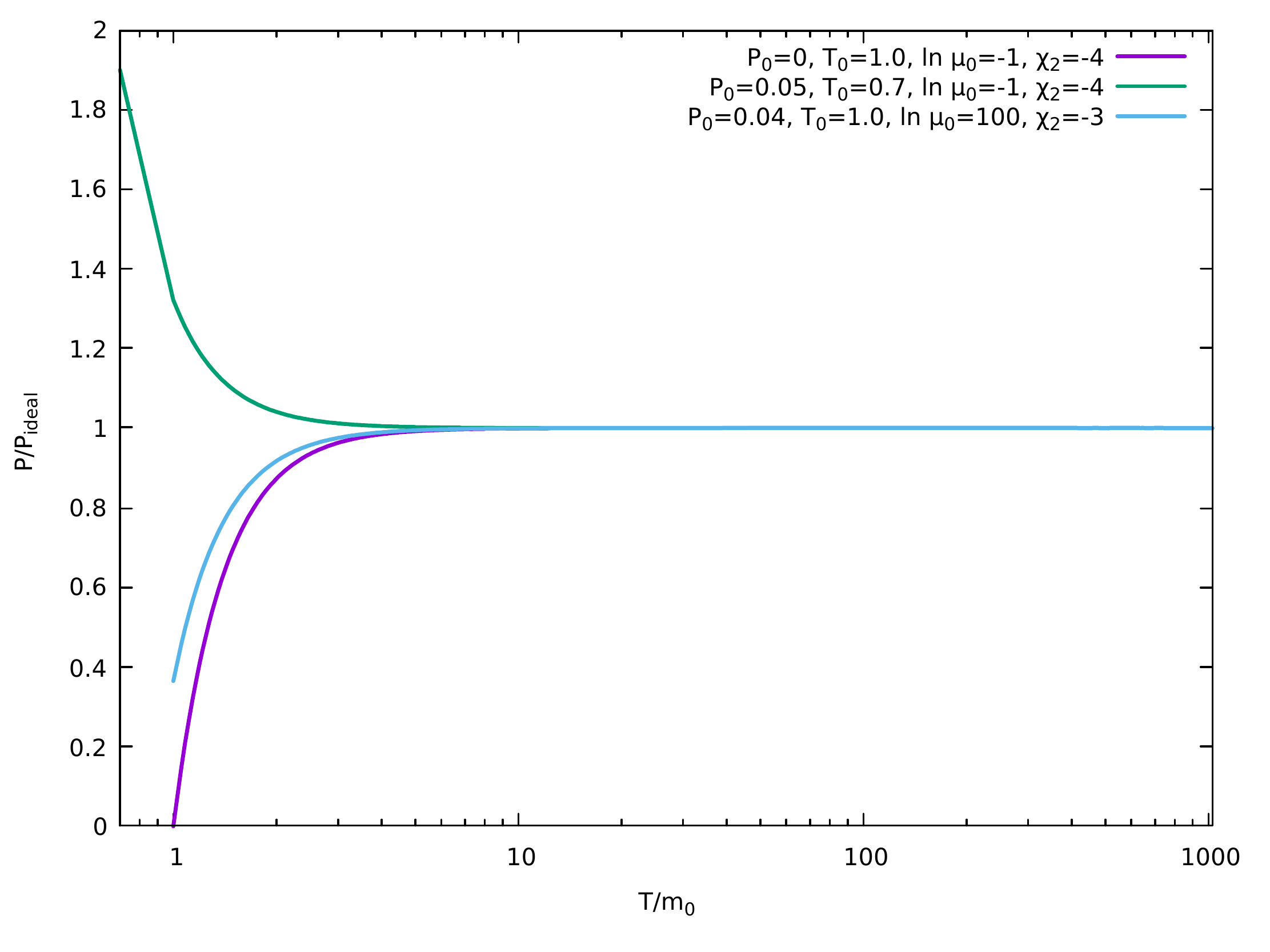}
\caption{Pressure scaled by $\frac{\pi^2}{90} T^4$  against $T/m_0$, with varying values of $T_0, P_0, \ln \mu_0, \chi_2$.}
 \end{figure}
   \begin{figure}[h]\label{fig4}
\includegraphics[scale=0.4]{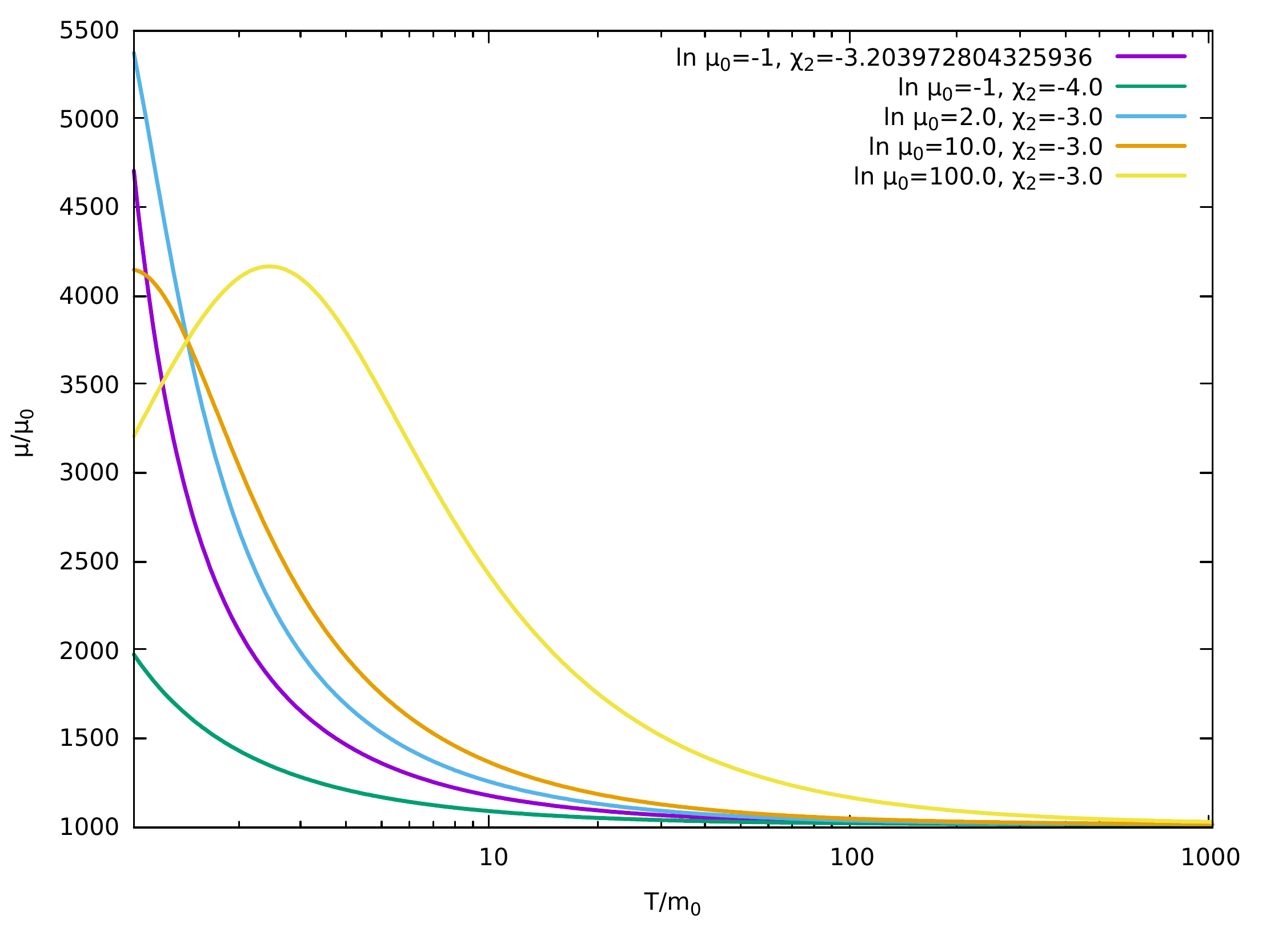}
\caption{$  \frac{\mu}{\mu_0} $ plotted against $T/m_0$ with varying integration constants $\ln(\mu_0)$ and $\chi_2$ with $m_0 \approx  1$}
 \end{figure}
We plot the main results in Fig.1-4, with varying integration constants. Two loop coupling constant results and two loop running mass results are shown in Fig. 1 and Fig. 2, respectively. It is evident from Fig. 2 that as temperature goes to infinity, the running mass per temperature goes to zero. To understand the nature of pressure proposed by the model, we plot pressure against temperature with different values of $T_0$, $P_0$ in Fig. 3. These results show similar trends, i.e., $\beta  \to 0$, i.e., $T \to \infty$ pressure reaches the ideal limit $P_{SB}=P_{ideal}=\frac{\pi^2}{90} T^4$, irrespective of the initial value. $P_{SB}$ is Stefan Boltzmann's limit of pressure. In Fig. 4, with a given $\ln \mu_0$ and $\chi_2$, at high temperatures, the ratio $\mu/\mu_0$ goes to a constant value. The same trend can be seen in Fig. 1, i.e., going to a constant value at a high-temperature limit. Fitting with lattice data can be used to find appropriate integration constants $\mu_0$ and $\chi_2$.
When $T \to 0$, as per \hyperref[couplinglimit]{Appendix D-1}, $g \approx -(\gamma_{m1}/\gamma_{m2})$. As a result, at T=0, $\ln(\mu/\mu_0)$ becomes a complex number. However, one can make the $\mu$ a real number by choosing $\mu_0$, a negative number/complex number/complex function that can cancel the complex factor appearing on the right-hand side \cref{mass-scale}.
The same holds for running mass. So by choosing the appropriate integration constants/functions at zero temperature limit, one can still make the mass scale and running mass real or complex. These integral constants used in our model make it flexible. However, the best way to determine the value of integration constants will be to fit the actual data with the corresponding function once it becomes available. In our model rather than selecting the mass scale ourselves, we set the mass scale $\mu$ to be a parameter that obeys RGE. The temperature dependence came naturally. \\

The quasi-particle model\cite{Bannur2007a, Bannur2007b, Bannur2007} derives energy density from standard statistical mechanics at relativistic Bose-Einstein distribution. The pressure \cref{pressure} has the property that at $T \to T_0$, the integral part of the equation becomes zero. Hence, the pressure converges to the $P_0$ value. From \hyperref[limitpressure]{Appendix D-2}, it is clear that as $T \to 0$, energy density $\varepsilon(T) \to 0$ and $\frac{\varepsilon(T)}{T^2} \to 0$.

\section{Conclusion}\label{conclusion}
We conclude that the renormalization constants in imaginary time formalism with one and two loop approximations are identical to those in non-thermal QFT. In our new approach, by  applying RGE on thermal and non-thermal vertex functions simultaneously, we derived the coupling constant and corresponding running mass at zero momentum limits, where self-energy is analytic \cite{Takao}. Running mass and coupling constant with temperature dependence with varying integration constants are plotted in figures for massive $\phi^4$ field theory. Combining these results with the quasi-particle model of Bannur \cite{Bannur2007,Bannur2007a,Bannur2007b} (\cref{energydensity,pressure}) and by choosing different values to $P_0$ and $T_0$, we have evaluated the pressure for the temperature dependent running mass. At higher temperatures, we observe pressure reaching its ideal limit $\frac{\pi^2}{90}T^4$ irrespective of its initial value of $P_ 0/T_0$. Our model is flexible such a way that once the experimental data or lattice data are available for some range, we will be able to tweak the integration constants in our model and compare. It will be interesting to know the results when we extend this model to QCD in the future.
 \section*{Acknowledgements}
 One of the authors (K. A) wishes to express gratitude to the Kerala State Council For Science, Technology and Environment for supporting the research by awarding KSCSTE Research Fellowship. 
\appendix 
\renewcommand{\theequation}{\thesection.\arabic{equation}}
\renewcommand{\thesubsection}{\thesection-\arabic{subsection}}
\begin{widetext}
\section{Integral equations ITF to Non thermal QFT Connections}
Here we derive the necessary diagrams for the calculation. Also, we used approximations $\lambda=g \mu^{\epsilon}$ on regularization, especially at the final stage. We used an operator $\mathcal{K}$, which picks up the diverging terms from the corresponding graphs it applied. i.e., $\mathcal{K} \left(\frac{A}{\epsilon^n}+B +c \epsilon \right)=\frac{A}{\epsilon^n}$.

\subsection{Diagram One loop A}\label{A-1}

We can write
\begin{align}\label{Eq. Tadpole2}
\left[\lbrace-\lambda \rbrace \begin{tikzpicture}
		\draw(-6,-11.75) circle(0.25);
		\draw(-6.5,-12) -- (-5.5,-12);
	\end{tikzpicture}_{ ITF} \right]&=\sumint \frac{1}{(2 \pi n T)^2+p^2+m^2}
	\\&=-\lambda \int T \sum_{n= -\infty}^\infty \frac{1}{\omega_n^2+\varepsilon_{p}^2} \frac{d^3p}{(2 \pi)^3} \nonumber \\
	&= -\lambda \int \frac{1}{P^2+m^2} \frac{d^4 P}{(2 \pi)^4} -\lambda S_1(m,T) \nonumber \\
	& = \left[\lbrace -\lambda \rbrace \begin{tikzpicture}
		\draw(-6,-11.75) circle(0.25);
		\draw(-6.5,-12) -- (-5.5,-12);
	\end{tikzpicture}_{ QFT} \right] -\lambda  S_{1}(m,T) \nonumber
\end{align}
where 
\begin{align}
S_1(m,T)=\int \frac{n_B(\beta \varepsilon_p)}{\varepsilon_p} \frac{d^3p}{(2 \pi)^3}=\frac{1}{\pi} \sum_{n=1}^\infty \left( \frac{m}{2 \pi n \beta} \right)K_1(n \beta m)
\end{align}
with $K(n,x)$ is the modified Bessel function of the second kind and $n_B(x)=(\exp(x)-1)^{-1}$. \\
The braces $\lbrace \lambda \rbrace$ serve as labels rather than a multiplication factor.
If we proceed with dimensional regularization, then Feynman diagram with substitution $\lambda=g \mu^\epsilon$ becomes
\begin{equation}\label{Eq.Tadpole2-1}
\left[\lbrace -g \mu^\epsilon \rbrace \begin{tikzpicture}
		\draw(-6,-11.75) circle(0.25);
		\draw(-6.5,-12) -- (-5.5,-12);
	\end{tikzpicture}_{ ITF} \right] = \frac{m^2g}{(4 \pi)^2}\left[ \frac{2}{\epsilon}+\psi(2)+\ln \left( \frac{4 \pi \mu^2}{m^2}  \right) \right]+\mathcal{O}(\epsilon) -g \mu^{\epsilon} S_{1}(m,T)
\end{equation}
where we used standard results \cite{Kleinert2001},
\begin{equation}
\left[\lbrace -g \mu^\epsilon \rbrace \begin{tikzpicture}
		\draw(-6,-11.75) circle(0.25);
		\draw(-6.5,-12) -- (-5.5,-12);
	\end{tikzpicture}_{QFT} \right] = \frac{m^2g}{(4 \pi)^2}\left[ \frac{2}{\epsilon}+\psi(2)+\ln \left( \frac{4 \pi \mu^2}{m^2}  \right) \right]+\mathcal{O}(\epsilon)
\end{equation}
\cref{Eq.Tadpole2-1} can be verified using a similar integral result \cite{Peshier1996,Andersen2001a}. The main difference is that we have defined 
\begin{equation}
\lambda \sumint f(p)=g \mu^{\epsilon} \sum_{n_p=-\infty}^\infty \int f(p) \frac{d^{N-\epsilon}p}{(2 \pi)^{N-\epsilon}}
\end{equation}
while \cite{Peshier1996,Andersen2001a} defined it as
\begin{equation}
g^2 \sumint_{\bar{MS}} f(p)= g^2 \left( \frac{e^{\gamma }\mu^2}{4 \pi}  \right)^{\epsilon} \sum_{n_p=-\infty}^\infty \int f(p) \frac{d^{N-2\epsilon}p}{(2 \pi)^{N-2\epsilon}}
\end{equation}
\subsection{Diagram One loop B}\label{A-2}
\begin{align}
\int \frac{d^3p}{(2 \pi)^3} \ T \sum_{n_p= -\infty}^\infty \frac{1}{\omega_{n_p}^2+\varepsilon_p^2} \frac{1}{\omega_{n_p-n_r}^2+\varepsilon_{p-r}^2} = \int \frac{1}{P^2+m^2}\frac{1}{(P-R)^2+m^2} \frac{d^4P}{(2 \pi)^4} \\
+\sum_{\sigma,\sigma_1=\pm 1} \int  \frac{n_B(\beta \varepsilon_p)}{2 \varepsilon_p \varepsilon_{p+r}} \frac{1}{\sigma_1 \varepsilon_p+\varepsilon_{p+r}+ i \sigma \omega_{n_r}} \frac{d^3p}{(2 \pi)^3} \nonumber
\end{align}
Corresponding diagrammatic expression is
\begin{align}
\lbrace (g \mu^\epsilon)^2 \rbrace \begin{tikzpicture}
		\draw(-6,-11.75) circle(0.25);
		\draw(-6.25,-11.75) -- (-6.43,-11.6);
		\draw(-6.25,-11.75) -- (-6.43,-11.92);
		\draw(-5.75,-11.75) -- (-5.57,-11.6);
		\draw(-5.75,-11.75) -- (-5.57,-11.92);
	\end{tikzpicture}_{ITF}= \lbrace (g \mu^\epsilon)^2 \rbrace
\begin{tikzpicture}
		\draw(-6,-11.75) circle(0.25);
		\draw(-6.25,-11.75) -- (-6.43,-11.6);
		\draw(-6.25,-11.75) -- (-6.43,-11.92);
		\draw(-5.75,-11.75) -- (-5.57,-11.6);
		\draw(-5.75,-11.75) -- (-5.57,-11.92);
	\end{tikzpicture}_{QFT,R_0=\omega_{n_r}} + (g \mu^\epsilon)^2 W(r,n_r)
\end{align}
with
\begin{align}
W(r,n_r)=\int \frac{d^3p}{(2 \pi)^3} \frac{2n_B(\beta \varepsilon_p)}{\varepsilon_p} \frac{r^2+ 2 pr \cos \theta+\omega_{n_r}^2}{(r^2+ 2 pr \cos \theta+\omega_{n_r}^2)^2+4 \varepsilon_p^2 \omega_{n_r}^2}
\end{align}
Applying dimensional regularization ($\lambda=g \mu^\epsilon$) and using the result from the standard textbook \cite{Kleinert2001}, we can write
\begin{align}
\lbrace g^2 \mu^\epsilon \rbrace \begin{tikzpicture}
		\draw(-6,-11.75) circle(0.25);
		\draw(-6.25,-11.75) -- (-6.43,-11.6);
		\draw(-6.25,-11.75) -- (-6.43,-11.92);
		\draw(-5.75,-11.75) -- (-5.57,-11.6);
		\draw(-5.75,-11.75) -- (-5.57,-11.92);
	\end{tikzpicture}_{ITF}= \frac{g^2\mu^\epsilon}{(4 \pi)^2} \left( \frac{2}{\epsilon}+\psi(1)+\int_0^1 dx \ln \left[ \frac{4 \pi \mu^2}{R^2x(1-x)+m^2} \right]|_{R_0=\omega_{n_r}}+\mathcal{O}(\epsilon) \right) + g^2 \mu^{2 \epsilon} W(r,n_r)
\end{align}
with $R=[\omega_{n_r},r]$
\subsection{Diagram Two loop C}\label{A-3}
\begin{align}
\lbrace \lambda^2 \rbrace\begin{tikzpicture}
		\draw(-6,-11.75) circle(0.25);
		\draw(-6.5,-12) -- (-5.5,-12);
		\draw(-6,-11.25) circle(0.25);
\end{tikzpicture}_{ITF}=\int \lambda^2 T^2 \sum_{n_{p1}=- \infty}^\infty \sum_{n_{p2}=- \infty}^\infty
\frac{1}{\omega_{n_{p1}}^2+\varepsilon_{p1}^2}  \left[\frac{1}{\omega_{n_{p2}}^2+\varepsilon_{p2}^2} \right]^2 \frac{d^3p_1}{(2 \pi)^3} \frac{d^3p_2}{(2 \pi)^3} \\
= -\lambda T \int \sum_{n_{p1}=- \infty}^\infty \frac{1}{\omega_{n_{p1}}^2+\varepsilon_{p1}^2} \frac{d^3 p_1}{(2 \pi)^3} \times -\frac{\partial}{\partial m^2}
\left[ -\lambda T \int \sum_{n_{p1}=- \infty}^\infty
\frac{1}{\omega_{n_{p1}}^2+\varepsilon_{p1}^2} \frac{d^3p_1}{(2 \pi)^3} \right] \nonumber
\end{align}
Using the results from \cite{Kleinert2001} and from \cref{Eq. Tadpole2} we can write
\begin{align}
	\begin{tikzpicture}
		\draw(-6,-11.75) circle(0.25);
		\draw(-6.5,-12) -- (-5.5,-12);
		\draw(-6,-11.25) circle(0.25);
\end{tikzpicture}_{ITF}&=	\begin{tikzpicture}
		\draw(-6,-11.75) circle(0.25);
		\draw(-6.5,-12) -- (-5.5,-12);
		\draw(-6,-11.25) circle(0.25);
\end{tikzpicture}_{QFT}
+\frac{g^2}{4 \pi}S_1(m,T)S_0(m,T) \\
&-\frac{g}{4 \pi}S_0(m,T) \left[  \begin{tikzpicture}
		\draw(-6,-11.75) circle(0.25);
		\draw(-6.5,-12) -- (-5.5,-12);
	\end{tikzpicture}_{QFT} \right]
	+ g S_1(m,T) \frac{\partial}{\partial m^2}  \begin{tikzpicture}
		\draw(-6,-11.75) circle(0.25);
		\draw(-6.5,-12) -- (-5.5,-12);
	\end{tikzpicture}_{QFT} \nonumber
	\end{align}
with
\begin{align}
S_N(m,T)=\frac{1}{\pi} \sum_{n=1}^\infty \left( \frac{m}{2 \pi n \beta} \right)^N \ K_N(nm\beta)
\end{align}
\begin{align}
	\begin{tikzpicture}
		\draw(-6,-11.75) circle(0.25);
		\draw(-6.5,-12) -- (-5.5,-12);
		\draw(-6,-11.25) circle(0.25);
\end{tikzpicture}_{QFT}=-\frac{m^2g^2}{(4 \pi)^4} \left[ \frac{4}{\epsilon^2}+2 \frac{\psi(1)+\psi(2)}{\epsilon}-\frac{4}{\epsilon} \ln \left( \frac{m^2}{4 \pi \mu^2} \right)+\mathcal{O}(\epsilon^0) \right]
\end{align}

\begin{equation}
\left[ \begin{tikzpicture}
		\draw(-6,-11.75) circle(0.25);
		\draw(-6.5,-12) -- (-5.5,-12);
	\end{tikzpicture}_{ QFT} \right] = \frac{m^2g}{(4 \pi)^2}\left[ \frac{2}{\epsilon}+\psi(2)+\ln \left( \frac{4 \pi \mu^2}{m^2}  \right) \right]+\mathcal{O}(\epsilon)
\end{equation}

\begin{equation}
\frac{\partial}{\partial m^2} \left[ \begin{tikzpicture}
		\draw(-6,-11.75) circle(0.25);
		\draw(-6.5,-12) -- (-5.5,-12);
	\end{tikzpicture}_{ QFT} \right] = \frac{g}{(4 \pi)^2}\left[ \frac{2}{\epsilon}+\psi(1)+\ln \left( \frac{4 \pi \mu^2}{m^2}  \right) \right]+\mathcal{O}(\epsilon)
\end{equation}

\subsection{Diagram Two loop D}\label{A-4}
\begin{equation}\label{sunset4}
I=\begin{tikzpicture}
		\draw(-6,-12) circle(0.25);
		\draw(-6.5,-12) -- (-5.5,-12);
	\end{tikzpicture}_{ITF}=\lambda^2 T^2\int \sum_{n_{p}=-\infty}^\infty \sum_{n_{q}=-\infty}^\infty
\frac{1}{\omega_{n_p}^2+\varepsilon_p^2} \frac{(2 \pi)^3}{\omega_{n_q}^2+\varepsilon_q^2}\frac{\delta^3(p+q+r+s)}{\omega_{n_p+n_q+n_s}^2+\varepsilon_{r}^2}  \frac{d^3p}{(2 \pi)^3} \frac{d^3q}{(2 \pi)^3}\frac{d^3r}{(2 \pi)^3} 
\end{equation}

From \cite{Bugrij1995,Andersen2001a} the integral can be expressed as $I=I_1+I_2+I_3$
where
\begin{align}
I_1&=\begin{tikzpicture}
		\draw(-6,-12) circle(0.25);
		\draw(-6.5,-12) -- (-5.5,-12);
	\end{tikzpicture}_{QFT}
=\int \frac{\lambda^2}{P^2+m^2} \frac{1}{Q^2+m^2} \frac{1}{R^2+m^2} (2 \pi)^4 \delta^4(P+Q+R+S) \frac{d^4P}{(2 \pi)^4} \frac{d^4Q}{(2 \pi)^4} \frac{d^4R}{(2 \pi)^4}
\end{align}
with $S=[\omega_{n_s},\vec{s}]$
\begin{align}
I_2=\int \frac{d^3p}{(2 \pi)^3}\frac{3n_B(\beta \varepsilon_p)}{2 \varepsilon_p} \sum_{\sigma_1=\pm 1} \begin{tikzpicture}
		\draw(-6,-11.75) circle(0.25);
		\draw(-6.25,-11.75) -- (-6.43,-11.6);
		\draw(-6.25,-11.75) -- (-6.43,-11.92);
		\draw(-5.75,-11.75) -- (-5.57,-11.6);
		\draw(-5.75,-11.75) -- (-5.57,-11.92);
	\end{tikzpicture}\left( (P+S)^2 \right)_{QFT} \nonumber
\end{align}
with 
\begin{align}
P=[i \sigma_1 \varepsilon_p,\vec{p}], \ S=[\omega_{n_s},\vec{s}] \nonumber
\end{align}
and
\begin{align}
\begin{tikzpicture}
		\draw(-6,-11.75) circle(0.25);
		\draw(-6.25,-11.75) -- (-6.43,-11.6);
		\draw(-6.25,-11.75) -- (-6.43,-11.92);
		\draw(-5.75,-11.75) -- (-5.57,-11.6);
		\draw(-5.75,-11.75) -- (-5.57,-11.92);
	\end{tikzpicture}_{QFT}(K^2)=\int \frac{\lambda^2}{R^2+m^2} \frac{1}{Q^2+m^2} (2 \pi)^4 \delta^4(R+Q+K) \frac{d^4R}{(2 \pi)^4} \frac{d^4Q}{(2 \pi)^4}
\end{align}
Similarly
\begin{align}
I_3=\lambda^2 \int \frac{d^3p}{(2 \pi)^3} \frac{d^3q}{(2 \pi)^3} \frac{3n_B(\beta \varepsilon_p) n_B(\beta \varepsilon_q)}{4 \varepsilon_p \varepsilon_q}  \times \sum_{\sigma_1, \sigma_2=\pm 1} \frac{1}{(i\sigma_1 \varepsilon_p+i\sigma_2 \varepsilon_q+\omega_{n_s})^2+(\vec{p}+\vec{q}+\vec{s})^2+m^2} \nonumber
\end{align}
with
\begin{align}
P=[i \sigma_1 \varepsilon_p, \vec{p}] \ Q=[i \sigma_2 \varepsilon_q, \vec{q}] , \ S=[\omega_{n_s},\vec{s}]
\end{align}
Now by taking the corresponding result from \cite{Kleinert2001} and Appendix \hyperref[A-2]{A-2}, we can write
\begin{align}
\begin{tikzpicture}
		\draw(-6,-12) circle(0.25);
		\draw(-6.5,-12) -- (-5.5,-12);
	\end{tikzpicture}_{ITF}=
 \begin{tikzpicture}
		\draw(-6,-12) circle(0.25);
		\draw(-6.5,-12) -- (-5.5,-12);
	\end{tikzpicture}_{QFT,S_0=\omega_{n_s}} 
	+\int \frac{d^3p}{(2 \pi)^3} \frac{3n_B(\beta \varepsilon_p)}{2 \varepsilon_p} \sum_{\sigma_1} \begin{tikzpicture}
		\draw(-6,-11.75) circle(0.25);
		\draw(-6.25,-11.75) -- (-6.43,-11.6);
		\draw(-6.25,-11.75) -- (-6.43,-11.92);
		\draw(-5.75,-11.75) -- (-5.57,-11.6);
		\draw(-5.75,-11.75) -- (-5.57,-11.92);
	\end{tikzpicture}_{QFT}(P+S)^2
	+ I_3
\end{align}
where
\begin{align}
\begin{tikzpicture}
		\draw(-6,-12) circle(0.25);
		\draw(-6.5,-12) -- (-5.5,-12);
	\end{tikzpicture}_{QFT,s_0=\omega_{n_s}}=-g^2 \frac{m^2}{(4 \pi)^4} \left( \frac{6}{\epsilon^2} +\frac{S^2}{2 m^2 \epsilon} \right)
	-g^2 \frac{m^2}{(4 \pi)^4} \frac{6}{\epsilon} \left[ \frac{3}{2}+\psi(1)+ \ln \left( \frac{4 \pi \mu^2}{m^2} \right) \right]
+\mathcal{O}(\epsilon)
\end{align}
with $S^2=\omega_{n_s}^2+s^2$
\begin{align}
\lbrace g^2\mu^\epsilon \rbrace
\sum_{\sigma_1} \begin{tikzpicture}
		\draw(-6,-11.75) circle(0.25);
		\draw(-6.25,-11.75) -- (-6.43,-11.6);
		\draw(-6.25,-11.75) -- (-6.43,-11.92);
		\draw(-5.75,-11.75) -- (-5.57,-11.6);
		\draw(-5.75,-11.75) -- (-5.57,-11.92);
	\end{tikzpicture}_{QFT}(P+S) &=\frac{2g^2\mu^\epsilon}{(4 \pi)^2} \left( \frac{2}{\epsilon}+\psi(1)\right) \\
	&+  \sum_{\sigma = \pm 1} \frac{g^2 \mu^\epsilon}{(4 \pi)^2} \left( \int_0^1 dx \ln \left[ \frac{4 \pi \mu^2}{[(i \sigma \varepsilon_p+\omega_{n_s})^2+(p+s)^2]x(1-x)+m^2} \right] \right) 
	+\mathcal{O}(\epsilon) \nonumber
\end{align}
When $S=0$
\begin{align}
\lbrace g^2\mu^\epsilon \rbrace
\sum_{\sigma_1} \begin{tikzpicture}
		\draw(-6,-11.75) circle(0.25);
		\draw(-6.25,-11.75) -- (-6.43,-11.6);
		\draw(-6.25,-11.75) -- (-6.43,-11.92);
		\draw(-5.75,-11.75) -- (-5.57,-11.6);
		\draw(-5.75,-11.75) -- (-5.57,-11.92);
	\end{tikzpicture}_{QFT}=\frac{2g^2\mu^\epsilon}{(4 \pi)^2} \left( \frac{2}{\epsilon}+\psi(1)\right)
	- \frac{2g^2 \mu^\epsilon}{(4 \pi)^2} \left( \int_0^1 dx \ln \left[ 1-x+x^2 \right] \right)
	+\frac{2g^2 \mu^\epsilon}{(4 \pi)^2}\ln \left( \frac{4 \pi \mu^2}{m^2} \right) 
	+\mathcal{O}(\epsilon)
\end{align}
Now combining results, we can write it as in the case of pole term, i.e.,
\begin{align}
 \mathcal{K} \left( \begin{tikzpicture}
		\draw(-6,-12) circle(0.25);
		\draw(-6.5,-12) -- (-5.5,-12);
	\end{tikzpicture}_{ITF} \right)= \mathcal{K} \left( \begin{tikzpicture}
		\draw(-6,-12) circle(0.25);
		\draw(-6.5,-12) -- (-5.5,-12);
	\end{tikzpicture}_{QFT,k_0=\omega_{n_k}} \right)+3 S_{1}(m,T) \ \mathcal{K}  \left(\begin{tikzpicture}
		\draw(-6,-11.75) circle(0.25);
		\draw(-6.25,-11.75) -- (-6.43,-11.6);
		\draw(-6.25,-11.75) -- (-6.43,-11.92);
		\draw(-5.75,-11.75) -- (-5.57,-11.6);
		\draw(-5.75,-11.75) -- (-5.57,-11.92);
\end{tikzpicture}_{QFT} \right) \\
= \mathcal{K} \left( \begin{tikzpicture}
		\draw(-6,-12) circle(0.25);
		\draw(-6.5,-12) -- (-5.5,-12);
	\end{tikzpicture}_{QFT,k_0=\omega_{n_k}} \right)+3gS_{1}(m,T) \ \frac{\partial}{\partial m^2} \mathcal{K} \left(  \begin{tikzpicture}
		\draw(-6,-11.75) circle(0.25);
		\draw(-6.5,-12) -- (-5.5,-12);
	\end{tikzpicture}_{QFT} \right)  \nonumber
\end{align}
When external momentum $S=0$, the integral result is
\begin{align}
 \begin{tikzpicture}
		\draw(-6,-12) circle(0.25);
		\draw(-6.5,-12) -- (-5.5,-12);
	\end{tikzpicture}_{ITF,S=0} &= \begin{tikzpicture}
		\draw(-6,-12) circle(0.25);
		\draw(-6.5,-12) -- (-5.5,-12);
	\end{tikzpicture}_{QFT,S=0}+3 S_{1}(m,T) \ \mathcal{K}  \left(\begin{tikzpicture}
		\draw(-6,-11.75) circle(0.25);
		\draw(-6.25,-11.75) -- (-6.43,-11.6);
		\draw(-6.25,-11.75) -- (-6.43,-11.92);
		\draw(-5.75,-11.75) -- (-5.57,-11.6);
		\draw(-5.75,-11.75) -- (-5.57,-11.92);
\end{tikzpicture}_{QFT} \right) 
 + 3S_1(m,T) \frac{g^2 \mu^\epsilon}{(4 \pi)^2} \left( \psi(1)+2-\frac{\sqrt{3} \pi}{3}+ \ln \left(\frac{4 \pi \mu^2}{m^2} \right) \right) \nonumber \\
	&+  \frac{3g^2m^2}{32 \pi^4}  \int_0^\infty \int_0^\infty U(x) U(y) G(x,y) \ dx \ dy
\end{align}
with
\begin{align}
U(x)=\frac{\sinh(x)}{\exp \left( \beta m \cosh(x) \right)-1}, \ G(x,y)=\ln \left( \frac{1+2 \cosh(x-y)}{1+2 \cosh(x+y)} \frac{1-2 \cosh(x+y)}{1-2 \cosh(x-y)} \right), \
\int_0^1 \ln(1-x+x^2) dx=\frac{\sqrt{3} \pi}{3}-2
\end{align}
The approximation can be verified using results from \cite{Andersen2001a}
\subsection{Diagram Two loop E}\label{A-5}
\begin{align}
\begin{tikzpicture}
		\draw(-6,-11.75) circle(0.25);
		\draw(-6.25,-11.75) -- (-6.43,-11.6);
		\draw(-6.25,-11.75) -- (-6.43,-11.92);
		\draw(-5.5,-11.75) circle(0.25);		
		\draw(-5.25,-11.75) -- (-5.07,-11.6);
		\draw(-5.25,-11.75) -- (-5.07,-11.92);
	\end{tikzpicture}_{ITF} &= -\frac{1}{\lambda} \left[\lambda^2 T \sum_{n_p=-\infty}^\infty \int  \frac{d^3 p}{(2 \pi)^3} \frac{1}{\varepsilon_{p-r}^2+{\omega_{n_p-n_r}^2}} \ \frac{1}{\varepsilon_p^2+\omega_{n_p}^2} \right]^2 \nonumber \\
	&=-\frac{1}{\lambda} \left[ \begin{tikzpicture}
		\draw(-6,-11.75) circle(0.25);
		\draw(-6.25,-11.75) -- (-6.43,-11.6);
		\draw(-6.25,-11.75) -- (-6.43,-11.92);
		\draw(-5.75,-11.75) -- (-5.57,-11.6);
		\draw(-5.75,-11.75) -- (-5.57,-11.92);
	\end{tikzpicture}_{ITF} \right]^2 \nonumber \\
	&\text{Using results from Appendix \hyperref[A-2]{A-2} }
	\nonumber \\
	&=-\frac{1}{\lambda} \left[  \begin{tikzpicture}
		\draw(-6,-11.75) circle(0.25);
		\draw(-6.25,-11.75) -- (-6.43,-11.6);
		\draw(-6.25,-11.75) -- (-6.43,-11.92);
		\draw(-5.75,-11.75) -- (-5.57,-11.6);
		\draw(-5.75,-11.75) -- (-5.57,-11.92);
	\end{tikzpicture}_{QFT}+ \lambda^2 \sum_{\sigma,\sigma_1=\pm 1} \int  \frac{n_B(\beta \varepsilon_p)}{2 \varepsilon_p \varepsilon_{p+r}} \frac{1}{\sigma_1 \varepsilon_p+\varepsilon_{p+r}+ i \sigma \omega_{n_r}} \frac{d^3p}{(2 \pi)^3} 
 \right]^2 \nonumber
\end{align}
If we put 
\begin{equation}
W(r,n_r)=\sum_{\sigma,\sigma_1=\pm 1} \int  \frac{n_B(\beta \varepsilon_p)}{2 \varepsilon_p \varepsilon_{p+r}} \frac{1}{\sigma_1 \varepsilon_p+\varepsilon_{p+r}+ i \sigma \omega_{n_r}} \frac{d^3p}{(2 \pi)^3} 
\end{equation}
Then 
\begin{align}
 \begin{tikzpicture}
		\draw(-6,-11.75) circle(0.25);
		\draw(-6.25,-11.75) -- (-6.43,-11.6);
		\draw(-6.25,-11.75) -- (-6.43,-11.92);
		\draw(-5.5,-11.75) circle(0.25);		
		\draw(-5.25,-11.75) -- (-5.07,-11.6);
		\draw(-5.25,-11.75) -- (-5.07,-11.92);
	\end{tikzpicture}_{ITF}=  \begin{tikzpicture}
		\draw(-6,-11.75) circle(0.25);
		\draw(-6.25,-11.75) -- (-6.43,-11.6);
		\draw(-6.25,-11.75) -- (-6.43,-11.92);
		\draw(-5.5,-11.75) circle(0.25);		
		\draw(-5.25,-11.75) -- (-5.07,-11.6);
		\draw(-5.25,-11.75) -- (-5.07,-11.92);
	\end{tikzpicture}_{QFT,r_0=\omega_{n_r}}-2 g W(r,n_r) \left[  \begin{tikzpicture}
		\draw(-6,-11.75) circle(0.25);
		\draw(-6.25,-11.75) -- (-6.43,-11.6);
		\draw(-6.25,-11.75) -- (-6.43,-11.92);
		\draw(-5.75,-11.75) -- (-5.57,-11.6);
		\draw(-5.75,-11.75) -- (-5.57,-11.92);
	\end{tikzpicture}_{QFT,r_0=\omega_{n_r}} \right] -\left( g \mu^\epsilon \right)^3W^2(r,n_r)
\end{align}
If we take results from \cite{Kleinert2001} and previous sections, we can write
\begin{align}
\begin{tikzpicture}
		\draw(-6,-11.75) circle(0.25);
		\draw(-6.25,-11.75) -- (-6.43,-11.6);
		\draw(-6.25,-11.75) -- (-6.43,-11.92);
		\draw(-5.5,-11.75) circle(0.25);		
		\draw(-5.25,-11.75) -- (-5.07,-11.6);
		\draw(-5.25,-11.75) -- (-5.07,-11.92);
	\end{tikzpicture}_{QFT,k_0=\omega_{n_k}}&= -\lambda^3 \int \frac{d^Np}{(2 \pi)^N} \frac{1}{(p-k)^2+m^2} \frac{1}{p^2+m^2} \int \frac{d^Nq}{(2 \pi)^N} \frac{1}{(q-k)^2+m^2} \frac{1}{q^2+m^2} \nonumber \\
	&\text{Setting $\lambda=g \mu^\epsilon$ , and as $N \to 4 - \epsilon$} \nonumber \\&=-g \mu^\epsilon \frac{g^2}{(4 \pi)^4} \left( \frac{4}{\epsilon^2}+\frac{4}{\epsilon} \psi(1) \right) 
	-g \mu^\epsilon \frac{g^2}{(4 \pi)^4} \frac{4}{\epsilon}\int_0^1 dx \ln \left[ \frac{4 \pi \mu^2}{K^2x(1-x)+m^2} \right]+\mathcal{O}(\epsilon^0)
\end{align}
and $K^2=\omega_{n_k}^2+k^2$
from Appendix \hyperref[A-2]{A-2}
\begin{align}
\begin{tikzpicture}
		\draw(-6,-11.75) circle(0.25);
		\draw(-6.25,-11.75) -- (-6.43,-11.6);
		\draw(-6.25,-11.75) -- (-6.43,-11.92);
		\draw(-5.75,-11.75) -- (-5.57,-11.6);
		\draw(-5.75,-11.75) -- (-5.57,-11.92);
	\end{tikzpicture}_{QFT}=\frac{g^2\mu^\epsilon}{(4 \pi)^2} \left( \frac{2}{\epsilon}+\psi(1)+\int_0^1 dx \ln \left[ \frac{4 \pi \mu^2}{K^2x(1-x)+m^2} \right] +\mathcal{O}(\epsilon)\right)
\end{align}
\subsection{Diagram Two loop F}\label{A-6}
\begin{align}
\lbrace - \lambda^3 \rbrace \begin{tikzpicture}
		\draw(-6,-11.375) circle(0.125);
		\draw(-6,-11.75) circle(0.25);
		\draw(-6.25,-11.75) -- (-6.43,-11.6);
		\draw(-6.25,-11.75) -- (-6.43,-11.92);
		\draw(-5.75,-11.75) -- (-5.57,-11.6);
		\draw(-5.75,-11.75) -- (-5.57,-11.92);
	\end{tikzpicture}_{ITF} &=-\lambda^3 T^2 \sum_{n_p=-\infty}^\infty \int \frac{d^3 p}{(2 \pi)^3} \frac{1}{\varepsilon_{p-r}^2+{\omega_{n_p-n_r}^2}} \ \frac{1}{(\varepsilon_p^2+\omega_{n_p}^2)^2} \sum_{n_q=-\infty}^\infty \int \frac{d^3 q}{(2 \pi)^3} \frac{1}{\varepsilon_q^2+\omega_{n_q}^2}  \\ \nonumber
=  & \left[  -\frac{1}{2} \frac{\partial}{\partial m^2} \sum_{n_p = -\infty}^\infty  \int \frac{d^3 p}{(2 \pi)^3} \frac{1}{\varepsilon_{p-r}^2+{\omega_{n_p-n_r}^2}} \ \frac{\lambda^2 T}{\varepsilon_p^2+\omega_{n_p}^2} \right] \times \sum_{n_q=-\infty}^\infty \int \frac{d^3 q}{(2 \pi)^3} \frac{-\lambda T}{\varepsilon_q^2+\omega_{n_q}^2} \\ \nonumber
&= \lbrace -\lambda\rbrace \begin{tikzpicture}
		\draw(-6,-11.75) circle(0.25);
		\draw(-6.5,-12) -- (-5.5,-12);
	\end{tikzpicture}_{ITF} \times -\frac{1}{2} \frac{\partial}{\partial m^2}\left[ \lbrace \lambda^2 \rbrace \begin{tikzpicture}
		\draw(-6,-11.75) circle(0.25);
		\draw(-6.25,-11.75) -- (-6.43,-11.6);
		\draw(-6.25,-11.75) -- (-6.43,-11.92);
		\draw(-5.75,-11.75) -- (-5.57,-11.6);
		\draw(-5.75,-11.75) -- (-5.57,-11.92);
	\end{tikzpicture}_{ITF} \right] \\ \nonumber
&	= \left( \lbrace -\lambda\rbrace \begin{tikzpicture}
		\draw(-6,-11.75) circle(0.25);
		\draw(-6.5,-12) -- (-5.5,-12);
	\end{tikzpicture}_{QFT}-\lambda S_1(m,T) \right)  \times -\frac{1}{2} \frac{\partial}{\partial m^2}\left[ \lbrace \lambda^2 \rbrace \begin{tikzpicture}
		\draw(-6,-11.75) circle(0.25);
		\draw(-6.25,-11.75) -- (-6.43,-11.6);
		\draw(-6.25,-11.75) -- (-6.43,-11.92);
		\draw(-5.75,-11.75) -- (-5.57,-11.6);
		\draw(-5.75,-11.75) -- (-5.57,-11.92);
	\end{tikzpicture}_{QFT}+\lambda^2 W(r,n_r) \right] \nonumber \\
&	=\lbrace -\lambda\rbrace \begin{tikzpicture}
		\draw(-6,-11.75) circle(0.25);
		\draw(-6.5,-12) -- (-5.5,-12);
	\end{tikzpicture}_{QFT} \times -\frac{1}{2} \frac{\partial}{\partial m^2} \left[ \lbrace \lambda^2 \rbrace \begin{tikzpicture}
		\draw(-6,-11.75) circle(0.25);
		\draw(-6.25,-11.75) -- (-6.43,-11.6);
		\draw(-6.25,-11.75) -- (-6.43,-11.92);
		\draw(-5.75,-11.75) -- (-5.57,-11.6);
		\draw(-5.75,-11.75) -- (-5.57,-11.92);
	\end{tikzpicture}_{QFT} \right]+(\dots) \nonumber
\end{align}
Solving we get
\begin{align}
 \begin{tikzpicture}
		\draw(-6,-11.375) circle(0.125);
		\draw(-6,-11.75) circle(0.25);
		\draw(-6.25,-11.75) -- (-6.43,-11.6);
		\draw(-6.25,-11.75) -- (-6.43,-11.92);
		\draw(-5.75,-11.75) -- (-5.57,-11.6);
		\draw(-5.75,-11.75) -- (-5.57,-11.92);
	\end{tikzpicture}_{ITF}&= \begin{tikzpicture}
		\draw(-6,-11.375) circle(0.125);
		\draw(-6,-11.75) circle(0.25);
		\draw(-6.25,-11.75) -- (-6.43,-11.6);
		\draw(-6.25,-11.75) -- (-6.43,-11.92);
		\draw(-5.75,-11.75) -- (-5.57,-11.6);
		\draw(-5.75,-11.75) -- (-5.57,-11.92);
	\end{tikzpicture}_{QFT,r_0=\omega_{n_r}}
+\frac{g S_1(m,T)}{2} \frac{\partial}{\partial m^2} \left[  \begin{tikzpicture}
		\draw(-6,-11.75) circle(0.25);
		\draw(-6.25,-11.75) -- (-6.43,-11.6);
		\draw(-6.25,-11.75) -- (-6.43,-11.92);
		\draw(-5.75,-11.75) -- (-5.57,-11.6);
		\draw(-5.75,-11.75) -- (-5.57,-11.92);
	\end{tikzpicture}_{QFT} \right] 
	-\frac{g^2}{2} \frac{\partial \ W(r,n_r)}{\partial m^2} \left[  \begin{tikzpicture}
		\draw(-6,-11.75) circle(0.25);
		\draw(-6.5,-12) -- (-5.5,-12);
	\end{tikzpicture}_{QFT}  \right]\\
	&+g^3 \frac{S_1(m,T)}{2}  \frac{\partial}{\partial m^2} W(r,n_r) \nonumber
\end{align}
Using \cite{Kleinert2001} and results from Appendices \hyperref[A-1]{A-1} and \hyperref[A-2]{A-2} we can write
\begin{align}
\mathcal{K} \left( \begin{tikzpicture}
		\draw(-6,-11.375) circle(0.125);
		\draw(-6,-11.75) circle(0.25);
		\draw(-6.25,-11.75) -- (-6.43,-11.6);
		\draw(-6.25,-11.75) -- (-6.43,-11.92);
		\draw(-5.75,-11.75) -- (-5.57,-11.6);
		\draw(-5.75,-11.75) -- (-5.57,-11.92);
	\end{tikzpicture}_{QFT,k_0=\omega_{n_k}} \right)=-2 \mathcal{K} \left[ \begin{tikzpicture}
		\draw(-6,-11.75) circle(0.25);
		\draw(-6.25,-11.75) -- (-6.43,-11.6);
		\draw(-6.25,-11.75) -- (-6.43,-11.92);
		\draw(-5.75,-11.75) -- (-5.57,-11.6);
		\draw(-5.75,-11.75) -- (-5.57,-11.92);
		\draw (-6,-11.5) node[cross,black]  {};
	\end{tikzpicture}_{QFT,k_0=\omega_{n_k}} \right] 
\end{align}
and $K^2=\omega_{n_k}^2+k^2$
\begin{align}
\frac{\partial}{\partial m^2} \left[  \begin{tikzpicture}
		\draw(-6,-11.75) circle(0.25);
		\draw(-6.25,-11.75) -- (-6.43,-11.6);
		\draw(-6.25,-11.75) -- (-6.43,-11.92);
		\draw(-5.75,-11.75) -- (-5.57,-11.6);
		\draw(-5.75,-11.75) -- (-5.57,-11.92);
	\end{tikzpicture}_{QFT} \right]=-\frac{g^2 \mu^\epsilon}{(4 \pi)^2}\int_0^1 \frac{1}{K^2 x(1-x)+m^2} dx
\end{align}
\begin{equation}
\left[ \begin{tikzpicture}
		\draw(-6,-11.75) circle(0.25);
		\draw(-6.5,-12) -- (-5.5,-12);
	\end{tikzpicture}_{ QFT} \right] = \frac{m^2g}{(4 \pi)^2}\left[ \frac{2}{\epsilon}+\psi(2)+\ln \left( \frac{4 \pi \mu^2}{m^2}  \right) \right]+\mathcal{O}(\epsilon)
\end{equation}
\subsection{Diagram Two loop G}\label{A-7}
We have to evaluate
\begin{align}\label{sunsethone1}
\begin{tikzpicture}
		\draw(-6,-12) circle(0.25);
		\draw(-6.5,-12) -- (-5.5,-12);
		\draw(-6,-11.75) -- (-6.2,-11.65);
		\draw(-6,-11.75) -- (-5.8,-11.65);
	\end{tikzpicture}=\int \sum_{n=-\infty}^{\infty} \sum_{\theta=-\infty}^{\infty} \frac{-\lambda^3}{\omega_n^2+\varepsilon_p^2} \ \frac{T^2}{\omega_\theta^2+\varepsilon_q^2} \ \frac{1}{\omega_{n-\alpha}^2+\varepsilon_r^2} \ \frac{(2 \pi)^6 \delta^6}{\omega_{n-\theta+\eta}^2+\varepsilon_s^2}   \ \frac{d^3r}{(2 \pi)^3} \ \frac{d^3s}{(2 \pi)^3} \frac{d^3p}{(2 \pi)^3} \frac{d^3q}{(2 \pi)^3}
\end{align}
where $\delta^6=\delta^3(\vec{r}+\vec{p}-\vec{k_1}-\vec{k_2}) \  \delta^3(\vec{s}+\vec{q}-\vec{p}-\vec{k}_3)$
. \\ The result expanded as the summation
\begin{align}
\begin{tikzpicture}
		\draw(-6,-12) circle(0.25);
		\draw(-6.5,-12) -- (-5.5,-12);
		\draw(-6,-11.75) -- (-6.2,-11.65);
		\draw(-6,-11.75) -- (-5.8,-11.65);
	\end{tikzpicture}=I_{11}+I_{21}+I_{22}+2I_{F1}+2I_{F2}+2I_{F3}+I_{F4}
\end{align}
where
\begin{align}
I_{11}=\int \frac{1}{P^2+m^2}\frac{1}{Q^2+m^2}\frac{1}{R^2+m^2}\frac{1}{S^2+m^2} \frac{d^4P}{(2 \pi)^4} \frac{d^4Q}{(2 \pi)^4} \frac{d^4R}{(2 \pi)^4} \frac{d^4S}{(2 \pi)^4}
 \\  \times (2 \pi)^4 \delta^4(R+P-K_1-K_2) (2 \pi)^4 \delta^4(S+Q-P-K_3) \nonumber
\end{align}
with
\begin{align}
R=\left[ r_0 \ , \ \vec{r} \right], \ P=\left[ p_0 \ , \ \vec{p} \right], \  K_1+K_2= \left[ \omega_\alpha \ , \ \vec{k}_1+\vec{k}_2 \right], \
K_3= \left[ \omega_\eta \ , \ \vec{k}_3 \right]
\end{align}
We define
\begin{align}
J(K^2)=\int \frac{1}{P^2+m^2} \frac{1}{(P-K)^2+m^2} \frac{d^4P}{(2 \pi)^4}
\end{align}
\begin{equation}
L(A,B,C)=\frac{1}{A^2+m^2} \frac{1}{B^2+m^2} \frac{1}{C^2+m^2}
\end{equation}
and
\begin{align}
G(A,B)=\frac{1}{A^2+m^2}\frac{1}{B^2+m^2}
\end{align}
Then 
\begin{align}
I_{21}= \int \frac{d^3p}{(2 \pi)^3} \frac{n_B(\beta \varepsilon_p)}{2 \varepsilon_p}  \sum_{\sigma = \pm 1} \left[ \frac{J \left[ (P+K_3)^2 \right]}{(P-K_1-K_2)^2+m^2} \right]_{p_0=- i \sigma \varepsilon_p}
\end{align}
Similarly
\small
\begin{align}
I_{22}= \int \frac{d^3r}{(2 \pi)^3} \frac{n_B(\beta \varepsilon_r)}{2 \varepsilon_r}  \sum_{\sigma = \pm 1} \left[ \frac{J \left[ (K_1+K_2+K_3-R)^2 \right]}{(R-K_1-K_2)^2+m^2} \right]_{r_0=- i \sigma \varepsilon_r}
\end{align}

\begin{equation}
I_{F1}=\int \frac{d^3q}{(2 \pi)^3} \frac{n_B(\beta \varepsilon_q)}{2 \varepsilon_q} \sum_{\sigma=\pm 1} L(R-K_1-K_2,R,R+Q-K_1-K_2-K_3) \frac{d^4R}{(2 \pi)^4}_{q_0=i \sigma \varepsilon_q}
\end{equation}

\begin{align}
I_{F2}= \int \frac{d^3p}{(2 \pi)^3} \frac{n_B(\beta \varepsilon_p)}{2 \varepsilon_p} \frac{d^3q}{(2 \pi)^3} \frac{n_B(\beta \varepsilon_q)}{2 \varepsilon_q}  \sum_{\sigma_1, \sigma_3 = \pm 1} G(P-K_1-K_2,Q-P-K_3)|_{p_0=i \sigma_1 \varepsilon_p}^{q_0=i \sigma_3 \varepsilon_q} 
\end{align}

\begin{align}
I_{F3}= \int \frac{d^3s}{(2 \pi)^3}\frac{d^3r}{(2 \pi)^3} \frac{n_B(\beta \varepsilon_s) n_B(\beta \varepsilon_r)}{4 \varepsilon_p  \varepsilon_s} \sum_{\sigma_1, \sigma_3 = \pm 1} G(R-K_1-K_2,S+R-K_1-K_2-K_3)|_{s_0=i \sigma_1 \varepsilon_s}^{r_0=i \sigma_3 \varepsilon_r} 
\end{align}

\begin{align}
I_{F4}=\int \frac{d^3q}{(2 \pi)^3} \frac{d^3s}{(2 \pi)^3} \frac{n_B(\beta \varepsilon_s)n_B(\beta \varepsilon_q)}{4 \varepsilon_s  \varepsilon_q}   \sum_{\sigma_1, \sigma_3 = \pm 1} G(S+Q-K_3,Q+S+K_1+K_2-K_3)|_{q_0=i \sigma_1 \varepsilon_q}^{s_0=i \sigma_3 \varepsilon_s} 
\end{align}
\normalsize

If we look at the integral, we can find one thing: the first three terms, $I_{11}$, $I_{21}$, and $I_{22}$ diverge, the rest becomes finite. i.e.,
I=$I_{11}+I_{21}+I_{22}$+ Finite terms ($2I_{F1}+2I_{F2}+2I_{F3}+I_{F4}$
).
If we define the pole finding operator $\mathcal{K}$, then by the structure, we can write
\small
\begin{align}
\begin{tikzpicture}
		\draw(-6,-12) circle(0.25);
		\draw(-6.5,-12) -- (-5.5,-12);
		\draw(-6,-11.75) -- (-6.2,-11.65);
		\draw(-6,-11.75) -- (-5.8,-11.65);
	\end{tikzpicture}_{ITF}=\begin{tikzpicture}
		\draw(-6,-12) circle(0.25);
		\draw(-6.5,-12) -- (-5.5,-12);
		\draw(-6,-11.75) -- (-6.2,-11.65);
		\draw(-6,-11.75) -- (-5.8,-11.65);
	\end{tikzpicture}_{QFT}+I_{21}+I_{22}+2(I_{F1}+I_{F2}+I_{F3})+I_{F4}
\end{align}
\begin{align}
&\mathcal{K} \left[ \lbrace -\lambda^3 \rbrace \begin{tikzpicture}
		\draw(-6,-12) circle(0.25);
		\draw(-6.5,-12) -- (-5.5,-12);
		\draw(-6,-11.75) -- (-6.2,-11.65);
		\draw(-6,-11.75) -- (-5.8,-11.65);
	\end{tikzpicture}_{ITF} \right]=\mathcal{K} \left[ \lbrace -\lambda^3 \rbrace \begin{tikzpicture}
		\draw(-6,-12) circle(0.25);
		\draw(-6.5,-12) -- (-5.5,-12);
		\draw(-6,-11.75) -- (-6.2,-11.65);
		\draw(-6,-11.75) -- (-5.8,-11.65);
	\end{tikzpicture}_{QFT, k_0= \omega_{n_k}} \right]  \\
	&-\left(\lambda \int \frac{d^3p}{(2 \pi)^3} \frac{n_B(\beta \varepsilon_p)}{2 \varepsilon_p} \times  \mathcal{K} \left[ \lbrace  \lambda^2 \rbrace  \sum_{\sigma =\pm 1} \frac{\begin{tikzpicture}
		\draw(-6,-11.75) circle(0.25);
		\draw(-6.25,-11.75) -- (-6.43,-11.6);
		\draw(-6.25,-11.75) -- (-6.43,-11.92);
		\draw(-5.75,-11.75) -- (-5.57,-11.6);
		\draw(-5.75,-11.75) -- (-5.57,-11.92);
	\end{tikzpicture}(P+K_3)+\begin{tikzpicture}
		\draw(-6,-11.75) circle(0.25);
		\draw(-6.25,-11.75) -- (-6.43,-11.6);
		\draw(-6.25,-11.75) -- (-6.43,-11.92);
		\draw(-5.75,-11.75) -- (-5.57,-11.6);
		\draw(-5.75,-11.75) -- (-5.57,-11.92);
	\end{tikzpicture}(-P+K_1+K_2+K_3)}{(P-K_1-K_2)^2+m^2}|_{p_0=- i \sigma \varepsilon_p} \right]\right) \nonumber
\end{align}
\normalsize
We rewrite
\begin{align}
\sum_{\sigma = \pm 1}\frac{1}{(P-K)^2+m^2} =  \sum_{\sigma = \pm 1}\frac{1}{(p-k)^2+(i \sigma \varepsilon_p + \omega_\alpha)^2+m^2} \\
= \frac{2 (k^2-2pk \cos \theta+\omega_{\alpha}^2)}{(k^2-2pk \cos \theta+\omega_{\alpha}^2)^2+4 \varepsilon_p^2 \omega_\alpha^2} \nonumber \\
=2 \frac{ (k^2+2pk \cos \theta+\omega_{\alpha}^2)}{(k^2+2pk \cos \theta+\omega_{\alpha}^2)^2+4 \varepsilon_p^2 \omega_\alpha^2} \nonumber
\end{align}
\begin{align}
W(r,n_r)=\int \frac{d^3p}{(2 \pi)^3} \frac{2n_B(\beta \varepsilon_p)}{\varepsilon_p} \frac{r^2+ 2 pr \cos \theta+\omega_{n_r}^2}{(r^2+ 2 pr \cos \theta+\omega_{n_r}^2)^2+4 \varepsilon_p^2 \omega_{n_r}^2}
\end{align}
We know that from Appendix \hyperref[A-2]{A-2}
\begin{align}
\mathcal{K} \left(
\begin{tikzpicture}
		\draw(-6,-11.75) circle(0.25);
		\draw(-6.25,-11.75) -- (-6.43,-11.6);
		\draw(-6.25,-11.75) -- (-6.43,-11.92);
		\draw(-5.75,-11.75) -- (-5.57,-11.6);
		\draw(-5.75,-11.75) -- (-5.57,-11.92);
	\end{tikzpicture}_{QFT}\right)=\frac{g^2\mu^\epsilon}{(4 \pi)^2} \left( \frac{2}{\epsilon}\right)
\end{align}
So
\begin{align}
\mathcal{K} \left[ \begin{tikzpicture}
		\draw(-6,-12) circle(0.25);
		\draw(-6.5,-12) -- (-5.5,-12);
		\draw(-6,-11.75) -- (-6.2,-11.65);
		\draw(-6,-11.75) -- (-5.8,-11.65);
	\end{tikzpicture}_{ITF} \right]&=\mathcal{K} \left[ \begin{tikzpicture}
		\draw(-6,-12) circle(0.25);
		\draw(-6.5,-12) -- (-5.5,-12);
		\draw(-6,-11.75) -- (-6.2,-11.65);
		\draw(-6,-11.75) -- (-5.8,-11.65);
	\end{tikzpicture}_{QFT,k_{0i}=\omega_{n_i}} \right] \nonumber \\
	&-g\mu^\epsilon \left(\int \frac{d^3p}{(2 \pi)^3} \frac{2 n_B(\beta \varepsilon_p)}{ \varepsilon_p}  \frac{ (k^2+2pk \cos \theta+\omega_{\alpha}^2)}{(k^2+2pk \cos \theta+\omega_{\alpha}^2)^2+4 \varepsilon_p^2 \omega_\alpha^2} \mathcal{K} \left(
\begin{tikzpicture}
		\draw(-6,-11.75) circle(0.25);
		\draw(-6.25,-11.75) -- (-6.43,-11.6);
		\draw(-6.25,-11.75) -- (-6.43,-11.92);
		\draw(-5.75,-11.75) -- (-5.57,-11.6);
		\draw(-5.75,-11.75) -- (-5.57,-11.92);
	\end{tikzpicture}_{QFT}\right)\right)
\end{align}
So the pole term relation can be written as
\begin{align}
\mathcal{K} \left[ \begin{tikzpicture}
		\draw(-6,-12) circle(0.25);
		\draw(-6.5,-12) -- (-5.5,-12);
		\draw(-6,-11.75) -- (-6.2,-11.65);
		\draw(-6,-11.75) -- (-5.8,-11.65);
	\end{tikzpicture}_{ITF} \right]&=\mathcal{K} \left[ \begin{tikzpicture}
		\draw(-6,-12) circle(0.25);
		\draw(-6.5,-12) -- (-5.5,-12);
		\draw(-6,-11.75) -- (-6.2,-11.65);
		\draw(-6,-11.75) -- (-5.8,-11.65);
	\end{tikzpicture}_{QFT,k_{0i}=\omega_{n_i}} \right]-g\mu^{\epsilon} W(k_i,n_{k_i}) \mathcal{K} \left(
\begin{tikzpicture}
		\draw(-6,-11.75) circle(0.25);
		\draw(-6.25,-11.75) -- (-6.43,-11.6);
		\draw(-6.25,-11.75) -- (-6.43,-11.92);
		\draw(-5.75,-11.75) -- (-5.57,-11.6);
		\draw(-5.75,-11.75) -- (-5.57,-11.92);
	\end{tikzpicture}_{QFT}\right)
\end{align}
where
\begin{align}
\mathcal{K} \left[ \begin{tikzpicture}
		\draw(-6,-12) circle(0.25);
		\draw(-6.5,-12) -- (-5.5,-12);
		\draw(-6,-11.75) -- (-6.2,-11.65);
		\draw(-6,-11.75) -- (-5.8,-11.65);
	\end{tikzpicture}_{QFT} \right]&=g \mu^\epsilon \frac{g^2}{(4 \pi)^4} \frac{2}{\epsilon^2} \left(1+\frac{\epsilon}{2}+\epsilon \ \psi(1) \right) -g \mu^\epsilon \frac{g^2}{(4 \pi)^4} \frac{2}{\epsilon}\int_0^1 dx \ln \left[ \frac{(K_1+K_2)^2 x(1-x)+m^2}{4 \pi \mu^2} \right]
\end{align}
with $K_i=[\omega_{n_{k_i}},\vec{k}_i]$
\section{Counter term diagrams}
\subsection{Counter term 1}\label{B-1}
From \cite{Kleinert2001}, the counter term for divergence for the four-point function derived is
\begin{align}
\begin{tikzpicture}
\draw[black,fill=black] (-6,-12) circle(0.5ex);
\draw (-6,-12) node[cross,rotate=0] {};
\end{tikzpicture}_{QFT}=- \mu^\epsilon g c_g^1=-\frac{3}{2} \mathcal{K} \left(\begin{tikzpicture}
		\draw(-6,-11.75) circle(0.25);
		\draw(-6.25,-11.75) -- (-6.43,-11.6);
		\draw(-6.25,-11.75) -- (-6.43,-11.92);
		\draw(-5.75,-11.75) -- (-5.57,-11.6);
		\draw(-5.75,-11.75) -- (-5.57,-11.92);
\end{tikzpicture}_{QFT} \right)
\end{align}
The corresponding diagram in imaginary time formalism is
\begin{align}
\begin{tikzpicture}
\draw[black,fill=black] (-6,-12) circle(0.5ex);
\draw (-6,-12) node[cross,rotate=0] {};
\end{tikzpicture}_{ITF}=- \mu^\epsilon g c_g^1&=-\frac{3}{2} \mathcal{K} \left(\begin{tikzpicture}
		\draw(-6,-11.75) circle(0.25);
		\draw(-6.25,-11.75) -- (-6.43,-11.6);
		\draw(-6.25,-11.75) -- (-6.43,-11.92);
		\draw(-5.75,-11.75) -- (-5.57,-11.6);
		\draw(-5.75,-11.75) -- (-5.57,-11.92);
\end{tikzpicture}_{ITF} \right) 
\end{align}

From Appendix \hyperref[A-2]{A-2}, one can find that the diverging term is the same for the diagram in imaginary time formalism and non-thermal QFT; thus, we can write 
\begin{align}
\begin{tikzpicture}
\draw[black,fill=black] (-6,-12) circle(0.5ex);
\draw (-6,-12) node[cross,rotate=0] {};
\end{tikzpicture}_{ITF}=-\frac{3}{2} \mathcal{K} \left(\begin{tikzpicture}
		\draw(-6,-11.75) circle(0.25);
		\draw(-6.25,-11.75) -- (-6.43,-11.6);
		\draw(-6.25,-11.75) -- (-6.43,-11.92);
		\draw(-5.75,-11.75) -- (-5.57,-11.6);
		\draw(-5.75,-11.75) -- (-5.57,-11.92);
\end{tikzpicture}_{QFT} \right)\\
=\begin{tikzpicture}
\draw[black,fill=black] (-6,-12) circle(0.5ex);
\draw (-6,-12) node[cross,rotate=0] {};
\end{tikzpicture}_{QFT}=-\mu^\epsilon g \frac{3g}{(4 \pi)^2}\frac{1}{\epsilon} \nonumber
\end{align}
\subsection{Counter term 2}\label{B-2}
From \cite{Kleinert2001} 
Defining $*$ operator the substitution of the counter term $-m^2c_{m^2}$ or $-\mu^\epsilon g c_g$, we can express counter terms as
\begin{align}
\begin{tikzpicture}
		\draw(-6,-11.75) circle(0.25);
		\draw(-6.5,-12) -- (-5.5,-12);
		\draw (-6,-11.5) node[cross,rotate=0] {};
\end{tikzpicture}_{QFT}={\begin{tikzpicture}
		\draw(-6,-11.75) circle(0.25);
		\draw(-6.5,-12) -- (-5.5,-12);
		\draw[black,fill=black] (-6,-11.5) circle(0.5ex);
	\end{tikzpicture}}_{QFT} * -\frac{1}{2} \mathcal{K} \left[  \begin{tikzpicture}
		\draw(-6,-11.75) circle(0.25);
		\draw(-6.5,-12) -- (-5.5,-12);
	\end{tikzpicture}_{QFT}
 \right] \\
=-g \mu^\epsilon \frac{-\partial}{\partial m^2}{\begin{tikzpicture}
		\draw(-6,-11.75) circle(0.25);
		\draw(-6.5,-12) -- (-5.5,-12);
	\end{tikzpicture}}_{QFT} \times \frac{1}{2 g \mu^\epsilon} \mathcal{K} \left[  \begin{tikzpicture}
		\draw(-6,-11.75) circle(0.25);
		\draw(-6.5,-12) -- (-5.5,-12);
	\end{tikzpicture}_{QFT}
 \right] \nonumber
\end{align}

We have from Appendix \hyperref[A-1]{A-1} the relation

\begin{align}
\mathcal{K} \left[  \begin{tikzpicture}
		\draw(-6,-11.75) circle(0.25);
		\draw(-6.5,-12) -- (-5.5,-12);
	\end{tikzpicture}_{ITF} \right]=\mathcal{K} \left[  \begin{tikzpicture}
		\draw(-6,-11.75) circle(0.25);
		\draw(-6.5,-12) -- (-5.5,-12);
	\end{tikzpicture}_{QFT}
 \right]
\end{align}
So, for ITF, the corresponding derivation is
\begin{align}
\begin{tikzpicture}
		\draw(-6,-11.75) circle(0.25);
		\draw(-6.5,-12) -- (-5.5,-12);
		\draw (-6,-11.5) node[cross,rotate=0] {};
\end{tikzpicture}_{ITF}&={\begin{tikzpicture}
		\draw(-6,-11.75) circle(0.25);
		\draw(-6.5,-12) -- (-5.5,-12);
		\draw[black,fill=black] (-6,-11.5) circle(0.5ex);
	\end{tikzpicture}}_{ITF}* -\frac{1}{2} \mathcal{K} \left[  \begin{tikzpicture}
		\draw(-6,-11.75) circle(0.25);
		\draw(-6.5,-12) -- (-5.5,-12);
	\end{tikzpicture}_{QFT}
 \right] \\
&=- g \mu^\epsilon \frac{-\partial}{\partial m^2}{ \begin{tikzpicture}
		\draw(-6,-11.75) circle(0.25);
		\draw(-6.5,-12) -- (-5.5,-12);
	\end{tikzpicture}}_{ITF} \times \frac{1}{2 g \mu^\epsilon} \mathcal{K} \left[  \begin{tikzpicture}
		\draw(-6,-11.75) circle(0.25);
		\draw(-6.5,-12) -- (-5.5,-12);
	\end{tikzpicture}_{QFT}
 \right] \nonumber \\
&=-g \mu^\epsilon \left(\frac{-\partial}{\partial m^2}{ \begin{tikzpicture}
		\draw(-6,-11.75) circle(0.25);
		\draw(-6.5,-12) -- (-5.5,-12);
	\end{tikzpicture}}_{QFT}-\frac{g S_0(m,T)}{4 \pi}\right) \times  \frac{1}{2g \mu^\epsilon} \mathcal{K} \left[  \begin{tikzpicture}
		\draw(-6,-11.75) circle(0.25);
		\draw(-6.5,-12) -- (-5.5,-12);
	\end{tikzpicture}_{QFT} \right] \nonumber
	\end{align}
So
\begin{align}
\begin{tikzpicture}
		\draw(-6,-11.75) circle(0.25);
		\draw(-6.5,-12) -- (-5.5,-12);
		\draw (-6,-11.5) node[cross,rotate=0] {};
\end{tikzpicture}_{ITF}&=\begin{tikzpicture}
		\draw(-6,-11.75) circle(0.25);
		\draw(-6.5,-12) -- (-5.5,-12);
		\draw (-6,-11.5) node[cross,rotate=0] {};
\end{tikzpicture}_{QFT}+\frac{g}{4 \pi} \frac{S_0(m,T)}{2} \mathcal{K}\left[  \begin{tikzpicture}
		\draw(-6,-11.75) circle(0.25);
		\draw(-6.5,-12) -- (-5.5,-12);
	\end{tikzpicture}_{QFT} \right] \nonumber \\
	&=\frac{2m^2g^2}{(4 \pi)^4} \left[ \frac{1}{\epsilon^2}+\frac{\psi(1)}{2 \epsilon}-\frac{1}{2 \epsilon} \ln \left( \frac{m^2}{4 \pi \mu^2} \right)+\mathcal{O}(\epsilon^0) \right] +\frac{g^2 m^2 S_0(m,T)}{(4 \pi)^3} \frac{1}{\epsilon} 
\end{align}
\subsection{Counter term 3}\label{B-3}
From \cite{Kleinert2001}, the calculation proceeds as
\begin{align}
\begin{tikzpicture}
		\draw(-6,-11.75) circle(0.25);
		\draw(-6.5,-12) -- (-5.5,-12);
		\draw[black,fill=black] (-6,-12) circle(0.5ex);
	\end{tikzpicture}_{QFT}&=\begin{tikzpicture}
		\draw(-6,-11.75) circle(0.25);
		\draw(-6.5,-12) -- (-5.5,-12);
	\end{tikzpicture}_{QFT} \times \frac{-1}{g \mu^\epsilon}\times - \mu^\epsilon g c_g^1 \nonumber \\
	&=\begin{tikzpicture}
		\draw(-6,-11.75) circle(0.25);
		\draw(-6.5,-12) -- (-5.5,-12);
	\end{tikzpicture}_{QFT} \times \frac{-1}{g \mu^\epsilon} \times -\frac{3}{2} \mathcal{K} \left(\begin{tikzpicture}
		\draw(-6,-11.75) circle(0.25);
		\draw(-6.25,-11.75) -- (-6.43,-11.6);
		\draw(-6.25,-11.75) -- (-6.43,-11.92);
		\draw(-5.75,-11.75) -- (-5.57,-11.6);
		\draw(-5.75,-11.75) -- (-5.57,-11.92);
\end{tikzpicture}_{QFT} \right)
\end{align}
the corresponding diagram made with results from Appendices \hyperref[A-1]{A-1} and \hyperref[A-2]{A-2} is
\begin{align}
\begin{tikzpicture}
		\draw(-6,-11.75) circle(0.25);
		\draw(-6.5,-12) -- (-5.5,-12);
		\draw[black,fill=black] (-6,-12) circle(0.5ex);
	\end{tikzpicture}_{ITF}&=\begin{tikzpicture}
		\draw(-6,-11.75) circle(0.25);
		\draw(-6.5,-12) -- (-5.5,-12);
	\end{tikzpicture}_{ITF} \times \frac{-1}{g \mu^\epsilon} \times -\frac{3}{2} \mathcal{K} \left(\begin{tikzpicture}
		\draw(-6,-11.75) circle(0.25);
		\draw(-6.25,-11.75) -- (-6.43,-11.6);
		\draw(-6.25,-11.75) -- (-6.43,-11.92);
		\draw(-5.75,-11.75) -- (-5.57,-11.6);
		\draw(-5.75,-11.75) -- (-5.57,-11.92);
\end{tikzpicture}_{ITF} \right)\\
&=\left(\begin{tikzpicture}
		\draw(-6,-11.75) circle(0.25);
		\draw(-6.5,-12) -- (-5.5,-12);
	\end{tikzpicture}_{QFT}-g \mu^\epsilon S_1(m,T) \right) \times \frac{-1}{g \mu^\epsilon} \times -\frac{3}{2} \mathcal{K} \left(\begin{tikzpicture}
		\draw(-6,-11.75) circle(0.25);
		\draw(-6.25,-11.75) -- (-6.43,-11.6);
		\draw(-6.25,-11.75) -- (-6.43,-11.92);
		\draw(-5.75,-11.75) -- (-5.57,-11.6);
		\draw(-5.75,-11.75) -- (-5.57,-11.92);
\end{tikzpicture}_{ITF} \right) \nonumber \\
&=\begin{tikzpicture}
		\draw(-6,-11.75) circle(0.25);
		\draw(-6.5,-12) -- (-5.5,-12);
		\draw[black,fill=black] (-6,-12) circle(0.5ex);
	\end{tikzpicture}_{QFT}-\frac{3}{2}S_1(m,T) \mathcal{K} \left(\begin{tikzpicture}
		\draw(-6,-11.75) circle(0.25);
		\draw(-6.25,-11.75) -- (-6.43,-11.6);
		\draw(-6.25,-11.75) -- (-6.43,-11.92);
		\draw(-5.75,-11.75) -- (-5.57,-11.6);
		\draw(-5.75,-11.75) -- (-5.57,-11.92);
\end{tikzpicture}_{ITF} \right) \nonumber \\
&=\begin{tikzpicture}
		\draw(-6,-11.75) circle(0.25);
		\draw(-6.5,-12) -- (-5.5,-12);
		\draw[black,fill=black] (-6,-12) circle(0.5ex);
	\end{tikzpicture}_{QFT}-\frac{3}{2}S_1(m,T) \mathcal{K} \left(\begin{tikzpicture}
		\draw(-6,-11.75) circle(0.25);
		\draw(-6.25,-11.75) -- (-6.43,-11.6);
		\draw(-6.25,-11.75) -- (-6.43,-11.92);
		\draw(-5.75,-11.75) -- (-5.57,-11.6);
		\draw(-5.75,-11.75) -- (-5.57,-11.92);
\end{tikzpicture}_{QFT} \right) \nonumber \\
&=\frac{6m^2g^2}{(4 \pi)^4} \left[ \frac{1}{\epsilon^2}+\frac{\psi(2)}{2 \epsilon}-\frac{1}{2 \epsilon} \ln \left( \frac{m^2}{4 \pi \mu^2} \right)+\mathcal{O}(\epsilon^0) \right] \nonumber \\
&-\frac{3 \mu^\epsilon g^2}{(4 \pi)^2}\frac{S_1(m,T)}{\epsilon}  \nonumber
\end{align}
\subsection{Counter term 4}\label{B-4}
From \cite{Kleinert2001}, the diagram evaluated is
\begin{align}
\mathcal{K} \left[\begin{tikzpicture}
		\draw(-6,-11.75) circle(0.25);
		\draw(-6.25,-11.75) -- (-6.43,-11.6);
		\draw(-6.25,-11.75) -- (-6.43,-11.92);
		\draw(-5.75,-11.75) -- (-5.57,-11.6);
		\draw(-5.75,-11.75) -- (-5.57,-11.92);
		\draw[black,fill=black] (-5.75,-11.75) circle(0.5ex);
	\end{tikzpicture}\right]_{QFT}=\mathcal{K} \left[\begin{tikzpicture}
		\draw(-6,-11.75) circle(0.25);
		\draw(-6.25,-11.75) -- (-6.43,-11.6);
		\draw(-6.25,-11.75) -- (-6.43,-11.92);
		\draw(-5.75,-11.75) -- (-5.57,-11.6);
		\draw(-5.75,-11.75) -- (-5.57,-11.92);
	\end{tikzpicture}_{QFT} * -\frac{3}{2} \mathcal{K} \left( \begin{tikzpicture}
		\draw(-6,-11.75) circle(0.25);
		\draw(-6.25,-11.75) -- (-6.43,-11.6);
		\draw(-6.25,-11.75) -- (-6.43,-11.92);
		\draw(-5.75,-11.75) -- (-5.57,-11.6);
		\draw(-5.75,-11.75) -- (-5.57,-11.92);
	\end{tikzpicture}\right)_{QFT} \right] 
\end{align}
Using the results of Appendix \hyperref[A-2]{A-2} corresponding diagram in ITF can be written as
\begin{align}
\mathcal{K} \left[\begin{tikzpicture}
		\draw(-6,-11.75) circle(0.25);
		\draw(-6.25,-11.75) -- (-6.43,-11.6);
		\draw(-6.25,-11.75) -- (-6.43,-11.92);
		\draw(-5.75,-11.75) -- (-5.57,-11.6);
		\draw(-5.75,-11.75) -- (-5.57,-11.92);
		\draw[black,fill=black] (-5.75,-11.75) circle(0.5ex);
	\end{tikzpicture}\right]_{ITF}&=\mathcal{K} \left[\begin{tikzpicture}
		\draw(-6,-11.75) circle(0.25);
		\draw(-6.25,-11.75) -- (-6.43,-11.6);
		\draw(-6.25,-11.75) -- (-6.43,-11.92);
		\draw(-5.75,-11.75) -- (-5.57,-11.6);
		\draw(-5.75,-11.75) -- (-5.57,-11.92);
	\end{tikzpicture}_{ITF} * -\frac{3}{2} \mathcal{K} \left( \begin{tikzpicture}
		\draw(-6,-11.75) circle(0.25);
		\draw(-6.25,-11.75) -- (-6.43,-11.6);
		\draw(-6.25,-11.75) -- (-6.43,-11.92);
		\draw(-5.75,-11.75) -- (-5.57,-11.6);
		\draw(-5.75,-11.75) -- (-5.57,-11.92);
	\end{tikzpicture}\right)_{ITF} \right] \\
&=\mathcal{K} \left[\begin{tikzpicture}
		\draw(-6,-11.75) circle(0.25);
		\draw(-6.25,-11.75) -- (-6.43,-11.6);
		\draw(-6.25,-11.75) -- (-6.43,-11.92);
		\draw(-5.75,-11.75) -- (-5.57,-11.6);
		\draw(-5.75,-11.75) -- (-5.57,-11.92);
		\draw[black,fill=black] (-5.75,-11.75) circle(0.5ex);
	\end{tikzpicture}\right]_{QFT,k_0=\omega_{n_k}}+\frac{3 g W(k,n_k)}{2} \mathcal{K} \left( \begin{tikzpicture}
		\draw(-6,-11.75) circle(0.25);
		\draw(-6.25,-11.75) -- (-6.43,-11.6);
		\draw(-6.25,-11.75) -- (-6.43,-11.92);
		\draw(-5.75,-11.75) -- (-5.57,-11.6);
		\draw(-5.75,-11.75) -- (-5.57,-11.92);
	\end{tikzpicture}\right)_{ITF}   \nonumber \\
&=\mathcal{K} \left[\begin{tikzpicture}
		\draw(-6,-11.75) circle(0.25);
		\draw(-6.25,-11.75) -- (-6.43,-11.6);
		\draw(-6.25,-11.75) -- (-6.43,-11.92);
		\draw(-5.75,-11.75) -- (-5.57,-11.6);
		\draw(-5.75,-11.75) -- (-5.57,-11.92);
		\draw[black,fill=black] (-5.75,-11.75) circle(0.5ex);
	\end{tikzpicture}\right]_{QFT,k_0=\omega_{n_k}}+W(k,n_k)\frac{3g^3}{(4 \pi)^2} \frac{1}{\epsilon} \nonumber
\end{align}
with
\begin{align}
\mathcal{K} \left[\begin{tikzpicture}
		\draw(-6,-11.75) circle(0.25);
		\draw(-6.25,-11.75) -- (-6.43,-11.6);
		\draw(-6.25,-11.75) -- (-6.43,-11.92);
		\draw(-5.75,-11.75) -- (-5.57,-11.6);
		\draw(-5.75,-11.75) -- (-5.57,-11.92);
		\draw[black,fill=black] (-5.75,-11.75) circle(0.5ex);
	\end{tikzpicture}\right]_{QFT,k_0=\omega_{n_k}}=\frac{\mu^\epsilon g^3}{(4 \pi)^4} \left[ \frac{6}{\epsilon^2}+\frac{3 \psi(1)}{\epsilon}-\frac{3}{\epsilon}\int_0^1 \ln \left[ \frac{m^2+K^2(x(1-x))}{4 \pi \mu^2} \right] dx \right]
\end{align}
where $K^2=k^2+\omega_{n_k}^2$
\subsection{Counter term 5}\label{B-5}
From \cite{Kleinert2001}, one can derive the diagram as
\begin{align}
\mathcal{K} \left[ \begin{tikzpicture}
		\draw(-6,-11.75) circle(0.25);
		\draw(-6.25,-11.75) -- (-6.43,-11.6);
		\draw(-6.25,-11.75) -- (-6.43,-11.92);
		\draw(-5.75,-11.75) -- (-5.57,-11.6);
		\draw(-5.75,-11.75) -- (-5.57,-11.92);
		\draw (-6,-11.5) node[cross,black]  {};
	\end{tikzpicture}_{QFT} \right]&=\mathcal{K} \left( {-\lambda}^3 \int \frac{1}{(P^2+m^2)^2} \frac{1}{(P-K)^2+m^2} \frac{d^4P}{(2 \pi)^4} \times -m^2c^1_{m^2} \times \frac{-1}{g \mu^\epsilon} \right) \nonumber \\
&=	\mathcal{K} \left( \frac{\lambda}{2} \frac{\partial }{\partial m^2} \lambda^2 \int \frac{1}{(P^2+m^2)} \frac{1}{(P-K)^2+m^2} \frac{d^4P}{(2 \pi)^4} \times -m^2c^1_{m^2} \times \frac{-1}{g \mu^\epsilon} \right)  \nonumber \\
&=\mathcal{K} \left(-g \mu^\epsilon \times -\frac{1}{2} \frac{\partial }{\partial m^2} \begin{tikzpicture}
		\draw(-6,-11.75) circle(0.25);
		\draw(-6.25,-11.75) -- (-6.43,-11.6);
		\draw(-6.25,-11.75) -- (-6.43,-11.92);
		\draw(-5.75,-11.75) -- (-5.57,-11.6);
		\draw(-5.75,-11.75) -- (-5.57,-11.92);
	\end{tikzpicture}_{QFT} \times \frac{-1}{\mu^\epsilon g} \times -\frac{1}{2} \mathcal{K} \left[  \begin{tikzpicture}
		\draw(-6,-11.75) circle(0.25);
		\draw(-6.5,-12) -- (-5.5,-12);
	\end{tikzpicture}_{QFT} \right] \right) \nonumber \\
	&=\mathcal{K} \left(- \frac{g^2 \mu^\epsilon}{(4 \pi)^2} \left[\frac{1}{2} \int_0^1 \frac{1}{K^2x(1-x)+m^2} \ dx \right] \times \frac{m^2g}{(4 \pi)^2} \frac{1}{\epsilon} \right) \nonumber \\
	&=-\frac{1}{2} \mathcal{K} \left( \begin{tikzpicture}
		\draw(-6,-11.375) circle(0.125);
		\draw(-6,-11.75) circle(0.25);
		\draw(-6.25,-11.75) -- (-6.43,-11.6);
		\draw(-6.25,-11.75) -- (-6.43,-11.92);
		\draw(-5.75,-11.75) -- (-5.57,-11.6);
		\draw(-5.75,-11.75) -- (-5.57,-11.92);
	\end{tikzpicture}_{QFT,k_0=\omega_{n_k}} \right)
\end{align} The corresponding counter term in imaginary time formalism can be written with Appendices \hyperref[A-2]{A-2} and \hyperref[A-1]{A-1}  as
\begin{align}
\mathcal{K} \left[ \begin{tikzpicture}
		\draw(-6,-11.75) circle(0.25);
		\draw(-6.25,-11.75) -- (-6.43,-11.6);
		\draw(-6.25,-11.75) -- (-6.43,-11.92);
		\draw(-5.75,-11.75) -- (-5.57,-11.6);
		\draw(-5.75,-11.75) -- (-5.57,-11.92);
		\draw (-6,-11.5) node[cross,black]  {};
	\end{tikzpicture}_{ITF} \right]&=\mathcal{K} \left(-g \mu^\epsilon \times -\frac{1}{2} \frac{\partial }{\partial m^2} \begin{tikzpicture}
		\draw(-6,-11.75) circle(0.25);
		\draw(-6.25,-11.75) -- (-6.43,-11.6);
		\draw(-6.25,-11.75) -- (-6.43,-11.92);
		\draw(-5.75,-11.75) -- (-5.57,-11.6);
		\draw(-5.75,-11.75) -- (-5.57,-11.92);
	\end{tikzpicture}_{ITF} \times \frac{-1}{\mu^\epsilon g} \times -\frac{1}{2} \mathcal{K} \left[  \begin{tikzpicture}
		\draw(-6,-11.75) circle(0.25);
		\draw(-6.5,-12) -- (-5.5,-12);
	\end{tikzpicture}_{ITF} \right] \right)  \\ 
&=	\mathcal{K} \left(-g \mu^\epsilon  \left( -\frac{1}{2} \frac{\partial }{\partial m^2} \begin{tikzpicture}
		\draw(-6,-11.75) circle(0.25);
		\draw(-6.25,-11.75) -- (-6.43,-11.6);
		\draw(-6.25,-11.75) -- (-6.43,-11.92);
		\draw(-5.75,-11.75) -- (-5.57,-11.6);
		\draw(-5.75,-11.75) -- (-5.57,-11.92);
	\end{tikzpicture}_{QFT} -\frac{g^2}{2} \frac{\partial W(k,n_k)}{\partial m^2} \right) \frac{-1}{\mu^\epsilon g} \times \frac{-1}{2} \mathcal{K} \left[  \begin{tikzpicture}
		\draw(-6,-11.75) circle(0.25);
		\draw(-6.5,-12) -- (-5.5,-12);
	\end{tikzpicture}_{QFT} \right] \right) \nonumber \\ 
&=	 \mathcal{K} \left( \begin{tikzpicture}
		\draw(-6,-11.75) circle(0.25);
		\draw(-6.25,-11.75) -- (-6.43,-11.6);
		\draw(-6.25,-11.75) -- (-6.43,-11.92);
		\draw(-5.75,-11.75) -- (-5.57,-11.6);
		\draw(-5.75,-11.75) -- (-5.57,-11.92);
		\draw (-6,-11.5) node[cross,black]  {};
	\end{tikzpicture}_{QFT,k_0=\omega_{n_k}} \right) + \frac{1}{4} g^2  \left( \frac{\partial W(k,n_k)}{\partial m^2} \right)  \mathcal{K}\left[  \begin{tikzpicture}
		\draw(-6,-11.75) circle(0.25);
		\draw(-6.5,-12) -- (-5.5,-12);
\end{tikzpicture}_{QFT}  \right] = - \frac{1}{2} \mathcal{K} \left( \begin{tikzpicture}
		\draw(-6,-11.375) circle(0.125);
		\draw(-6,-11.75) circle(0.25);
		\draw(-6.25,-11.75) -- (-6.43,-11.6);
		\draw(-6.25,-11.75) -- (-6.43,-11.92);
		\draw(-5.75,-11.75) -- (-5.57,-11.6);
		\draw(-5.75,-11.75) -- (-5.57,-11.92);
	\end{tikzpicture}_{ITF} \right) \nonumber
\end{align}
It can be also derived using * operation \cite{Kleinert2001} i.e,
\begin{align}
\mathcal{K} \left[ \begin{tikzpicture}
		\draw(-6,-11.75) circle(0.25);
		\draw(-6.25,-11.75) -- (-6.43,-11.6);
		\draw(-6.25,-11.75) -- (-6.43,-11.92);
		\draw(-5.75,-11.75) -- (-5.57,-11.6);
		\draw(-5.75,-11.75) -- (-5.57,-11.92);
		\draw (-6,-11.5) node[cross,black]  {};
	\end{tikzpicture}_{ITF} \right]=\mathcal{K} \left(\begin{tikzpicture}
		\draw(-6,-11.75) circle(0.25);
		\draw(-6.25,-11.75) -- (-6.43,-11.6);
		\draw(-6.25,-11.75) -- (-6.43,-11.92);
		\draw(-5.75,-11.75) -- (-5.57,-11.6);
		\draw(-5.75,-11.75) -- (-5.57,-11.92);
		\draw[black,fill=black] (-6,-11.5) circle(0.5ex);
	\end{tikzpicture}_{ITF} * -\frac{1}{2} \mathcal{K} \left[  \begin{tikzpicture}
		\draw(-6,-11.75) circle(0.25);
		\draw(-6.5,-12) -- (-5.5,-12);
	\end{tikzpicture}_{ITF} \right] \right)
\end{align}
\section{Renormalization constants}\label{rmcoeff}
We have from \cite{Kleinert2001}
\begin{align} \label{eq-imp-1}
\gamma(g)&=-Z_{\phi,1} = - \epsilon c_{\phi}\\
\gamma_m(g)&=\frac{1}{2} \frac{g}{(4 \pi)^2}-\frac{1}{2} \frac{g^2}{(4 \pi)^4}+ \gamma(g) \label{eq-imp-2} \\
\beta(g)&=-\epsilon g +\frac{3g^2}{(4 \pi)^2}-\frac{6g^3}{(4 \pi)^4}+4g \gamma(g ) \label{eq-imp-3}
\end{align} 
\subsection{Case 1: $K \neq 0$}
In this case as per \cref{cphi} 
\begin{align}
K^2c_{\phi}=-\frac{g^2}{(4 \pi)^4} \frac{K^2}{12} \frac{1}{\epsilon}
\end{align}
\cref{eq-imp-1,eq-imp-2,eq-imp-3} becomes
\begin{align}\label{gammamg}
\gamma(g)&=\frac{g^2}{(4 \pi)^4} \frac{1}{12}\\
\gamma_m(g)&=\frac{1}{2} \frac{g}{(4 \pi)^2}-\frac{5}{12} \frac{g^2}{(4 \pi)^4}  \\
\beta(g)&=-\epsilon g +\frac{3g^2}{(4 \pi)^2}-\frac{17g^3}{3(4 \pi)^4} 
\end{align}
\subsection{Case 2: $K = 0$}
Now \cref{eq-imp-1,eq-imp-2,eq-imp-3} changes because of
\begin{align}
K^2c_{\phi}=0
\end{align}
So
\begin{align}
\gamma(g)&=0 \\
\gamma_m(g)&=\frac{1}{2} \frac{g}{(4 \pi)^2}-\frac{1}{2} \frac{g^2}{(4 \pi)^4}  \\
\beta(g)&=-\epsilon g +\frac{3g^2}{(4 \pi)^2}-\frac{6g^3}{(4 \pi)^4} 
\end{align}
\subsection{Relation}
We can relate them as
\begin{align}
\gamma(g)_{k =0}&=\gamma(g)_{k \neq 0}-\frac{g^2}{(4 \pi)^4} \frac{1}{12} =0\\
{\gamma_m(g)}_{k=0}&={\gamma_{m}(g)}_{k \neq 0}-\frac{g^2}{(4 \pi)^4} \frac{1}{12}  \\
\beta(g)_{k=0}&=\beta(g)_{k \neq 0} - \frac{1}{3} \frac{g^3}{(4 \pi)^4}
\end{align}
\section{Coupling constant form derivation}\label{coupling derivation}
We have
\begin{align}
\begin{split}
\Gamma^{\text{diff}}_{n_k,\vec{k}=0}&= \  \frac{g}{2} \ S_1(m,T) - \frac{3g^2}{4} \frac{S_1(m,T)}{(4 \pi)^2}   \left[ \psi(1)+\ln\left( \frac{4 \pi \mu^2}{m^2} \right) \right]  \\
  &- \frac{g^2}{4(4\pi)}S_0(m,T) S_1(m,T) + \frac{g^2 m^2}{4 (4 \pi)^3} S_0(m,T) \left[ \psi(2)+\ln \left( \frac{4 \pi \mu^2}{m^2} \right) \right] \\ 
	&- \frac{g^2m^2}{64 \pi^4} Y(m,T)  -\frac{g^2}{32 \pi^2} S_1(m,T)  \left[ 2-\frac{\pi}{\sqrt{3}}   \right] \text{.} 
\end{split}
\end{align}
Let us express
\begin{align}
	\Gamma^{\text{diff}}_{n_k,\vec{k}=0} &= g T_1+ g^2 T_2 \\
	 \frac{\partial}{\partial \ln \mu} \Gamma^{\text{diff}}_{n_k,\vec{k}=0} &= g T_{1,\ln \mu} + g^2 T_{2,\ln \mu}  \\
	 \frac{\partial}{\partial \ln m} \Gamma^{\text{diff}}_{n_k,\vec{k}=0} &= g  T_{1,\ln m} + g^2 T_{2,\ln m}
\end{align}
with
\begin{align}
	S_N(m,T) &= \frac{1}{\pi} \sum_{n=1}^\infty \left( \frac{m}{2 \pi n \beta } \right)^N K_N(nm \beta) \\
	T_1 &= \frac{1}{2} S_1(m,T) \label{T1} \\
	T_2 &= V_1(m,T)+ V_2(m,T) \ln (\mu) \nonumber 
\end{align}
\begin{align}
	T_{1, \ln \mu} &= 0 \label{T1mu}\\
	T_{1, \ln m} &= -\frac{m^2}{4 \pi} S_0(m,T) \label{T1m}\\
	T_{2, \ln \mu} &= V_2(m,T) \label{T2mu}\\
	T_{2, \ln m} & = V_{1, \ln m}+V_{2, \ln m} \ln \mu
	\end{align}
\begin{align}
\begin{split}
	V_1(m,T) &= \frac{m^2}{4(4 \pi)^3}S_0(m,T) \left[ \psi(2)+ \ln \left( \frac{4 \pi}{m^2}\right) \right]  -\frac{3}{4} \frac{S_1(m,T)}{(4 \pi)^2} \left[ \psi(1)+ \ln \left( \frac{4 \pi}{m^2} \right)  \right] \\
&-\frac{1}{4(4 \pi)} S_0(m,T) S_1(m,T) -\frac{m^2}{64 \pi^4} Y(m,T) -\frac{S_1(m,T)}{32 \pi^2} \left[ 2- \frac{\sqrt{3} \pi}{3} \right]  \\
\end{split}
\end{align}
\begin{equation}
V_2(m,T) = \left( \frac{m^2}{2(4 \pi)^3}S_0(m,T) - \frac{3S_1(m,T)}{2(4 \pi)^2} \right) 
\end{equation}
\begin{equation}
	V_{2, \ln m} = \frac{4m^2 S_0(m,T)}{(4 \pi)^3} - \frac{m^4 S_{-1}(m,T)}{(4 \pi)^4}
\end{equation}
\begin{align}
	V_{1,\ln m} &=\frac{2 m^2}{(4 \pi)^3} S_0(m,T) \left[ \psi(1)+\ln \left( \frac{4 \pi}{m^2} \right) \right]  -\frac{m^4}{2(4 \pi)^4} S_{-1}(m,T) \left[ \psi(2)+\ln \left(\frac{4 \pi}{m^2} \right) \right]  \\
	& + \frac{3}{2} \frac{S_1(m,T)}{(4 \pi)^2}+\frac{m^2 S_0^2(m,T)}{2(4 \pi)^2}+ \frac{m^2}{2(4 \pi)^2}S_1(m,T)S_{-1}(m,T) \nonumber \\
	&+\frac{m^2 S_0(m,T)}{(4 \pi)^3} \left[ 2 - \frac{\sqrt{3} \pi}{3} \right]  - \frac{m^2}{32 \pi^4} Y(m,T)-\frac{m^4}{32 \pi^4} \frac{\partial Y(m,T)}{\partial m^2} \nonumber
\end{align}
Let us define the RGE as
\begin{align}
\frac{d}{d \ln \mu} = \widehat{RGE}=\mu \frac{\partial}{\partial \mu}+ \beta(g) \frac{\partial}{\partial g}-n \gamma(g) + \gamma_m m \frac{\partial}{\partial m}
\end{align}
with 
\begin{align}
\beta(g) &= \beta_2 g^2+ \beta_3 g^3 \\
\gamma(g) & = \gamma_2 g^2 \\
\gamma_m(g) & = \gamma_{m1}g+ \gamma_{m2} g^2
\end{align}
We have to evaluate
\begin{align}\label{EQ. 3-2}
	\widehat{RGE} \ \widetilde{\Gamma} \left( m, g, T, \mu \right) \approx_{TLA} 0 \text{.}
\end{align}
Term by term results are
\begin{align}
	\beta(g) \frac{\partial}{\partial g} \Gamma^{\text{diff}}_{n_k,\vec{k}=0} = g^4 \left\lbrace 2 \beta_3 T_2 \right\rbrace + g^3 \left\lbrace 2 \beta_2 T_2 + \beta_3 T_1 \right\rbrace + g^2 \left\lbrace \beta_2 T_1 \right\rbrace
\end{align}
\begin{align}
	\gamma_m (g) \frac{\partial}{\partial \ln m} \Gamma^{\text{diff}}_{n_k,\vec{k}=0} = g^4 \left\lbrace \gamma_{m2} T_{2,\ln m} \right\rbrace + g^3 \left\lbrace \gamma_{m1}T_{2, \ln m}+\gamma_{m2}T_{1,\ln m} \right\rbrace + g^2 \left\lbrace \gamma_{m1}T_{1,\ln m} \right\rbrace
\end{align}
\begin{align}
	-2 \gamma(g) \Gamma^{\text{diff}}_{n_k,\vec{k}=0} = g^4 \left\lbrace -2 \gamma_2 T_2 \right\rbrace+ g^3 \left\lbrace - 2 \gamma_2 T_1 \right\rbrace
\end{align}
\begin{align}
	\frac{\partial }{\partial \ln \mu} \Gamma^{\text{diff}}_{n_k,\vec{k}=0} = g^2 T_{2,\ln \mu} + g T_{1, \ln \mu}
\end{align}
\begin{align}
	\widehat{RGE} \ \Gamma^{\text{diff}}_{n_k,\vec{k}=0} &= g^4 \left\lbrace 2 \left( \beta_3 - \gamma_2 \right) T_2 + \gamma_{m2} T_{2, \ln m} \right\rbrace \\
	 &+ g^3 \left\lbrace 2 \beta_2 T_2 +(\beta_3 -2 \gamma_2)T_1 + \gamma_{m1} T_{2, \ln m} +\gamma_{m2}T_{1,\ln m} \right\rbrace \nonumber \\
	 &+ g^2 \left\lbrace \beta_2 T_1 + \gamma_{m1} T_{1, \ln m} +T_{2, \ln \mu}\right\rbrace + g T_{1, \ln \mu} \nonumber
\end{align}
On applying \cref{T1,T1m,T1mu,T2mu} on the above equation, it is clear that terms that have coefficient \emph{g}, and $g^2$ are zero.
\begin{align}\label{coupling-2}
\widehat{RGE} \ \Gamma^{\text{diff}} &= 0 \implies g = \frac{A(m,T) \ln \mu + B(m,T)}{C(m,T) \ln \mu + D(m,T)} 
\end{align}
and
\begin{align}
A&=\left[-\gamma_{m1}V_{2, \ln m} -2 \beta_2 V_2(m,T) \right] \\
C&=\left[2(\beta_3-\gamma_2)V_2+\gamma_{m2}V_{2,\ln m} \right] \\
B&=(2 \gamma_2-\beta_3)T_1- 2 \beta_2 V_1(m,T) - \gamma_{m1} V_{1, \ln m} - \gamma_{m2}T_{1, \ln m} \\
D&=2(\beta_3-\gamma_2)V_1+\gamma_{m2}V_{1,\ln m}
\end{align}
We combine with beta coupling constant relation such as

\begin{equation}
\frac{d \ g(\mu)}{d \ \ln(\mu)} = \beta_2 g^2+ \beta_3 g^3
\end{equation}

give rise to the result

\begin{align}\label{mass-scale-2}
\ln(\mu) &=\int^{g} \frac{1}{\beta_2 t^2+ \beta_3 t^3} dt  = - \frac{1}{\beta_2 \ g}+ \frac{\beta_3}{\beta_2^2} \ln \left(\beta_3+ \frac{\beta_2}{g} \right)+\ln \mu_0
\end{align}
Similarly, the corresponding running mass coupling relation is
\begin{equation}\label{runningmass-2}
\frac{d \ \ln(m)}{d \ \ln(\mu)}=\gamma_m (g)
\end{equation}
Combining with the above
\begin{equation}
\frac{\partial  \ \ln(m)}{\partial g } \frac{d g}{d \ \ln(\mu)}=\gamma_m(g) \implies \frac{\partial \ \ln(m)}{\partial g}=\frac{\gamma_m(g)}{\beta(g)}
\end{equation}
Solving by substituting
\begin{equation}
 \frac{\partial \ \ln(m)}{\partial g}= \frac{\gamma_{m1}+\gamma_{m2} g}{\beta_2 g+ \beta_3 g^2}  
\end{equation} 
\begin{align}\label{mass-coupling-2}
\ln \left( \frac{m}{m_0} \right) &=\chi_2 + \frac{\gamma_{m1}}{\beta_2} \ln (g)   + \left( \frac{\gamma_{m2}}{\beta_3}-\frac{\gamma_{m1}}{\beta_2} \right) \ln(\beta_3 \ g+\beta_2) 
\end{align}
$\ln \mu_0$, $m_0$ and $\chi_2$ are the respective integration constants.
Now we have three equations containing mass scale $\mu$, coupling constant $g$,and running mass $m$, so combining \cref{mass-coupling-2,mass-scale-2,coupling-2} and solving simultaneously, we get temperature dependent running mass and coupling constant. \\
\subsection{Coupling $g$ Limit Case T $\to$ 0}\label{couplinglimit}
Here we consider limit case $T \to 0$ at $m \neq 0$.
In order to find the coupling nature of $\mu$ at T=0, $\beta m \to \infty$, we have to find the rate of convergence of $A(m,T)$, $B(m,T)$, $C(m,T)$, $D(m,T)$ as $T \to 0$, because of
\begin{equation}\label{coupling2}
g= \frac{A(m,T) \ln \mu + B(m,T)}{C(m,T) \ln \mu + D(m,T)}
\end{equation}
Here
\begin{equation}
\lim_{\beta m \to \infty} S_N(m,T) \to 0
\end{equation}
Since both numerator and denominator of \cref{coupling2} contain $S_N(m,T)$ with varying N. The rate of convergence of the ratios is important. So
\begin{equation}
\lim_{\beta m \to \infty} \frac{S_{N+1}(m,T)}{S_{N}(m,T)} \to 0
\end{equation}
As N increases, the rate of $S_N(m,T)$ convergence also increases.
The convergence of $Y(m,T)$ can be found and is
\begin{align}
Y(m,T)&=\int_0^\infty \int_0^\infty U(x) U(y) G(x,y) \ dx \ dy \\
U(x)&=\frac{\sinh(x)}{\exp \left( \beta m \cosh(x) \right)-1}\\
	G(x,y)&=\ln \left( \frac{1+2 \cosh(x-y)}{1+2 \cosh(x+y)} \frac{2 \cosh(x+y)-1}{2 \cosh(x-y)-1} \right)  \\
S_N(m,T) &= \frac{1}{\pi}   \sum_{j=1}^\infty \left(\frac{m}{2 \pi j \beta}\right)^{N} K_N(j \beta m)
\end{align}
From the expression of $G(x,y)$ itself, it is clear that $G(x,y)< \ln(3)$. 
So,
\begin{align}
Y(m,T)&< \ln (3) \left[ \int_0^\infty \frac{\sinh(x)}{\exp(\beta m \cosh(x))-1} dx \right]^2 \\
& < \ln (3) \left[ \frac{2 \pi}{m} S_{\frac{1}{2}}(m,T) \right]^2
\end{align}
Therefore
\begin{align}
\lim_{\beta m \to \infty} \frac{Y(m,T)}{S_{-1}(m,T)} \to 0
\end{align}
As a result dividing both numerator and denominator by $S_{-1}(m,T)$  in \emph{g}

\begin{equation} \label{coupling-limit}
\lim_{\beta m \to \infty} g = \lim_{\beta m \to \infty} \frac{\frac{A(m,T)}{S_{-1}(m,T)} \ln \mu +\frac{B(m,T)}{S_{-1}(m,T)}}{\frac{C(m,T)}{S_{-1}(m,T)} \ln \mu +\frac{D(m,T)}{S_{-1}(m,T)}} \approx - \left( \frac{\gamma_{m1}}{\gamma_{m2}} \right)
\end{equation}
where
\begin{align}
\lim_{\beta m \to \infty} \frac{A(m,T)}{S_{-1}(m,T)} \approx & -\gamma_{m1} \times -\frac{m^4}{2(4 \pi)^4} \left[ \psi(2)+ \ln \left( \frac{4 \pi}{m^2} \right) \right] \\
\lim_{\beta m \to \infty} \frac{B(m,T)}{S_{-1}(m,T)} \approx & -\gamma_{m1} \times -\frac{m^4}{(4 \pi)^4} \\
\lim_{\beta m \to \infty} \frac{C(m,T)}{S_{-1}(m,T)} \approx & \gamma_{m2} \times -\frac{m^4}{2(4 \pi)^4} \left[ \psi(2)+ \ln \left( \frac{4 \pi}{m^2} \right) \right] \\
\lim_{\beta m \to \infty} \frac{D(m,T)}{S_{-1}(m,T)} \approx & \gamma_{m2} \times -\frac{m^4}{(4 \pi)^4} 
\end{align}
Now we have other equations relating $\mu$ and $g$ as 
\begin{align}
\ln \left( \frac{\mu}{\mu_0} \right)= \frac{-1}{\beta_2 g}+ \frac{\beta_3}{\beta_2^2} \ln \left( \beta_3+\frac{\beta_2}{g} \right)
\end{align}
When we apply the result \cref{coupling-limit} on the above equations, we get
\begin{equation} \label{mu0}
\lim_{\beta m \to \infty} \ln \left( \frac{\mu}{\mu_0} \right)= \frac{\gamma_{m2}}{\beta_2 \gamma_{m1}}+ \frac{\beta_3}{\beta_2^2} \ln \left( \beta_3-\frac{\gamma_{m2} \beta_2}{\gamma_{m1}} \right) 
\end{equation}
At this approximation, at zero momentum limit, RHS of \cref{mu0} goes to a complex number. If we choose $\mu_0$ in LHS accordingly (i.e., to a complex number/complex function approximation at T=0), one can still make $\mu(T=0)$ a real number. \\

The relation between running mass and coupling relation at $T \to 0$ approximated is
\begin{equation}\label{limitrunningmass}
\lim_{\beta m \to \infty} \ln \left( \frac{m}{m_0} \right)= \chi_2 + \frac{\gamma_{m1}}{\beta_2} \ln \left( \frac{-\gamma_{m1}}{\gamma_{m2}} \right) + \left(\frac{\gamma_{m2}}{\beta_3}-\frac{\gamma_{m1}}{\beta_2} \right) \ln \left(\beta_2 - \frac{\gamma_{m1} \beta_3}{\gamma_{m2}}  \right)
\end{equation} 
At this approximation one can choose running mass $m(T=0)$, as real or complex by intentionally choosing $\chi_2$ and $\ln \mu_0$ accordingly.
\subsection{Pressure $P$ Limit Case T $\to$ 0}\label{limitpressure}

In Fig. 3, we have chosen different values as $T_0$, $P_0$, $\chi_2$, $\ln \mu$. All those results show a similar course. i.e., $T \to \infty$, $P \to P_{ideal}$, irrespective of the initial value.

The quasiparticle model derives energy density from standard statistical mechanics at relativistic Bose-Einstein distribution.
i.e., 
\begin{align}
<\varepsilon>=\int \frac{\sqrt{p^2+m^2}}{\exp(\beta \sqrt{p^2+m^2})-1} \frac{d^3p}{(2 \pi)^3} 
\end{align}
\begin{align}
< \varepsilon > &= \frac{1}{2 \pi^2} \int_0^\infty \frac{p^2 \sqrt{p^2+m^2}}{\exp(\beta \sqrt{p^2+m^2})-1} dp \\
&\text{Put $p=m \sinh x$} \\
&= \frac{m^4}{16 \pi^2} \int_0^\infty \frac{\cosh(4x)-1}{\exp(\beta m \cosh(x))-1} \ dx 
\end{align}
At $T \to 0$, i.e, $\beta m \to \infty$ we have
\begin{align}
\lim_{\beta m \to \infty} <\varepsilon>&= \frac{m^4}{16 \pi^2} \int_0^\infty \left( \cosh(4x)-1 \right) \exp(-\beta m \cosh(x)) \\
&=\frac{m^4}{16 \pi^2}\left(  K_4(\beta m)-K_0(\beta m) \right) \label{zeroenergy}
\end{align}
We have 
\begin{equation}
\lim_{x \to \infty} K(N,x) \to 0 \therefore \lim_{\beta m \to \infty} <\varepsilon>=0
\end{equation}
The equation relating pressure with energy is
\begin{equation}\label{pressureexp}
P(T)=\frac{T}{T_0}P_0+ T \int_{T_0}^T \frac{\varepsilon(T)}{T^2} dT
\end{equation}
In \cref{pressureexp} as $T \to T_0$, the integral becomes zero. (Integral becomes a zero width integral). So at $T \to T_0$, $P \to P_0$. One can choose $P_0$ as negative or positive or zero. We have shown in the figure that irrespective of that value pressure goes to the ideal limit.
In the case of the integrand $\frac{\varepsilon(T)}{T^2}$ at zero temperature limit,
\\Assume $m \neq 0$, and $T \to 0 \implies \beta m \to \infty$, according to \cref{zeroenergy}
\begin{equation}
\lim_{\beta m \to \infty} (\beta m)^2 K_N(\beta m) \to 0
\end{equation}
Since energy density at zero temperature limits reaches zero value. There is no point on  choosing anything other than zero for $P_0$ at $T \to 0$. The value of pressure can be made negative at some points if one chooses $P_0$ as negative at appropriate $T_0$. In our case, we found that it reaches the ideal value at a high-temperature limit, whatever may be the initial value.
\section{Matsubara summation results}
Here we introduce specific Matsubara summation results involved in calculations we have described earlier, and symbols have the usual meaning. Here $\omega_n$ denotes the $2 \pi nT$ where $n$ is the Matsubara frequency, and T is the temperature. $\beta$ indicates the inverse temperature in natural units, i.e., 1/T.
\begin{align}\label{Eq:2}
&T \sum_{n= -\infty}^\infty \frac{1}{\omega_n^2+\varepsilon_{p}^2}=\frac{1}{2 \varepsilon_{p}}+\frac{n_B(\beta \varepsilon_{p})}{\varepsilon_{p}}\\
&T \sum_{n_p= -\infty}^\infty \frac{1}{\omega_{n_p}^2+\varepsilon_p^2} \frac{1}{\omega_{n_p-n_q}^2+\varepsilon^2_{q}}= \left[ t_1(p,q,n_q)+t_2(p,q,n_q)+t_2(q,p,n_q)\right] \\
&T^2\sum_{n_{p}=-\infty}^\infty \sum_{n_{q}=-\infty}^\infty
\frac{1}{\omega_{n_p}^2+\varepsilon_p^2} \frac{1}{\omega_{n_q}^2+\varepsilon_q^2}\frac{1}{\omega_{n_p+n_q+n_r}^2+\varepsilon_{r}^2} =S_1+S_2+S_3 \\
& T^2\sum_{n=-\infty}^{\infty} \sum_{\theta=-\infty}^{\infty} \frac{1}{\omega_n^2+\varepsilon_p^2} \ \frac{1}{\omega_\theta^2+\varepsilon_q^2} \ \frac{1}{\omega_{n-\alpha}^2+\varepsilon_r^2} \ \frac{1}{\omega_{n-\theta+\eta}^2+\varepsilon_s^2} = T^2\sum_{n=-\infty}^{\infty} \sum_{\theta=-\infty}^{\infty} \frac{1}{\omega_n^2+\varepsilon_p^2} \ \frac{1}{\omega_\theta^2+\varepsilon_s^2} \ \frac{1}{\omega_{n-\alpha}^2+\varepsilon_r^2} \ \frac{1}{\omega_{n-\theta+\eta}^2+\varepsilon_q^2}  \\
&=\frac{T^1_{s2}+T^2_{s2}+T^3_{s2}}{16 \varepsilon_p \varepsilon_r \varepsilon_q \varepsilon_s} 
\end{align}
with
\begin{align}
&t_1(p,q,n_r)=\sum_{\sigma = \pm1} \frac{1}{4 \varepsilon_p \varepsilon_{q}} \left( \frac{1}{\varepsilon_p+\varepsilon_{q}+i \sigma \omega_{n_r}} \right) \\
&t_2(p,q,n_r)=\sum_{\sigma, \sigma_1 = \pm 1}\frac{1}{4 \varepsilon_p \varepsilon_{q}} \frac{1}{\sigma_1 \varepsilon_p+\varepsilon_{q}+i \sigma \omega_{n_r}} n_B(\beta \varepsilon_p) \\
\end{align}
\begin{align}
&S_1=\frac{1}{8 \varepsilon_p \varepsilon_q \varepsilon_r} \left(\sum_{\sigma=\pm 1}\frac{1}{\varepsilon_p+\varepsilon_q+\varepsilon_r+i \sigma \omega_{n_{r}}} \right) \\
&S_2= \frac{1}{8 \epsilon_p \epsilon_q \epsilon_r} \sum_{\sigma,\sigma_1=\pm 1} \left(  \frac{n_B(\beta \varepsilon_p)}{\sigma_1 \varepsilon_p+\varepsilon_q+\varepsilon_r+i \sigma \omega_{n_{r}}} 
+  \frac{n_B(\beta \varepsilon_q)}{ \varepsilon_p+\sigma_1 \varepsilon_q+\varepsilon_r+i \sigma \omega_{n_{r}}}
+  \frac{n_B(\beta \varepsilon_r)}{ \varepsilon_p+ \varepsilon_q+\sigma_1 \varepsilon_r+i \sigma \omega_{n_{r}}} \right) \\
&S_3=\frac{1}{8 \varepsilon_p \varepsilon_q \varepsilon_r} \sum_{\sigma,\sigma_1,\sigma_2=\pm 1} \left(  \frac{n_B(\beta \varepsilon_q) n_B(\beta \varepsilon_r)}{ \varepsilon_p+ \sigma_1 \varepsilon_q+\sigma_2 \varepsilon_r+i \sigma \omega_{n_{r}}}
+ \frac{n_B(\beta \varepsilon_p) n_B(\beta \varepsilon_q)}{ \sigma_1 \varepsilon_p+ \sigma_2 \varepsilon_q+ \varepsilon_r+i \sigma \omega_{n_{r}}}
 +  \frac{n_B(\beta \varepsilon_p) n_B(\beta \varepsilon_r)}{ \sigma_1 \varepsilon_p+ \varepsilon_q+ \sigma_2 \varepsilon_r+i \sigma \omega_{n_{r}}} 
 \right) 
 \end{align}
 \begin{align}
	&T_{s2}^1=\sum_{\sigma=\pm 1}\frac{1}{\varepsilon_r+\varepsilon_p-i \sigma \omega_\alpha} \frac{1}{\varepsilon_p+\varepsilon_q+\varepsilon_s+i \sigma \omega_\eta}
	+\sum_{\sigma=\pm 1} \frac{1}{\varepsilon_r+\varepsilon_p+i \sigma \omega_\alpha} \frac{1}{\varepsilon_r+\varepsilon_q+\varepsilon_s+i \sigma \omega_{\eta+\alpha}}  \\
	& \ \ \ \ +\sum_{\sigma=\pm 1} \frac{1}{\varepsilon_r+\varepsilon_q+\varepsilon_s+i \sigma \omega_{\alpha+\eta}} \frac{1}{\varepsilon_p+\varepsilon_q+\varepsilon_s+i \sigma \omega_\eta} \nonumber \\
&T_{s2}^2=t_{21}+t_{22}+t_{23}+t_{24}+t_{25}+t_{26}+t_{27}+t_{28} \\
&{T_{s2}}^3=t_{31}+t_{32}+t_{33}+t_{34}+t_{35}\\
&t_{21}=\sum_{\sigma,\sigma_1,\sigma_2=\pm 1}\frac{1}{\varepsilon_r+\sigma_1(\varepsilon_p-i \sigma \omega_\alpha)} \frac{1}{\varepsilon_q+\varepsilon_s+\sigma_2(\varepsilon_p+i \sigma \omega_\eta)} \ n_B(\beta \varepsilon_p) \\
&t_{22}=\sum_{\sigma,\sigma_1,\sigma_2=\pm 1}\frac{1}{\varepsilon_p+\sigma_1(\varepsilon_r+i \sigma \omega_\alpha)} \frac{1}{\varepsilon_q+\varepsilon_s+\sigma_2(\varepsilon_r+i \sigma \omega_{\eta+\alpha})} \ n_B(\beta \varepsilon_r) \\
&t_{23}=\sum_{\sigma,\sigma_2=\pm 1}\frac{1}{\varepsilon_s+\varepsilon_r+\sigma_2 \varepsilon_q+i \sigma(\omega_{\eta+\alpha})} \frac{1}{\varepsilon_p+\varepsilon_s+\sigma_2 \varepsilon_q+i \sigma \omega_\eta} \ n_B(\beta \varepsilon_q) \\
&t_{24}=\sum_{\sigma,\sigma_2=\pm 1} \frac{1}{\varepsilon_r+\varepsilon_p+i \sigma \omega_\alpha} \frac{1}{\varepsilon_p+\varepsilon_s+\sigma_2 \varepsilon_q-i \sigma \omega_\eta}n_B(\beta \varepsilon_q) \\
&t_{25}=\sum_{\sigma,\sigma_2=\pm 1} \frac{1}{\varepsilon_r+\varepsilon_p+i \sigma \omega_\alpha} \frac{1}{\varepsilon_r+\varepsilon_s+\sigma_2 \varepsilon_q+i \sigma \omega_{\eta+\alpha}}n_B(\beta \varepsilon_q) \\
&t_{26}=\sum_{\sigma,\sigma_2=\pm 1} \frac{1}{\varepsilon_q+\varepsilon_r+\sigma_2 \varepsilon_s+i \sigma \omega_{\eta+\alpha}} \frac{1}{\varepsilon_p+\varepsilon_q+\sigma_2 \varepsilon_s+i \sigma \omega_\eta}n_B(\beta \varepsilon_s)\\
&t_{27}=\sum_{\sigma,\sigma_2=\pm 1} \frac{1}{\varepsilon_r+\varepsilon_p+i \sigma \omega_\alpha} \frac{1}{\varepsilon_p+\varepsilon_q+\sigma_2 \varepsilon_s-i \sigma \omega_\eta}n_B(\beta \varepsilon_s) \\
&t_{28}=\sum_{\sigma,\sigma_2=\pm 1} \frac{1}{\varepsilon_r+\varepsilon_p+i \sigma \omega_\alpha} \frac{1}{\varepsilon_r+\varepsilon_q+\sigma_2 \varepsilon_s+i \sigma \omega_{\eta+\alpha}}n_B(\beta \varepsilon_s)\\
&t_{31}=\sum_{\sigma,\sigma_1,\sigma_2,\sigma_3=\pm 1} \frac{1}{\varepsilon_r+\sigma_1(\varepsilon_p-i \sigma \omega_\alpha)} \frac{1}{\varepsilon_s+\sigma_2 \varepsilon_q+\sigma_3(\varepsilon_p+i\sigma \omega_\eta)}n_B(\beta \varepsilon_p)n_B(\beta \varepsilon_q) \\
&t_{32}=\sum_{\sigma,\sigma_1,\sigma_2,\sigma_3=\pm 1} \frac{1}{\varepsilon_r+\sigma_1(\varepsilon_p-i \sigma \omega_\alpha)} \frac{1}{\varepsilon_q+\sigma_2 \varepsilon_s+\sigma_3(\varepsilon_p+i\sigma \omega_\eta)}n_B(\beta \varepsilon_p)n_B(\beta \varepsilon_s)\\
&t_{33}=\sum_{\sigma,\sigma_1,\sigma_2,\sigma_3=\pm 1} \frac{1}{\varepsilon_p+\sigma_1(\varepsilon_r+i \sigma \omega_\alpha)} \frac{1}{\varepsilon_q+\sigma_2 \varepsilon_s+\sigma_3(\varepsilon_r+i\sigma \omega_{\eta+\alpha})}n_B(\beta \varepsilon_r)n_B(\beta \varepsilon_s)\\
&t_{34}=\sum_{\sigma,\sigma_1,\sigma_2,\sigma_3=\pm 1} \frac{1}{\varepsilon_p+\sigma_1(\varepsilon_r-i \sigma \omega_\alpha)} \frac{1}{\varepsilon_s+\sigma_2 \varepsilon_q+\sigma_3(\varepsilon_r-i\sigma \omega_{\eta+\alpha})}n_B(\beta \varepsilon_r)n_B(\beta \varepsilon_q) \\
&t_{35}=\sum_{\sigma,\sigma_1,\sigma_2,\sigma_3=\pm 1} \frac{1}{\varepsilon_p+\sigma_2 \varepsilon_s+\sigma_1 \varepsilon_q+i \sigma \omega_{\eta}} \frac{1}{\varepsilon_r+\sigma_3(\sigma_2 \varepsilon_s+\sigma_1 \varepsilon_q+i\sigma \omega_{\eta+\alpha})}n_B(\beta \varepsilon_s)n_B(\beta \varepsilon_q) \\
&n_B(x)=(e^x-1)^{-1} \\
&\beta=1/T
\end{align}
\end{widetext}
\bibliographystyle{apsrev4-2}
\bibliography{main}

\end{document}